# Thermodynamic extension of density-functional theory.
# III. Zero-temperature limit of the ensemble spin-density functional theory

Robert Balawender[a]

Institute of Physical Chemistry of Polish Academy of Sciences, Kasprzaka 44/52, PL-01-224 Warsaw, Poland

Abstract.

In this work, the zero-temperature limit of the thermodynamic spin-density functional theory is investigated. The coarse-grained approach to the equilibrium density operator is used to describe the equilibrium state. The characteristic functions of a macrostate are introduced and their zero-temperature limits are investigated. A detailed discussion of the spin-grand-canonical ensemble in the entropy and energy representations is performed. The maps between the state function variables at 0K limit are rigorously studied for both representations. In the spin-canonical ensemble, the energy surface and the discontinuity pattern are investigated. Finally, based on the maps between the state function variables at 0K limit, the Hohenberg-Kohn theorem for the systems with non-integer electron and spin numbers at zero-temperature limit is formulated.

[a] rbalawender@ichf.edu.pl

# I. INTRODUCTION

This paper concludes a three-part series dedicated to the thermodynamic extension of the spin density functional theory (SDFT). In the first paper[1] of the series, hereafter referred to as Paper I, the formulation of the equilibrium state of a many-electron system in terms of an ensemble (mixed-state) density matrix operator, which applies the maximum entropy principle combined with the use of the basic Massieu function, was presented. Based on the effective action formalism, the convexity and concavity properties of the basic Massieu function, its Legendre transforms (Massieu functions) and the corresponding Massieu-Planck transforms (Gibbs-Helmholtz functions) for various ensembles were determined.

In the second paper[2], hereafter referred to as Paper II, the formalism developed in Paper I is applied to two thermodynamic systems: (i) of three global observables (the energy, the total electron number and the spin number), (ii) of one global observable (the internal electron energy) and two local (position-dependent) observables (the total electron density and the spin density). The two-component external potential of the many-electron system of interest consists of a scalar external potential and a collinear magnetic field (coupled in the Hamiltonian with the spin operator only). Various equilibrium characteristics of two systems are defined and investigated. Conditions for the equivalence between two systems (the same equilibrium density matrix for them is required) are derived and thoroughly discussed. The applicability of the Hohenberg-Kohn theorem is extended to the thermodynamic spin-density functional theory.

In the present (third) paper, the results obtained in Paper II are used to provide a rigorous mathematical derivation of the zero-temperature limit of thermodynamic extension of SDFT (TSDFT). Recognizing the behavior of the exact state functions at zero-temperature limit is very helpful for exploring the mathematical properties of exact functionals, as they yield constraints that approximate functionals should satisfy. The major failures in DFT calculations are connected with incorrect behavior of approximate functionals for fractional charges and fractional spins. In the case of the energy functional, the exact condition was derived in a form of the so called flat-plane condition. The exact functional should give piecewise-flat-plane energies as a function of a fractional electron number and spin number.[3] When the fulfilling of the flat-plane condition becomes an important criterion in the development of functionals, its correct formulation is essential. The answer to the question "how the energy depends on the electron number?" can be done in two ways. One basic idea considers $Q$ replicas of an $N$-electron molecule, with the replicas spaced far enough apart so that the interaction between them can be neglected. When all of the replicas are identical (they have the same energy), after addition of $P$ extra electrons, the energy of a replica can be viewed as the energy of the system with $N + P/Q$ electrons.[4] This analysis was extended to irrational numbers.[5] Based on this idea, certain desired properties which an approximate functional should obey, were formulated.[4,6] This methodology was successfully used in the examination of the dissociation limit for the hydrogen molecule[3] and in the construction of new self-interaction free functionals.[7,8] These results are consistent with the zero-temperature limit of the ensemble theory introduced by Perdew et al.[9], the second way. In this approach the fractional electron number is treated as the average value of the electron number operator in the Fock space. Extension to simultaneously occurring a fractional charge and a fractional spin was proposed.[10-12] In this paper we focus on the zero-temperature limit of the TSDFT for interacting electron system in the spirit of the second way. However, the subjects of the foundation,[13] exact conditions,[14] scaling properties,[15] and the existence of the functional derivative in a Kohn-Sham-type formulation of TSDFT are not considered.

This paper is organized as follows. The properties of the functionals defined in Paper II are discussed using the coarse-grained approach. In Sec.II, microstate, mesostate and macrostate resolutions are introduced and followed by a discussion of the entropy properties in these resolutions. In Sec.III, the macrostate characteristic functions are defined and their zero-temperature limits are investigated. The contact with the traditional DFT formulation is established. In Sec.IV, the spin-grand-canonical ensemble at 0K limit in the entropy and the energy representations is discussed. Derivative discontinuities of the state function at 0K limit are investigated. In Sec.V, the zero-temperature limit of the spin-canonical ensemble is intensively studied. The energy surface, discontinuity pattern, external sources conjugate to the electron and spin numbers and macrostate weights at 0K limit are thoroughly investigated. Finally, the maps between different independent variables are set up to reformulate the Hohenberg-Kohn theorem for the systems with non-integer electron and spin numbers at zero-temperature limit.

Note: Atomic units are used throughout the paper. The notation for specific vectors is also introduced (see Section II): a two-component column vector (2vec) of quantities like the density, the external potential, electron number, sources etc., is denoted by a bold italic letter: $\boldsymbol{\rho}$, $\boldsymbol{v}$, $\boldsymbol{\mathcal{N}}$, $\boldsymbol{\mu}$, etc.; the "transpose" mark at 2vec is omitted, e.g., the scalar product denoted $\boldsymbol{\mu}\boldsymbol{\mathcal{N}}$ means $\boldsymbol{\mu}^{\mathrm{T}}\boldsymbol{\mathcal{N}}$ or $\boldsymbol{\mathcal{N}}^{\mathrm{T}}\boldsymbol{\mu}$; a vector in the configuration space is denoted by a non-italic, bold letter, e.g. $\mathbf{r}$, $\mathbf{R}_a$. The integral functional $\int[...]$ is introduced to represent the integration over the whole space of a function of a spatial variable, $\int[f] \equiv \int d\mathbf{r}\, f(\mathbf{r})$. The symbol $\doteq$ is used to denote the zero-temperature limit, $f[\beta] \doteq g$ or $g \doteq f[\beta]$ means $\lim_{\beta\to\infty} f[\beta] = \lim_{T\to 0} f[1/T] = g$. The abbreviation "w.r.t." means "with respect to", "fn." means "function", "fnl." means "functional".

As an illustrative example, the carbon atom and its ions were chosen. All calculations were performed using the GAMESS program with the diffuse, aug-cc-pVTZ[16] basis set and applying PBE[17] exchange-correlation functional. Since bound results obtained for some anions are inconsistent with the real (unbound) carbon anions, our illustrative example should be treated as a carbon-like model.

## II. COARSE-GRAINED APPROACH TO THE EQUILIBRIUM DENSITY MATRIX OPERATOR

A general formulation of the equilibrium state of a many-electron system defined by its Hamiltonian $\hat{H}$, in terms of an ensemble density matrix (DM) operator in the Fock space was introduced in Paper I. The equilibrium density matrix (eq-DM) operator $\hat{\Gamma}_{\mathrm{eq}}$, which completely describes the equilibrium state, was determined by the principle of the maximum for the informational entropy, based on the phenomenological description.[18,19] The eq-DM operator, $\hat{\Gamma}_{\text{eq}}$, can be expressed in terms of its eigenvalues and eigenvectors as

$$\begin{aligned}\hat{\Gamma}_{\text{eq}} &\equiv \hat{\Gamma}\left[\left\{p_K^{\text{eq}}\right\},\left\{\Psi_K\right\}\right] \\ &= \left\{\sum_K \left|\Psi_K\right\rangle p_K^{\text{eq}} \left\langle\Psi_K\right| \,\middle|\, 0 \le p_K^{\text{eq}},\, \left\langle\Psi_K\middle|\Psi_L\right\rangle = \delta_{KL},\, \text{Tr}\,\hat{\Gamma}_{\text{eq}} = \sum_K p_K^{\text{eq}} = 1\right\}.\end{aligned} \tag{1}$$

The set of eigenvectors $\left\{\left|\Psi_K\right\rangle\right\}$ is complete in the Fock-space (this set was denoted as $\left\{\left|\Psi_K^{\text{cev}}\right\rangle\right\}$ in Paper I, see the text above Eq.(I.18), i.e., Eq.(18) of Paper I).

The pure state defined by the eigenvector $\left|\Psi_K\right\rangle$ can be, in fact, labeled with a set of eigenvalues $\left\{o_j^K\right\}$ of a complete set of compatible observables $\left\{\hat{O}_j\right\}^{\text{full}}$ – the generating operators that specify the system of interest (all these observables are assumed to be time-independent, hermitian, local (**r**-dependent) or global (**r**-independent), and mutually commuting $\left[\hat{O}_i,\hat{O}_j\right]_- = 0$). Such pure states are called also microstates. Knowing that the system is found in a state $K$ requires very detailed information about its internal state. In practice, we possess limited information or we restrict ourselves only to some chosen subset of the complete set of observables, $\left\{\hat{O}_j\right\} \equiv \left\{\hat{O}_j\right\}^{\text{subset}}$, which incorporates the sufficient amount of information necessary for a mesoscopic characterization of the equilibrium state, e.g. an electronic term groups all those microstates which have the same energy, the total-angular and total-spin momentum, etc. In Paper II, one of the considered thermodynamic systems was characterized by the subset $\left\{\hat{O}_j\right\} = \left\{\hat{H},\hat{\boldsymbol{\mathcal{N}}}\right\}$ (a set of three global observables), where $\hat{H}[\boldsymbol{v}]$ is a non-relativistic, time-independent Hamiltonian (see Eq.(II.1)), the 2vec $\hat{\boldsymbol{\mathcal{N}}} = \left(\hat{\mathcal{N}},\hat{\mathcal{S}}\right)$ $= \left(\hat{\mathcal{N}}_\uparrow + \hat{\mathcal{N}}_\downarrow, \hat{\mathcal{N}}_\uparrow - \hat{\mathcal{N}}_\downarrow\right)$; its components are: $\hat{\mathcal{N}}$, the (total) particle-number operator, and $\hat{\mathcal{S}} = 2\hat{S}_{\mathrm{z}}$, termed the spin-number operator ($\hat{S}_{\mathrm{z}}$ is the $z$-component of the total-spin vector operator $\hat{\mathbf{S}} = \left(\hat{S}_{\mathrm{x}}, \hat{S}_{\mathrm{y}}, \hat{S}_{\mathrm{z}}\right)$ of the many-electron system). The $\left\{\hat{H}, \hat{\boldsymbol{\mathcal{N}}}\right\}$ system satisfies conditions $\mathrm{Tr}\,\hat{\Gamma}_{\mathrm{eq}}\hat{H} = E$ and $\mathrm{Tr}\,\hat{\Gamma}_{\mathrm{eq}}\hat{\boldsymbol{\mathcal{N}}} = \boldsymbol{\mathcal{N}}$ when it is fully isolated. Here $E$ and $\boldsymbol{\mathcal{N}} = (\mathcal{N}, \mathcal{S})$ are the average energy and the average electron number 2vec, respectively. When fully open, the system is characterized by the conditions operator (Eq.(II.4)): $\hat{O} \equiv \hat{O}[\beta, \boldsymbol{\alpha}; \boldsymbol{v}] = -\beta\hat{H}[\boldsymbol{v}] - \boldsymbol{\alpha}\hat{\boldsymbol{\mathcal{N}}}$, where $\beta$ and $\boldsymbol{\alpha} = (\alpha_{\mathrm{N}}, \alpha_{\mathrm{S}})$ are the Lagrange multipliers (sources) conjugate to $\hat{H}$ and $\hat{\boldsymbol{\mathcal{N}}}$. We conveniently reduced the general dependence of $\hat{O}$ on $\left\{\hat{H}, \hat{\boldsymbol{\mathcal{N}}}\right\}$ to the dependence on the 2vec $\boldsymbol{v}(\mathbf{r})$ — the only system-specific characteristic of the observable operators $\left\{\hat{H}, \hat{\boldsymbol{\mathcal{N}}}\right\}$. Besides the system characterized by $\left\{\hat{H}, \hat{\boldsymbol{\mathcal{N}}}\right\}$, the second thermodynamic system characterized by $\left\{\hat{F}_{\mathrm{int}}, \hat{\boldsymbol{\rho}}(\mathbf{r})\right\}$, suitable for spin-density description, was introduced in Paper II, where $\hat{F}_{\mathrm{int}}$ and $\hat{\boldsymbol{\rho}}(\mathbf{r}) = \left(\hat{\rho}_{\mathrm{N}}(\mathbf{r}), \hat{\rho}_{\mathrm{S}}(\mathbf{r})\right)$ are the internal electron energy operator and the density 2vec operator, respectively.

The microstate $|\Psi_K\rangle$, convenient for description of the above mentioned system $\left\{\hat{H}, \hat{\boldsymbol{\mathcal{N}}}\right\}$, besides being an eigenfunction of the Hamiltonian $\hat{H}$, is simultaneously an eigenfunction of the total particle-number operator $\hat{\mathcal{N}}$, the spin-number operator $\hat{\mathcal{S}} = 2\hat{S}_{\mathrm{z}}$, and the squared spin momentum operator $\hat{\mathbf{S}}^2 = \hat{S}_{\mathrm{x}}^2 + \hat{S}_{\mathrm{y}}^2 + \hat{S}_{\mathrm{z}}^2$. In addition, $|\Psi_K\rangle$ is a symmetry partner of some irreducible representation of the symmetry group (in the configuration space). This group is dictated by the symmetry of the potential $\boldsymbol{v}(\mathbf{r}) = \left(v_{\mathrm{ext}}(\mathbf{r}), B_{\mathrm{z}}(\mathbf{r})\right)$.[20] In the case of atoms at $B_{\mathrm{z}}(\mathbf{r}) \equiv 0$, due to the spherical symmetry of $\boldsymbol{v}(\mathbf{r})$, $|\Psi_K\rangle$ is also an eigenfunction of the angular momentum operators $\left(\hat{\mathbf{L}}^2, \hat{L}_z\right)$ (for the non-relativistic Hamiltonian used here, spin momentum operators commutate with the angular momentum operators).

The microstates can be grouped into non-overlapping subsets named mesostates: the *mesostate* is a set of microstates which satisfy the condition

$$(i) \equiv \left\{ |\Psi_K\rangle \middle| \ \hat{O}_{\mathsf{F}} |\Psi_K\rangle = o_{\mathsf{F}}^{(i)} |\Psi_K\rangle \right\}. \tag{2}$$

Here $o_{\mathsf{F}}^{(i)} = \left(E_{(i)}^{[v]}, \mathcal{N}_{(i)}, \mathcal{S}_{(i)}\right)$, which is specific for the $(i)$ mesostate, denotes a vector of particular eigenvalues of the $\hat{O}_{\mathsf{F}} = \left(\hat{H}[\boldsymbol{v}], \hat{\mathcal{N}}, \hat{\mathcal{S}}\right)$ — the set of generating operators of the considered thermodynamic system (see Appendix A of Paper I for the notation $x_{\mathsf{F}}$).

It is convenient also to use a resolution coarser than mesostate. The mesostates can be grouped into non-overlapping sets named macrostates: the *macrostate* is a set of mesostates, which satisfy the condition

$$(I) \equiv \left\{ (i) \middle| \left(\mathcal{N}_{(i)}, \mathcal{S}_{(i)}\right) = \left(\mathcal{N}_{(I)}, \mathcal{S}_{(I)}\right) \right\}. \tag{3}$$

Here $\boldsymbol{\mathcal{N}}_{(I)} = \left(\mathcal{N}_{(I)}, \mathcal{S}_{(I)}\right)$, which is specific for the $(I)$ macrostate, denotes a vector of particular eigenvalues of $\hat{\boldsymbol{\mathcal{N}}} = \left(\hat{\mathcal{N}}, \hat{\mathcal{S}}\right) \subset \left\{\hat{H}, \hat{\mathcal{N}}, \hat{\mathcal{S}}\right\}$. It should be noted that for a given particle number eigenvalue $\mathcal{N}_{(I)} \in \{0, 1, 2, ...\}$, the allowed values of $\left(\mathcal{N}_{(I)} + 1\right)$ spin numbers are $\mathcal{S}_{(I)} \in \left\{-\mathcal{N}_{(I)}, -\mathcal{N}_{(I)} + 2, \ldots, \mathcal{N}_{(I)} - 2, \mathcal{N}_{(I)}\right\}$. The macrostate $(I)$ characterized by such allowed values $\boldsymbol{\mathcal{N}}_{(I)} = \left(\mathcal{N}_{(I)}, \mathcal{S}_{(I)}\right)$ will be termed *admissible*. An alternative definition of the $(I)$ macrostate, equivalent to one given in Eq.(3), is a set of microstates that satisfy

$$(I) \equiv \left\{ |\Psi_K\rangle \middle| \ \left(\hat{\mathcal{N}}, \hat{\mathcal{S}}\right) |\Psi_K\rangle = \left(\mathcal{N}_{(I)}, \mathcal{S}_{(I)}\right) |\Psi_K\rangle \right\}. \tag{4}$$

To establish connection between the thermodynamic description in various resolutions (in terms of macrostates, mesostates or microstates), we introduce a macrostate projection operator and the macrostate DM operator. The $(I)$ macrostate projection operator defined as

$$\hat{g}_{(I)} \equiv \sum_{K\in(I)} |\Psi_K\rangle\langle\Psi_K| \tag{5}$$

projects the Fock space onto the subspace of state vectors with the particular eigenvalues of $\hat{\mathcal{N}}$ equal to $\left(\mathcal{N}_{(I)},\mathcal{S}_{(I)}\right)$. The projectors $\left\{\hat{g}_{(I)}\right\}$ are pairwise orthogonal and idempotent: $\hat{g}_{(I)}\hat{g}_{(J)} = \delta_{(I)(J)}\hat{g}_{(I)}$. They are mutually commuting with $\hat{H},\hat{\mathcal{N}}$. The sum of all projection operators equals the identity operator, $\sum_I \hat{g}_{(I)} = \hat{1}$. The $(I)$ macrostate set of DM operators is defined as

$$\left\{\hat{\Gamma}_{(I)}\right\} \equiv \left\{ \sum_{K\in(I)} p_K^{(I)} |\Psi_K\rangle\langle\Psi_K| \;\middle|\; 0 \le p_K^{(I)}, \sum_{K\in(I)} p_K^{(I)} = 1 \right\} \tag{6}$$

(the curly brackets denote a set). Eq.(6) defines a set of operators because various sets of $\left\{p_K^{(I)}\right\}$ can be chosen and the subspace $\left\{|\Psi_K\rangle, K\in(I)\right\}$, Eq.(4), is defined with accuracy to a unitary transformation. The macrostate DM operators are pairwise orthogonal, but not idempotent: $\hat{\Gamma}_{(I)}\hat{\Gamma}_{(J)} = \delta_{IJ}\hat{\Gamma}_{(I)}\hat{\Gamma}_{(I)}$, $\hat{\Gamma}_{(I)}\hat{\Gamma}_{(I)} \ne \hat{\Gamma}_{(I)}$. The product of the macrostate projection operator and the DM operator is $\hat{g}_{(I)}\hat{\Gamma}_{(J)} = \delta_{IJ}\hat{\Gamma}_{(J)}$. An arbitrary DM operator can be resolved into the macrostate DM operators as $\hat{\Gamma} = \hat{1}\hat{\Gamma} = \sum_I \hat{g}_{(I)}\hat{\Gamma} = \sum_I p_{(I)}\hat{\Gamma}_{(I)}$ with $p_{(I)} \ge 0$ and $\sum_I p_{(I)} = 1$.This resolving can be applied to a wide class of thermodynamic systems: their generating operators need to commute with $\hat{\mathcal{N}}$. Analogous definitions and relations can be written for mesostates and microstates.

We can describe the considered system at equilibrium with the highest resolution as being in any one of its microstates (pure states) $K$ with probability

$$p_K^{\text{eq}} \equiv p_K\left[\{\Psi_L\},\hat{O}\right] = \exp\left\langle \Psi_K \left|\hat{O}\right|\Psi_K \right\rangle \Big/ \sum_L \exp\left\langle \Psi_L \left|\hat{O}\right|\Psi_L \right\rangle \tag{7}$$

(see the text above Eq.(I.18)), or, alternatively, with a lower resolution, as being in any one of the macrostates $(I)$ with probability $p_{(I)}^{\text{eq}} \equiv \omega_{(I)}$, or with an intermediate resolution in terms of mesostates.

The connection between the thermodynamic descriptions on different levels can be established using the maximum entropy principle by rearranging eq-DM operator (determined by this principle), Eq.(1) with Eq.(7), namely

$$\begin{aligned}\hat{\Gamma}_{\text{eq}} &= \sum_K \left|\Psi_K\right\rangle p_K^{\text{eq}} \left\langle \Psi_K\right| = \sum_I \omega_{(I)} \sum_i p_{(i|I)} \sum_K \left|\Psi_K\right\rangle p_{(K|i)} \left\langle \Psi_K\right| = \\ &= \sum_I \omega_{(I)} \sum_i p_{(i|I)} \hat{\Gamma}_{(i)} = \sum_i p_{(i)} \hat{\Gamma}_{(i)} = \sum_I \omega_{(I)} \hat{\Gamma}_{(I)},\end{aligned} \tag{8}$$

All objects in the above equation correspond to equilibrium state, but their label "eq" is now suppressed for brevity. Here $p_K^{\text{eq}}$, $p_{(i)}$, $\omega_{(I)}$ are the probabilities that the system is found in the $K$ microstate (pure state), the $(i)$ mesostate, and the $(I)$ macrostate, respectively. Note that $\omega_{(I)} = \operatorname{Tr} \hat{g}_{(I)} \hat{\Gamma}_{\text{eq}} = \sum_{K\in(I)} p_K^{\text{eq}}$ satisfies $\sum_I \omega_{(I)} = 1$.

If $p_{(i)} > 0$, $p_{(K|i)}$ denotes the conditional probability that the system is in the microstate (pure state) $K$ provided we know it is in the mesostate $(i)$:

$$p_{(K|i)} \equiv \frac{\operatorname{Tr} \hat{g}_K \hat{g}_{(i)} \hat{\Gamma}_{\text{eq}}}{\operatorname{Tr} \hat{g}_{(i)} \hat{\Gamma}_{\text{eq}}} = \begin{cases} p_K^{\text{eq}} \big/ p_{(i)} & \text{for } K \in (i) \\ 0 & \text{for } K \notin (i) \end{cases}, p_{(i)} = \sum_{K\in(i)} p_K^{\text{eq}}, \sum_K p_{(K|i)} = 1. \tag{9}$$

We used here the $K$ microstate projection operator $\hat{g}_K = \left|\Psi_K\right\rangle\left\langle\Psi_K\right|$, and the $(i)$ mesostate projection operator $\hat{g}_{(i)} = \sum_{K\in(i)} \hat{g}_K$.

If $\omega_{(I)} > 0$, the probability $p_{(i|I)}$ denotes the conditional probability that the system is in the $(i)$ mesostate provided we know it to be in the $(I)$ macrostate:

$$p_{(i|I)} \equiv \frac{\mathrm{Tr}\,\hat{g}_{(i)}\hat{g}_{(I)}\hat{\Gamma}_{\mathrm{eq}}}{\mathrm{Tr}\,\hat{g}_{(I)}\hat{\Gamma}_{\mathrm{eq}}} = \begin{cases} p_{(i)}/\omega_{(I)} & \text{for}\,(i)\in(I) \\ 0 & \text{for}\,(i)\notin(I) \end{cases}, \ \omega_{(I)} = \sum_{(i)\in(I)} p_{(i)} = 1 \ \ \sum_i p_{(i|I)} = 1\,. \quad (10)$$

Thus, the eq-DM operators describing the system on the meso and macro level are $\hat{\Gamma}_{(i)} = \sum_K p_{(K|i)}\hat{g}_K$ and $\hat{\Gamma}_{(I)} = \sum_i p_{(i|I)}\hat{\Gamma}_{(i)}$. Note that $\omega_{(I)}p_{(i|I)}p_{(K|i)} = p_K^{\mathrm{eq}},\ \forall K\in(i), \forall(i)\in(I)$.

The Boltzmann-Gibbs-Shannon (BGS) entropy — the expectation value of the natural logarithm of the DM operator $\hat{\Gamma}$ (with a minus sign; the Boltzmann constant is omitted because we use temperature expressed in energy unit) — $S^{\mathrm{ent}}\left[\hat{\Gamma}\right] = S^{\mathrm{BGS}}\left[\hat{\Gamma}\right] \equiv -\mathrm{Tr}\,\hat{\Gamma}\ln\hat{\Gamma}$, can be rearranged into (see Eq.(8))

$$\begin{aligned} S^{\mathrm{BGS}}\left[\hat{\Gamma}_{\mathrm{eq}}\right] &= -\sum_K p_K^{\mathrm{eq}}\ln p_K^{\mathrm{eq}} = -\sum_I\sum_i\sum_K \left(\omega_{(I)}p_{(i|I)}p_{(K|i)}\right)\ln\left(\omega_{(I)}p_{(i|I)}p_{(K|i)}\right) \\ &= -\sum_I \omega_{(I)}\left(\ln\omega_{(I)} + \sum_i p_{(i|I)}\left(\ln p_{(i|I)} + \sum_K p_{(K|i)}\ln p_{(K|i)}\right)\right) \\ &= -\sum_I \omega_{(I)}\left(\ln\omega_{(I)} + \sum_i p_{(i|I)}\left(\ln p_{(i|I)} - S^{\mathrm{BGS}}\left[\hat{\Gamma}_{(i)}\right]\right)\right) \\ &= -\sum_I \omega_{(I)}\left(\ln\omega_{(I)} - S^{\mathrm{BGS}}\left[\hat{\Gamma}_{(I)}\right]\right), \end{aligned} \quad (11)$$

where $S^{\mathrm{BGS}}\left[\hat{\Gamma}_{(i)}\right] = -\sum_K p_{(K|i)}\ln p_{(K|i)}$ and $S^{\mathrm{BGS}}\left[\hat{\Gamma}_{(I)}\right] = -\sum_i p_{(i|I)}\left(\ln p_{(i|I)} - S^{\mathrm{BGS}}\left[\hat{\Gamma}_{(i)}\right]\right)$ are the mesostate and macrostate entropy, respectively.

Microstates which belong to the mesostate $(i)$ have the same value of $\alpha_{\mathsf{F}}^{\mathrm{T}}o_{\mathsf{F}}^{(i)} = (\beta, \alpha_{\mathrm{N}}, \alpha_{\mathrm{S}})\left(E_{(i)}^{[\nu]}, N_{(i)}, S_{(i)}\right)^{\mathrm{T}}$ (the sources $\alpha_j$, corresponding to the remaining quantum numbers of $K$ microstate, $o_j^K$, are taken zero). Therefore these microstates must have equal weights (conditional probabilities)

$$p_{(K|i)} = 1 \big/ d_{(i)}^{[v]}\,, \tag{12}$$

and $d_{(i)}^{[v]} = \sum_{K\in(i)} 1$ is the number of microstates belonging to the given mesostate (i.e., a degeneracy of $(i)$ mesostate energy). It is equal to the dimension of the irreducible representation of the symmetry group connected with the given $E_{(i)}^{[v]}$. The entropy assigned to the $(i)$ mesostate is equal to the Boltzmann entropy (note that value of this entropy depends neither on $\alpha_\mathsf{F}$ nor on $o_\mathsf{F}$)

$$\begin{aligned} S_{(i)}^{[v]} = S^{\mathrm{BGS}}\left[\hat{\Gamma}_{(i)}\right] &\equiv -\mathrm{Tr}\,\hat{\Gamma}_{(i)} \ln \hat{\Gamma}_{(i)} = -\sum_{K\in(i)} p_{(K|i)} \ln p_{(K|i)} \\ &= -\frac{1}{d_{(i)}^{[v]}} \ln\left(\frac{1}{d_{(i)}}\right) \sum_{K\in(i)} 1 = \ln d_{(i)}^{[v]}. \end{aligned} \tag{13}$$

For the carbon atom, when the magnetic field is switched off, the $(i)$ mesostate connected with the lowest-energy solution of the Schrödinger equation with $\mathcal{N}_{(i)} = (6,2)$, includes the following microstates: ${}^{3}\mathrm{P}_{-1,1}$, ${}^{3}\mathrm{P}_{0,1}$, ${}^{3}\mathrm{P}_{+1,1}$ (${}^{2S+1}L_{L_\mathrm{z},S_\mathrm{z}}$ notation is used). Due to the spherical symmetry of the 2vec potential, the ground state energy is the same for these microstates, so $d_{(i)}$ is equal to the orbital multiplicity $\left(2L_{(i)}+1\right)=3$, where for $K\in(i)$, $\hat{\mathbf{L}}^2\left|\Psi_K\right\rangle = L_{(i)}\left(L_{(i)}+1\right)\left|\Psi_K\right\rangle$). Moreover, this energy is the same also for states with $S_\mathrm{z}=0$ and $S_\mathrm{z}=-1$, i.e., mesostates with $\mathcal{N}_{(i)}=(6,0)$ and $(6,-2)$. The degeneracy of the mesostate can be changed by magnetic field. The magnetic field introduced in Paper II is assumed to be weak, nonuniform, linear (along *z*-direction) field, $B_\mathrm{x}(\mathbf{r})=0$, $B_\mathrm{y}(\mathbf{r})=0$, $B_\mathrm{z}(\mathbf{r})\neq \mathrm{const.}$ With a cylindrical magnetic field, $B_\mathrm{z}(\mathbf{r}) = \tilde{B}_\mathrm{z}\left(\sqrt{x^2+y^2}, z\right)$, switched on, the threefold degenerate mesostate splits into twofold mesostates: the nondegenerate one (from the former ${}^{3}\mathrm{P}_{0,1}$ microstate) and the degenerate one (from two former ${}^3\mathrm{P}_{-1,1}$ and ${}^3\mathrm{P}_{+1,1}$ microstates). The energy ordering of the new mesostates depends on the shape of the spatial dependence of $\tilde{B}_\mathrm{z}$. For $B_\mathrm{z}(\mathbf{r})$ of lower symmetry, the previous degeneracy is removed.

Using the macrostate projection operators and applying $\hat{\Gamma}_\mathrm{eq} = \sum_I \omega_{(I)} \hat{\Gamma}_{(I)}^{[\beta,v]}$ to $S^\mathrm{BGS}\left[\hat{\Gamma}\right] = -\mathrm{Tr}\,\hat{\Gamma}\ln\hat{\Gamma}$, the entropy Eq.(11), can be rewritten as

$$S^\mathrm{BGS}\left[\hat{\Gamma}_\mathrm{eq}\right] = -\sum_I \omega_{(I)} \ln \omega_{(I)} + \sum_I \omega_{(I)} S_{(I)}^{[\beta,v]}, \tag{14}$$

where the entropy assigned to the $(I)$ macrostate is

$$S_{(I)}^{[\beta,v]} \equiv S^\mathrm{BGS}\left[\hat{\Gamma}_{(I)}^{[\beta,v]}\right] = -\mathrm{Tr}\,\hat{\Gamma}_{(I)}^{[\beta,v]} \ln \hat{\Gamma}_{(I)}^{[\beta,v]}. \tag{15}$$

The above result, Eq.(14), is consistent with Araki-Lieb inequality[21]: $S^\mathrm{BGS}\left[\sum \omega_{(I)} \hat{\Gamma}_{(I)}\right] - \sum \omega_{(I)} S^\mathrm{BGS}\left[\hat{\Gamma}_{(I)}\right] \geq 0$, and with the conclusion that the coarse partition has always smaller Shannon entropy. The "grouping" properties, Eq.(11), play an important role because they establish a relation between different descriptions and, in doing so, they invoke different entropies.

# III. CHARACTERISTIC FUNCTIONS OF THE MACROSTATE AND THEIR ZERO-TEMPERATURE LIMIT IN TSDFT

## A. Macrostate equilibrium DM operators and weight functions

In analogy with the state functions and functionals defined for the system in the whole Fock space (see Fig. 1), we will define the state functions and functionals for a particular $(I)$ macrostate. Such macrostate is characterized by the set of $(I)$ macrostate (non-equilibrium) DM operators, Eq.(6), rewritten in mesostate resolution as

$$\left\{\hat{\Gamma}_{(I)}\right\} \equiv \left\{ \sum_i p_{(i,I)} \hat{\Gamma}_{(i,I)} \middle| 0 \le p_{(i,I)}, \mathrm{Tr}\,\hat{\Gamma}_{(I)} = \sum_i p_{(i,I)} = 1 \right\}, \tag{16}$$

where the label $(i)$, numbering previously all mesostates in Eq.(2) (irrespectively of macrostate numbering), is now replaced by the pair $(i,I)$ which indicates the $(i)$ mesostate belonging to the $(I)$ macrostate

$$(i,I) \equiv \left\{ \left|\Psi_K^{[\nu]}\right\rangle \,\middle|\, \left(\hat{H},\hat{\mathcal{N}},\hat{\mathcal{S}}\right)\left|\Psi_K^{[\nu]}\right\rangle = \left(E_{(i,I)}^{[\nu]},\mathcal{N}_{(I)},\mathcal{S}_{(I)}\right)\left|\Psi_K^{[\nu]}\right\rangle \right\}. \tag{17}$$

For each $(I)$, the label $i \in \{0,1,2,...\}$ numbers energy levels so that $E_{(0,I)}^{[\nu]} < E_{(1,I)}^{[\nu]} < E_{(2,I)}^{[\nu]} < \cdots$. Thus, under the new labeling, the three specific eigenvalues $\left(E_{(i,I)}^{[\nu]},\mathcal{N}_{(I)},\mathcal{S}_{(I)}\right)$ characterize the $(i,I)$ mesostate. In Eq.(16), $\hat{\Gamma}_{(i,I)}$ belongs to the set of $(i,I)$ mesostate DM operators, defined similarly to the set of $(I)$ macrostate DM operators, Eq.(6), but with $(I)$ replaced by $(i,I)$.

The (global) eq-DM operator, Eq.(I.8) with Eq.(I.9), can be rearranged as follows

$$\hat{\Gamma}_{\mathrm{eq}}^{[\hat{o}]} = \frac{\hat{1}\exp\left(\hat{O}\right)}{\mathrm{Tr}\exp\left(\hat{O}\right)} = \sum_I \frac{\hat{g}_{(I)}\exp\left(\hat{O}\right)}{\mathrm{Tr}\exp\left(\hat{O}\right)} = \sum_I \frac{\mathrm{Tr}\,\hat{g}_{(I)}\exp\left(\hat{O}\right)}{\mathrm{Tr}\exp\left(\hat{O}\right)} \frac{\hat{g}_{(I)}\exp\left(\hat{O}\right)}{\mathrm{Tr}\,\hat{g}_{(I)}\exp\left(\hat{O}\right)} = \sum_I \omega_{(I)}^{[\hat{o}]} \hat{\Gamma}_{(I)}^{[\hat{o}]} \tag{18}$$

(see Eq.(5) for $\hat{g}_{(I)}$). According to Eq.(18), the *macrostate eq-DM* is

$$\hat{\Gamma}_{(I)}^{[\hat{o}]} = \frac{\hat{g}_{(I)}\exp\left(\hat{O}\right)}{\mathrm{Tr}\,\hat{g}_{(I)}\exp\left(\hat{O}\right)}, \tag{19}$$

and *the macrostate equilibrium weight* is

$$\omega_{(I)}^{[\hat{o}]} = \frac{\mathrm{Tr}\,\hat{g}_{(I)}\exp\left(\hat{O}\right)}{\mathrm{Tr}\exp\left(\hat{O}\right)}. \tag{20}$$

The operator $\hat{O}$ can be the spin-grand-canonical or spin-canonical conditions operator.

In the entropy representation for the $\left\{\hat{H},\hat{\mathcal{N}}\right\}$ system, the above mentioned operator $\hat{O}$ is $\hat{O}[\beta,\boldsymbol{\alpha};\boldsymbol{v}]=\left(-\beta\hat{H}-\boldsymbol{\alpha}\hat{\mathcal{N}}\right)$ or $\hat{O}\left[\beta,\mathcal{N};\boldsymbol{v}\right]=\left(-\beta\hat{H}-\breve{\boldsymbol{\alpha}}^{\left[\beta,\mathcal{N};\boldsymbol{v}\right]}\left(\hat{\mathcal{N}}-\mathcal{N}\right)\right)$, respectively (see Eqs.(II.4) and (II.17)).

When $\hat{\Gamma}_{(I)}^{[\hat{O}]}$ is evaluated from Eq.(18), cancellation of the exponential factor containing $\boldsymbol{\alpha}$ or $\breve{\boldsymbol{\alpha}}$ in the numerator with the same factor in the denominator for both conditions operators occurs, so the result is

$$\hat{\Gamma}_{(I)}^{[\beta,\boldsymbol{v}]}=\frac{\hat{g}_{(I)}\exp\left(-\beta\hat{H}[\boldsymbol{v}]\right)}{\mathrm{Tr}\,\hat{g}_{(I)}\exp\left(-\beta\hat{H}[\boldsymbol{v}]\right)}. \tag{21}$$

This means that $\hat{\Gamma}_{(I)}^{[\beta,\boldsymbol{\alpha};\boldsymbol{v}]}=\hat{\Gamma}_{(I)}^{\left[\beta,\mathcal{N};\boldsymbol{v}\right]}=\hat{\Gamma}_{(I)}^{[\beta,\boldsymbol{v}]}$. These $(I)$ DM operators are equilibrium ones, while a non-equilibrium one is denoted $\hat{\Gamma}_{(I)}$, as in Eqs. (15) and (6). The corresponding macrostate equilibrium weights, Eq.(19) are

$$\omega_{(I)}[\beta,\boldsymbol{\alpha};\boldsymbol{v}]=\frac{\mathrm{Tr}\,\hat{g}_{(I)}\exp\left(-\beta\hat{H}[\boldsymbol{v}]-\boldsymbol{\alpha}\hat{\mathcal{N}}\right)}{\mathrm{Tr}\exp\left(-\beta\hat{H}[\boldsymbol{v}]-\boldsymbol{\alpha}\hat{\mathcal{N}}\right)}, \tag{22}$$

and

$$\omega_{(I)}\left[\beta,\mathcal{N};\boldsymbol{v}\right]=\frac{\mathrm{Tr}\,\hat{g}_{(I)}\exp\left(-\beta\hat{H}[\boldsymbol{v}]-\breve{\boldsymbol{\alpha}}^{\left[\beta,\mathcal{N};\boldsymbol{v}\right]}\hat{\mathcal{N}}\right)}{\mathrm{Tr}\exp\left(-\beta\hat{H}[\boldsymbol{v}]-\breve{\boldsymbol{\alpha}}^{\left[\beta,\mathcal{N};\boldsymbol{v}\right]}\hat{\mathcal{N}}\right)}. \tag{23}$$

With $\boldsymbol{\alpha}=-\beta\boldsymbol{\mu}$, these weights can be rewritten for the energy representation, Eqs.(89) and (107).

For the $\left\{\hat{F}_{\mathrm{int}},\hat{\boldsymbol{\rho}}\right\}$ system, in the spin-grand-canonical ensemble with $\hat{O}=\hat{O}[\beta,\boldsymbol{w}]=-\beta\hat{F}_{\mathrm{int}}-\int[\boldsymbol{w}\hat{\boldsymbol{\rho}}]$ we have the macrostate eq-DM

$$\hat{\Gamma}_{(I)}^{[\beta,\boldsymbol{w}]} = \frac{\hat{g}_{(I)} \exp\left(-\beta \hat{F}_{\text{int}} - \int [\boldsymbol{w}\hat{\boldsymbol{\rho}}]\right)}{\operatorname{Tr} \hat{g}_{(I)} \exp\left(-\beta \hat{F}_{\text{int}} - \int [\boldsymbol{w}\hat{\boldsymbol{\rho}}]\right)}, \tag{24}$$

and in the spin-canonical ensemble with $\hat{O} = \hat{O}[\beta, \boldsymbol{\rho}] = -\beta \hat{F}_{\text{int}} - \int \left[ \tilde{\boldsymbol{w}}^{[\beta,\boldsymbol{\rho}]} (\hat{\boldsymbol{\rho}} - \boldsymbol{\rho}) \right]$ we have

$$\hat{\Gamma}_{(I)}^{[\beta,\boldsymbol{\rho}]} = \frac{\hat{g}_{(I)} \exp\left(-\beta \hat{F}_{\text{int}} - \int \left[ \tilde{\boldsymbol{w}}^{[\beta,\boldsymbol{\rho}]} \hat{\boldsymbol{\rho}} \right]\right)}{\operatorname{Tr} \hat{g}_{(I)} \exp\left(-\beta \hat{F}_{\text{int}} - \int \left[ \tilde{\boldsymbol{w}}^{[\beta,\boldsymbol{\rho}]} \hat{\boldsymbol{\rho}} \right]\right)}. \tag{25}$$

The functionals having distinct arguments are considered to be different, e.g., $\hat{\Gamma}_{\text{eq}}^{[\beta,\boldsymbol{v}]}$, $\hat{\Gamma}_{\text{eq}}^{[\beta,\boldsymbol{w}]}$, $\hat{\Gamma}_{\text{eq}}^{[\beta,\boldsymbol{\rho}]}$.

Using the equivalence conditions between $\left\{\hat{H}, \hat{\boldsymbol{\mathcal{N}}}\right\}$ system and $\left\{\hat{F}_{\text{int}}, \hat{\boldsymbol{\rho}}(\mathbf{r})\right\}$ system (see Table III in Paper II), applied to the spin-grand-canonical ensemble, $\beta \boldsymbol{v}(\mathbf{r}) + \boldsymbol{\alpha} = \boldsymbol{w}(\mathbf{r})$, we find from Eq.(24)

$$\hat{\Gamma}_{(I)}^{[\beta,\boldsymbol{w}]} = \frac{\hat{g}_{(I)} \exp\left(-\beta \hat{F}_{\text{int}} - \int \left[ \left(\beta \boldsymbol{v}(\mathbf{r}) + \boldsymbol{\alpha}\right) \hat{\boldsymbol{\rho}} \right]\right)}{\operatorname{Tr} \hat{g}_{(I)} \exp\left(-\beta \hat{F}_{\text{int}} - \int \left[ \left(\beta \boldsymbol{v}(\mathbf{r}) + \boldsymbol{\alpha}\right) \hat{\boldsymbol{\rho}} \right]\right)} = \frac{\hat{g}_{(I)} \exp\left(-\beta \hat{H}[\boldsymbol{v}]\right)}{\operatorname{Tr} \hat{g}_{(I)} \exp\left(-\beta \hat{H}[\boldsymbol{v}]\right)} = \hat{\Gamma}_{(I)}^{[\beta,\boldsymbol{v}]}, \tag{26}$$

and from Eq.(24), $\omega_{(I)}^{[\beta,\boldsymbol{w}]} = \omega_{(I)}^{[\beta,\boldsymbol{\alpha}=\boldsymbol{0};\boldsymbol{v}=-\boldsymbol{w}/\beta]}$

With the equivalence condition applied to spin-canonical ensemble at given $\boldsymbol{\mathcal{N}} = \int [\boldsymbol{\rho}]$, $\beta \boldsymbol{v}(\mathbf{r}) + \breve{\boldsymbol{\alpha}}^{[\beta,\boldsymbol{\mathcal{N}};\boldsymbol{v}]} = \tilde{\boldsymbol{w}}^{[\beta,\boldsymbol{\rho}]}(\mathbf{r})$, Eq.(25) yields

$$\hat{\Gamma}_{(I)}^{[\beta,\boldsymbol{\rho}]} = \frac{\hat{g}_{(I)} \exp\left(-\beta \hat{F}_{\text{int}} - \int \left[ \left(\beta \boldsymbol{v}(\mathbf{r}) + \breve{\boldsymbol{\alpha}}^{[\beta,\boldsymbol{\mathcal{N}};\boldsymbol{v}]}\right) \hat{\boldsymbol{\rho}} \right]\right)}{\operatorname{Tr} \hat{g}_{(I)} \exp\left(-\beta \hat{F}_{\text{int}} - \int \left[ \left(\beta \boldsymbol{v}(\mathbf{r}) + \breve{\boldsymbol{\alpha}}^{[\beta,\boldsymbol{\mathcal{N}};\boldsymbol{v}]}\right) \hat{\boldsymbol{\rho}} \right]\right)} = \frac{\hat{g}_{(I)} \exp\left(-\beta \hat{H}[\boldsymbol{v}]\right)}{\operatorname{Tr} \hat{g}_{(I)} \exp\left(-\beta \hat{H}[\boldsymbol{v}]\right)} = \hat{\Gamma}_{(I)}^{[\beta,\boldsymbol{v}]}, \tag{27}$$

and Eq.(25) yields $\omega_{(I)}^{[\beta,\boldsymbol{\rho}]} = \omega_{(I)}^{[\beta,\boldsymbol{\alpha}=\boldsymbol{0};\boldsymbol{v}=-\tilde{\boldsymbol{w}}[\beta,\boldsymbol{\rho}]/\beta]}$.

Note $\int[\hat{\boldsymbol{\rho}}]=\hat{\boldsymbol{\mathcal{N}}}$, so the cancellation of $\boldsymbol{\alpha}$ – or $\breve{\boldsymbol{\alpha}}$ –containing exponential factors occured in Eqs. (26) and (27). As we see, $\hat{\Gamma}_{(I)}^{[\beta,\mathbf{v}]}$, Eq.(21), is the $(I)$ macrostate eq-DMs at finite temperature common for spin-grand-canonical and spin-canonical ensembles, both for $\{\hat{H},\hat{\boldsymbol{\mathcal{N}}}\}$ system and the equivalent $\{\hat{F}_{\text{int}},\hat{\boldsymbol{\rho}}\}$ system.

The weights, Eq.(20), for two ensembles of the $\{\hat{F}_{\text{int}},\hat{\boldsymbol{\rho}}\}$ systems, $\omega_{(I)}^{[\beta,\mathbf{w}]}$ and $\omega_{(I)}^{[\beta,\boldsymbol{\rho}]}$, become equal to $\omega_{(I)}^{[\beta,\boldsymbol{\alpha};\mathbf{v}]}$ and $\omega_{(I)}^{[\beta,\boldsymbol{\mathcal{N}};\mathbf{v}]}$, respectively, when equivalence conditions with the system $\{\hat{H},\hat{\boldsymbol{\mathcal{N}}}\}$ are imposed. Then, obviously, $\hat{\Gamma}_{\text{eq}}=\sum_I \omega_{(I)}^{[\hat{o}]}\hat{\Gamma}_{(I)}^{[\hat{o}]}$ is the same for both systems.

### B. Characteristic functions of the macrostate for the $\{\hat{H},\hat{\boldsymbol{\mathcal{N}}}\}$ system

The spin Kramer function $K[\beta,\boldsymbol{\alpha};\mathbf{v}]$ introduced in Paper II, Eq.(II.8), can be obtained via maximization of entropy (with constrains), Eq.(I.10) with Eqs.(I.4) and (I.7):

$$K[\beta,\boldsymbol{\alpha};\mathbf{v}]=\operatorname*{Max}_{\hat{\Gamma}}\left\{-\operatorname{Tr}\hat{\Gamma}\left(\ln\hat{\Gamma}+\beta\hat{H}[\mathbf{v}]+\boldsymbol{\alpha}\hat{\boldsymbol{\mathcal{N}}}\right)\right\}. \tag{28}$$

Therefore for the macrostate counterpart of this function, $K_{(I)}[\beta,\boldsymbol{\alpha};\mathbf{v}]$, the maximization should be confined to the $(I)$ macrostate set of DM operators, Eq.(16):

$$\begin{aligned}K_{(I)}[\beta,\boldsymbol{\alpha};\mathbf{v}]&=\operatorname*{Max}_{\hat{\Gamma}\in\{\hat{\Gamma}_{(I)}\}}\left\{-\operatorname{Tr}\hat{\Gamma}\left(\ln\hat{\Gamma}+\beta\hat{H}[\mathbf{v}]+\boldsymbol{\alpha}\hat{\boldsymbol{\mathcal{N}}}\right)\right\}\\&=Y_{(I)}[\beta;\mathbf{v}]-\boldsymbol{\alpha}\boldsymbol{\mathcal{N}}_{(I)},\end{aligned} \tag{29}$$

where

$$\begin{aligned}Y_{(I)}[\beta;\mathbf{v}]&\equiv\operatorname*{Max}_{\hat{\Gamma}\in\{\hat{\Gamma}_{(I)}\}}\left\{-\operatorname{Tr}\hat{\Gamma}\left(\ln\hat{\Gamma}+\beta\hat{H}[\mathbf{v}]\right)\right\}\\&=S_{(I)}^{[\beta;\mathbf{v}]}-\beta E_{(I)}^{[\beta;\mathbf{v}]},\end{aligned} \tag{30}$$

is the $(I)$ *macrostate spin Massieu function*. The macrostate eq-DM $\hat{\Gamma}_{(I)}^{[\beta,\mathbf{v}]}$, Eq.(21), is the maximizer in Eqs.(29) and (30), so the macrostate average energy and entropy are $E_{(I)}^{[\beta;\mathbf{v}]} \equiv \mathrm{Tr}\,\hat{\Gamma}_{(I)}^{[\beta;\mathbf{v}]}\hat{H}[\mathbf{v}]$ and $S_{(I)}^{[\beta;\mathbf{v}]}$, Eq.(15).

The spin Massieu fn. $Y$, Eq.(II.15), and the spin Kramer fn. $K$, Eq.(II.8), in the macrostate resolution can be written as

$$Y[\beta,\boldsymbol{\mathcal{N}};\mathbf{v}] = \sum_I \omega_{(I)}^{[\beta,\boldsymbol{\mathcal{N}};\mathbf{v}]}\left(Y_{(I)}^{[\beta;\mathbf{v}]} - \ln \omega_{(I)}^{[\beta,\boldsymbol{\mathcal{N}};\mathbf{v}]}\right) \tag{31}$$

and

$$\begin{aligned} K[\beta,\boldsymbol{\alpha};\mathbf{v}] &= \sum_I \omega_{(I)}^{[\beta,\boldsymbol{\alpha};\mathbf{v}]}\left(Y_{(I)}^{[\beta;\mathbf{v}]} - \boldsymbol{\alpha}\boldsymbol{\mathcal{N}}_{(I)} - \ln \omega_{(I)}^{[\beta,\boldsymbol{\alpha};\mathbf{v}]}\right) \\ &= \sum_I \omega_{(I)}^{[\beta,\boldsymbol{\alpha};\mathbf{v}]}\left(K_{(I)}^{[\beta,\boldsymbol{\alpha};\mathbf{v}]} - \ln \omega_{(I)}^{[\beta,\boldsymbol{\alpha};\mathbf{v}]}\right), \end{aligned} \tag{32}$$

because $S^{\mathrm{BGS}}$ can be resolved as in Eq.(14).

The Massieu-Planck transform of the spin Massieu fn. gives the spin Helmholtz function, Eq.(II.20), (use Eqs.(31) and (30))

$$\begin{aligned} A^{[\beta,\boldsymbol{\mathcal{N}};\mathbf{v}]} &\equiv -\beta^{-1}Y^{[\beta,\boldsymbol{\mathcal{N}};\mathbf{v}]} = E^{[\beta,\boldsymbol{\mathcal{N}};\mathbf{v}]} - \beta^{-1}S^{[\beta,\boldsymbol{\mathcal{N}};\mathbf{v}]} \\ &= \sum_I \omega_{(I)}^{[\beta,\boldsymbol{\mathcal{N}};\mathbf{v}]}\left(A_{(I)}^{[\beta;\mathbf{v}]} + \beta^{-1}\ln \omega_{(I)}^{[\beta,\boldsymbol{\mathcal{N}};\mathbf{v}]}\right) \end{aligned} \tag{33},$$

where *the macrostate spin Helmholtz function* is defined as

$$A_{(I)}^{[\beta;\mathbf{v}]} \equiv -\beta^{-1}Y_{(I)}^{[\beta;\mathbf{v}]} = E_{(I)}^{[\beta;\mathbf{v}]} - \beta^{-1}S_{(I)}^{[\beta;\mathbf{v}]}. \tag{34}$$

In the case of the spin Kramer fn. with $\boldsymbol{\alpha} = -\beta\boldsymbol{\mu}$, Eqs.(29) and (34) give *the macrostate spin-grand-potential*

$$\begin{aligned} \Omega_{(I)}[\beta,\boldsymbol{\mu};\mathbf{v}] &\equiv -\beta^{-1}K_{(I)}^{[\beta,-\beta\boldsymbol{\mu};\mathbf{v}]} = -\beta^{-1}\left(Y_{(I)}[\beta;\mathbf{v}] + \beta\boldsymbol{\mu}\boldsymbol{\mathcal{N}}_{(I)}\right) \\ &= E_{(I)}^{[\beta;\mathbf{v}]} - \beta^{-1}S_{(I)}^{[\beta;\mathbf{v}]} - \boldsymbol{\mu}\boldsymbol{\mathcal{N}}_{(I)}. \end{aligned} \tag{35}$$

These macrostate functions are related as follows

$$K_{(I)}[\beta,\boldsymbol{\alpha};\mathbf{v}] = Y_{(I)}\left[\beta;\left(\mathbf{v}+\frac{\boldsymbol{\alpha}}{\beta}\right)\right] = -\beta A_{(I)}\left[\beta;\left(\mathbf{v}+\frac{\boldsymbol{\alpha}}{\beta}\right)\right] = -\beta\Omega_{(I)}\left[\beta,-\frac{\boldsymbol{\alpha}}{\beta};\mathbf{v}\right]. \tag{36}$$

From Eqs.(29) and (30) follows that DM operators in the form $\hat{\varGamma} = \sum_I p_{(I)} \hat{\varGamma}_{(I)}^{[\beta,v]}$ (with some $\{p_{(I)} \neq 0\}$ chosen arbitrarily), maximize the entropy in each macrostate separately. But only $\hat{\varGamma}_{\rm eq}^{[\hat{o}]} = \sum_I \omega_{(I)}^{[\hat{o}]} \hat{\varGamma}_{{\rm eq},(I)}^{[\beta,v]}$ with the $(I)$ macrostate equilibrium weight $\omega_{(I)}^{[\hat{o}]}$, Eq.(22) or (23), is the maximizer in the whole Fock space.

## C. Characteristic functions of the macrostate for the $\{\hat{F}_{\rm int}, \hat{\boldsymbol{\rho}}\}$ system

The macrostate counterpart of the (global) spin Kramer fnl., $X[\beta, \boldsymbol{w}]$, Eq.(II.39) with Eqs.(II.37) and (II.36) inserted, *the macrostate spin Kramer functional* is defined (similarly to the macrostate spin Kramer fn., Eq.(29))

$$X_{(I)}[\beta, \boldsymbol{w}] \equiv \operatorname*{Max}_{\hat{\varGamma} \in \{\hat{\varGamma}_{(I)}\}} \left\{ -\operatorname{Tr} \hat{\varGamma} \left( \ln \hat{\varGamma} + \beta \hat{F}_{\rm int} + \int [\boldsymbol{w} \hat{\boldsymbol{\rho}}] \right) \right\}, \tag{37}$$

where $\{\hat{\varGamma}_{(I)}\}$ is defined in Eq.(6). To follow the constrained-search method proposed by Levy and Lieb,[22,23] we define the set of $(I)$ macrostate-representable densities

$$\{\boldsymbol{\rho}_{(I)}(\mathbf{r})\} \equiv \{\operatorname{Tr} \hat{\varGamma}_{(I)} \hat{\boldsymbol{\rho}}(\mathbf{r})\} = \left\{ \sum_{K \in (I)} p_K^{(I)} \boldsymbol{\rho}_K(\mathbf{r}) \middle| p_K^{(I)} \geq 0; \sum_K p_K^{(I)} = 1 \right\}, \tag{38}$$

see Eq.(6) with Eq.(4). Here the pure $K$ microstate expectation value of operator $\hat{O}$ is denoted as

$$O_K \equiv \left\langle \Psi_K \middle| \hat{O} \middle| \Psi_K \right\rangle, \tag{39}$$

and applied to $\hat{O} = \hat{\boldsymbol{\rho}}(\mathbf{r})$. The set $\{\hat{\varGamma}_{(I)}\}$ is discussed after Eq.(6). Next, we define also *the macrostate spin Massieu universal functional* (comp. Eq.(II.36))

$$G_{(I)}\left[\beta,\rho_{(I)}\right] \equiv \underset{\hat{\Gamma}\in\left\{\hat{\Gamma}_{(I)}\right\}}{\mathrm{Max}}\left\{-\mathrm{Tr}\,\hat{\Gamma}\left(\ln\hat{\Gamma}+\beta\hat{F}_{\mathrm{int}}\right)\middle|\mathrm{Tr}\,\hat{\Gamma}\hat{\rho}=\rho_{(I)}\right\}$$
$$= S^{\mathrm{BGS}}\left[\hat{\Gamma}_{(I)}^{\left[\beta,\rho_{(I)}\right]}\right]-\beta\,\mathrm{Tr}\,\hat{\Gamma}_{(I)}^{\left[\beta,\rho_{(I)}\right]}\hat{F}_{\mathrm{int}} = S_{(I)}^{\left[\beta,\rho_{(I)}\right]}-\beta F_{\mathrm{int},(I)}^{\left[\beta,\rho_{(I)}\right]}. \tag{40}$$

It should be stressed that $S_{(I)}$ and $F_{\mathrm{int},(I)}$ are defined only as components of $G_{(I)}$ functional, this means that their values are available if the maximizer $\hat{\Gamma}_{(I)}^{\left[\beta,\rho_{(I)}\right]}$ from Eq.(40) is known. In Eq.(40), a convention is applied: when $\rho_{(I)}$ is used as an argument of a functional, it means that any density belonging to $\left\{\rho_{(I)}\right\}$ can be used, e.g. $F_{\mathrm{int},(I)}\left[\beta,\rho_{(I)}\right]$ means $F_{\mathrm{int},(I)}\left[\beta,\rho\right]$ for $\rho\in\left\{\rho_{(I)}\right\}$, so $\left\{\rho_{(I)}\right\}$ is the domain of $F_{\mathrm{int},(I)}\left[\beta,\rho_{(I)}\right]$.

Now, we can perform the maximization in Eq.(37) in two steps

$$X_{(I)}\left[\beta,\boldsymbol{w}\right] = \underset{\rho\in\left\{\rho_{(I)}\right\}}{\mathrm{Max}}\left\{\underset{\hat{\Gamma}\in\left\{\hat{\Gamma}_{(I)}\right\}}{\mathrm{Max}}\left\{-\mathrm{Tr}\,\hat{\Gamma}\left(\ln\hat{\Gamma}+\beta\hat{F}_{\mathrm{int}}\right)\middle|\mathrm{Tr}\,\hat{\Gamma}\hat{\rho}=\rho\right\}-\int\left[\boldsymbol{w}\rho\right]\right\}$$
$$= \underset{\rho\in\left\{\rho_{(I)}\right\}}{\mathrm{Max}}\left\{G_{(I)}\left[\beta,\rho\right]-\int\left[\boldsymbol{w}\rho\right]\right\} \tag{41}$$
$$= G_{(I)}\left[\beta,\tilde{\rho}_{(I)}^{[\beta,w]}\right]-\int\left[\boldsymbol{w}\tilde{\rho}_{(I)}^{[\beta,w]}\right].$$

In terms of $G_{(I)}\left[\beta,\rho_{(I)}\right]$ and $X_{(I)}\left[\beta,\boldsymbol{w}\right]$ defined above with the maximizer $\rho(\mathbf{r})=\tilde{\rho}_{(I)}^{[\beta,w]}$, and taking into account $S^{\mathrm{BGS}}$ resolved as in Eq.(14), the spin Massieu universal functional $G\left[\beta,\rho\right]$ , Eq.(II.36), can be rewritten as

$$G\left[\beta,\rho\right] = X\left[\beta,\tilde{\boldsymbol{w}}^{[\beta,\rho]}\right]+\int\left[\tilde{\boldsymbol{w}}^{[\beta,\rho]}\rho\right]$$
$$= \sum_I \omega_{(I)}^{[\beta,\rho]}\left(X_{(I)}\left[\beta,\tilde{\boldsymbol{w}}^{[\beta,\rho]}\right]+\int\left[\tilde{\boldsymbol{w}}^{[\beta,\rho]}\tilde{\tilde{\rho}}_{(I)}^{[\beta,\rho]}\right]-\ln\omega_{(I)}^{[\beta,\rho]}\right) \tag{42}$$
$$= \sum_I \omega_{(I)}^{[\beta,\rho]}\left(G_{(I)}\left[\beta,\tilde{\tilde{\rho}}_{(I)}^{[\beta,\rho]}\right]-\ln\omega_{(I)}^{[\beta,\rho]}\right).$$

Here, in the first line, the mapping property $(\beta,\rho)\leftrightarrow(\beta,\boldsymbol{w})$, discussed in detail in Paper II, was used. The mapping "→" defines $\tilde{\boldsymbol{w}}\left[\beta,\rho\right]$ of Eq.(42), while "←" defines $\tilde{\rho}\left[\beta,\boldsymbol{w}\right]$ of Eq.(41), the functions satisfying identities $\tilde{\boldsymbol{w}}\left[\beta,\tilde{\rho}\left[\beta,\boldsymbol{w}\right]\right](\mathbf{r})=\boldsymbol{w}(\mathbf{r})$ and $\tilde{\boldsymbol{\rho}}\left[\beta,\tilde{\boldsymbol{w}}[\beta,\boldsymbol{\rho}]\right](\mathbf{r})=\boldsymbol{\rho}(\mathbf{r})$. The last line is the effect of inserting the relation from Eq.(41). The macrostate density $\tilde{\tilde{\boldsymbol{\rho}}}_{(I)}$ depends on $\boldsymbol{\rho}$ trough $\tilde{\boldsymbol{w}}[\beta,\boldsymbol{\rho}]$, namely $\tilde{\tilde{\boldsymbol{\rho}}}_{(I)}[\beta,\boldsymbol{\rho}](\mathbf{r})=\tilde{\boldsymbol{\rho}}_{(I)}\left[\beta,\boldsymbol{w}[\beta,\boldsymbol{\rho}]\right](\mathbf{r})$, satisfying the identity

$$\boldsymbol{\rho}(\mathbf{r})=\sum_I \omega_{(I)}^{[\beta,\boldsymbol{\rho}]}\tilde{\tilde{\boldsymbol{\rho}}}_{(I)}^{[\beta,\boldsymbol{\rho}]}(\mathbf{r}) . \tag{43}$$

The functional space of admissible $\boldsymbol{\rho}(\mathbf{r})$ are described after Eq.(II.35).

The Massieu-Planck transform of these functionals gives *the macrostate spin-canonical universal functional* (see Eq.(40)

$$F_{(I)}\left[\beta,\boldsymbol{\rho}_{(I)}\right]\equiv-\beta^{-1}G_{(I)}\left[\beta,\boldsymbol{\rho}_{(I)}\right]=F_{\mathrm{int},(I)}\left[\beta,\boldsymbol{\rho}_{(I)}\right]-\beta^{-1}S_{(I)}\left[\beta,\boldsymbol{\rho}_{(I)}\right], \tag{44}$$

and the *macrostate spin grand canonical functional*

$$B_{(I)}[\beta,\boldsymbol{u}]\equiv-\beta^{-1}X_{(I)}[\beta,-\beta\boldsymbol{u}]=F_{(I)}\left[\beta,\tilde{\boldsymbol{\rho}}_{(I)}^{[\beta,\boldsymbol{u}]}\right]-\int\left[\boldsymbol{u}\tilde{\boldsymbol{\rho}}_{(I)}^{[\beta,\boldsymbol{u}]}\right], \tag{45}$$

where $\boldsymbol{u}(\mathbf{r})=-\beta^{-1}\boldsymbol{w}(\mathbf{r})$ and $\tilde{\boldsymbol{\rho}}_{(I)}[\beta,\boldsymbol{u}]=\tilde{\boldsymbol{\rho}}_{(I)}[\beta,\boldsymbol{w}=-\beta\boldsymbol{u}]$. The functionals $F_{(I)}\left[\beta,\boldsymbol{\rho}_{(I)}\right]$ and $B_{(I)}[\beta,\boldsymbol{u}]$ are Legendre transformation equivalent (see Paper I, Appendix C). The functional $F_{(I)}\left[\beta,\boldsymbol{\rho}_{(I)}\right]$ can be viewed as the thermodynamic extension of the Lieb-Valone functional[13,24], when the search for the minimum (i.e., for the maximum in Eq.(40), $F_{(I)}=-\beta^{-1}G_{(I)}$) is performed only over the set of $(I)$ macrostate density matrices (comp. Eq.(II.57) and the text after it).

By inserting $\hat{H}[\boldsymbol{v}]=\hat{F}_{\mathrm{int}}+\int[\hat{\boldsymbol{\rho}}\boldsymbol{v}]$, the relation between the macrostate Massieu function and the macrostate spin Massieu universal functional is found from Eq.(28) (as in Eq.(37)) using next Eq.(40)

$$
\begin{aligned}
Y_{(I)}[\beta;\boldsymbol{v}] &= \operatorname*{Max}_{\boldsymbol{\rho}\in\{\boldsymbol{\rho}_{(I)}\}}\left\{\operatorname*{Max}_{\hat{\varGamma}\in\{\hat{\varGamma}_{(I)}\}}\left\{-\operatorname{Tr}\hat{\varGamma}\left(\ln\hat{\varGamma}+\beta\hat{F}_{\text{int}}\right)\middle|\operatorname{Tr}\hat{\varGamma}\hat{\boldsymbol{\rho}}=\boldsymbol{\rho}\right\}-\beta\int[\boldsymbol{v}\boldsymbol{\rho}]\right\} \\
&= \operatorname*{Max}_{\boldsymbol{\rho}\in\{\boldsymbol{\rho}_{(I)}\}}\left\{G_{(I)}[\beta,\boldsymbol{\rho}]-\beta\int[\boldsymbol{v}\boldsymbol{\rho}]\right\}=G_{(I)}\left[\beta,\breve{\boldsymbol{\rho}}_{(I)}^{[\beta;\boldsymbol{v}]}\right]-\beta\int\left[\boldsymbol{v}\breve{\boldsymbol{\rho}}_{(I)}^{[\beta;\boldsymbol{v}]}\right].
\end{aligned}
\tag{46}
$$

For their Massieu-Planck transform, we have similarly

$$
\begin{aligned}
A_{(I)}^{[\beta;\boldsymbol{v}]} &= \operatorname*{Min}_{\boldsymbol{\rho}\in\{\boldsymbol{\rho}_{(I)}\}}\left\{\operatorname*{Min}_{\hat{\varGamma}\in\{\hat{\varGamma}_{(I)}\}}\left\{\operatorname{Tr}\hat{\varGamma}\left(\hat{F}_{\text{int}}+\beta^{-1}\ln\hat{\varGamma}\right)\middle|\operatorname{Tr}\hat{\varGamma}\hat{\boldsymbol{\rho}}=\boldsymbol{\rho}\right\}+\int[\boldsymbol{v}\boldsymbol{\rho}]\right\} \\
&= \operatorname*{Min}_{\boldsymbol{\rho}\in\{\boldsymbol{\rho}_{(I)}\}}\left\{F_{(I)}[\beta,\boldsymbol{\rho}]+\int[\boldsymbol{v}\boldsymbol{\rho}]\right\}=F_{(I)}\left[\beta,\breve{\boldsymbol{\rho}}_{(I)}^{[\beta;\boldsymbol{v}]}\right]+\int\left[\boldsymbol{v}\breve{\boldsymbol{\rho}}_{(I)}^{[\beta;\boldsymbol{v}]}\right],
\end{aligned}
\tag{47}
$$

where the macrostate spin-canonical universal functional is defined via minimization w.r.t. $\hat{\varGamma}_{(I)}$ as

$$
\forall\boldsymbol{\rho}\in\left\{\boldsymbol{\rho}_{(I)}\right\},\ F_{(I)}[\beta,\boldsymbol{\rho}]\equiv\operatorname*{Min}_{\hat{\varGamma}\in\{\hat{\varGamma}_{(I)}\}}\left\{\operatorname{Tr}\hat{\varGamma}\left(\hat{F}_{\text{int}}+\beta^{-1}\ln\hat{\varGamma}\right)\middle|\operatorname{Tr}\hat{\varGamma}\hat{\boldsymbol{\rho}}=\boldsymbol{\rho}\right\}. \tag{48}
$$

It should be noted that the maximizer $\breve{\boldsymbol{\rho}}_{(I)}^{[\beta,\boldsymbol{v}]}$ in Eq.(46) is identical with the minimizer in Eq.(47), because the two extremized functionals are related by the fixed factor $\left(-\beta^{-1}\right)$. The final results in Eqs.(46) and (47) describe the equilibrium states, so

$$
\breve{\boldsymbol{\rho}}_{(I)}^{[\beta;\boldsymbol{v}]}(\mathbf{r})=\operatorname{Tr}\hat{\varGamma}_{(I)}^{[\beta;\boldsymbol{v}]}\hat{\boldsymbol{\rho}}(\mathbf{r}). \tag{49}
$$

Based on the demand that two systems are equivalent, the relation $\boldsymbol{w}(\mathbf{r})=\beta\boldsymbol{v}(\mathbf{r})+\boldsymbol{\alpha}$ was established previously, Eq.(II.72). Using it, the macrostate spin Massieu and Kramer functions can be written in terms of the macrostate spin Kramer fnl. as

$$
K_{(I)}[\beta,\boldsymbol{\alpha};\boldsymbol{v}]=Y_{(I)}\left[\beta;\left(\boldsymbol{v}+\beta^{-1}\boldsymbol{\alpha}\right)\right]=X_{(I)}\left[\beta,(\beta\mathbf{v}+\boldsymbol{\alpha})\right]. \tag{50}
$$

For Massieu-Planck transform, Eq.(50) with Eq.(35) can be rewritten as

$$
\Omega_{(I)}[\beta,\boldsymbol{\mu};\boldsymbol{v}]=A_{(I)}\left[\beta;(\boldsymbol{v}-\boldsymbol{\mu})\right]=B_{(I)}\left[\beta,(\boldsymbol{\mu}-\boldsymbol{v})\right]. \tag{51}
$$

## D. The zero-temperature limit of the macrostate functions

First, we analyze the macrostate eq-DM $\hat{\Gamma}_{(I)}^{[\beta,v]}$ at zero-temperature (0K) limit. The conditional probability that the system is in mesostate $(i)$ provided we know it to be in the macrostate $(I)$, Eq.(10) with Eq.(21), (using new $(i)$ numbering, Eq.(17)) can be rewritten as

$$p_{(i,I)}^{[\beta;v]} = \mathrm{Tr}\, \hat{g}_{(i)} \hat{\Gamma}_{(I)}^{[\beta;v]} = \frac{d_{(i,I)}^{[v]} \exp\left(-\beta E_{(i,I)}^{[v]}\right)}{\sum_k d_{(k,I)}^{[v]} \exp\left(-\beta E_{(k,I)}^{[v]}\right)} = \frac{\exp\left(-\beta E_{(i,I)}^{[v]} + S_{(i,I)}^{[v]}\right)}{\sum_k \exp\left(-\beta E_{(k,I)}^{[v]} + S_{(k,I)}^{[v]}\right)}, \tag{52}$$

because $\ln d_{(i,I)}^{[v]} = S^{\mathrm{BGS}}\left[\hat{\Gamma}_{\mathrm{eq},(i,I)}^{[v]}\right] = S_{(i,I)}^{[v]}$, see Eq.(13), $d_{(i,I)}^{[v]}$ is the degeneracy of the mesostate energy $E_{(i,I)}^{[v]}$, Eq.(17). At 0K limit, $(\beta \to \infty)$, only $p_{(i,I)}^{[\infty,v]}$ for the mesostate $i=0$ which describes the ground state, is different than zero:

$$\lim_{\beta\to\infty} p_{(i,I)}^{[\beta;v]} = \lim_{\beta\to\infty} \frac{\exp\left(-\beta\left(E_{(i,I)}^{[v]} - E_{(0,I)}^{[v]}\right) + S_{(i,I)}^{[v]} - S_{(0,I)}^{[v]}\right)}{1 + \sum_{k\geq 1} \exp\left(-\beta\left(E_{(k,I)}^{[v]} - E_{(0,I)}^{[v]}\right) + S_{(k,I)}^{[v]} - S_{(0,I)}^{[v]}\right)} \equiv p_{(i,I)}^{[\infty,v]} = \delta_{0,i}\,. \tag{53}$$

Note that the relations $E_{(0,I)}^{[v]} < E_{(1,I)}^{[v]} < E_{(2,I)}^{[v]} < \cdots$ were used. Notation for the result of the 0K limit are introduced: $\lim_{\beta\to\infty} f[\beta,x] \equiv f[\infty,x] \equiv f[x]$, abbreviated further to $f[\beta,x] \doteq f[x]$.

Using Eq.(16) with (53) and Eq.(17), the macrostate eq-DM operator at 0K limit is obtained as

$$\hat{\Gamma}_{(I)}^{[\beta;v]} \doteq \hat{\Gamma}_{(I)}^{[v]} = \hat{\Gamma}_{(0,I)}^{[v]} = \frac{1}{d_{(0,I)}^{[v]}} \sum_{K\in(0,I)} \left|\Psi_K^{[v]}\right\rangle\left\langle\Psi_K^{[v]}\right|, \qquad \mathrm{Tr}\,\hat{\Gamma}_{(0,I)}^{[v]} = 1, \tag{54}$$

where $d_{(0,I)}^{[v]}$ is a degeneracy of the ground-state energy (orbital multiplicity in the case of the absence of a magnetic field in our illustrative example) at given macrostate $(I)$, i.e., the number of terms in the summation. Using Eq.(54), we have the following 0K limits:

$$\begin{aligned} S_{(I)}^{[\beta,v]} &\doteq -\mathrm{Tr}\,\hat{\Gamma}_{(0,I)}^{[v]} \ln \hat{\Gamma}_{(0,I)}^{[v]} = S_{(0,I)}^{[v]} = \ln d_{(0,I)}^{[v]}, \\ E_{(I)}^{[\beta;v]} &\doteq \mathrm{Tr}\,\hat{\Gamma}_{(0,I)}^{[v]} \hat{H} = E_{(0,I)}^{[v]}. \end{aligned} \tag{55}$$

Based on these limits, the *asymptotic* forms (see Appendix, Eqs.(A6) and (A7)) of the macrostate spin Helmholtz fn. $A_{(I)}^{[\beta;v]}$, Eq.(34), and the macrostate spin Massieu fn. $Y_{(I)}^{[\beta;v]}$, Eq.(30), are

$$A_{(I)}^{[\beta;v]} \simeq E_{(0,I)}^{[v]} - \beta^{-1}\ln d_{(0,I)}^{[v]} \doteq E_{(0,I)}^{[v]}, \tag{56}$$

and

$$Y_{(I)}^{[\beta;v]} \simeq -\beta E_{(0,I)}^{[v]} + S_{(0,I)}^{[v]} = -\beta E_{(0,I)}^{[v]} + \ln d_{(0,I)}^{[v]}, \tag{57}$$

respectively. The macrostate spin Massieu fn. has in the large-$\beta$ region a slant asymptote, while the macrostate spin Helmholtz fn. — a horizontal asymptote. These asymptotic expressions and the relations (55) demonstrate that at the zero-temperature limit the mesostate description and macrostate description provide the same information about the equilibrium state. For $\hat{\Gamma}_{(I)}^{[\beta,w]}$ (with an independent argument $\boldsymbol{w}$), we cannot repeat this procedure, because the states $\left\{\left|\Psi_K^{[v]}\right\rangle\right\}_{K\in(i,I)}$ are eigenstates neither of $\hat{F}_{\text{int}}$ nor $\hat{\boldsymbol{\rho}}$. Therefore, we are not able to predict the mesostate $(i,I)$ giving the dominant contribution when $\beta \to \infty$.

To calculate $\hat{\Gamma}_{(I)}^{[\beta,\rho]}$ in the large-$\beta$ region (when $\left\{\hat{H},\hat{\boldsymbol{\mathcal{N}}}\right\}$ and $\left\{\hat{F}_{\text{int}},\hat{\boldsymbol{\rho}}\right\}$ systems are required to be equivalent), we should examine first the behavior of the sources $\breve{\boldsymbol{\alpha}}\left[\beta,\boldsymbol{\mathcal{N}};\boldsymbol{v}\right]$ which occur in Eq.(27). They are related by the Massieu-Planck transform and the equivalence conditions

$$\tilde{\boldsymbol{w}}^{[\beta,\rho]} = -\beta\tilde{\boldsymbol{u}}^{[\beta,\rho]} = \beta\boldsymbol{v}(\mathbf{r}) + \breve{\boldsymbol{\alpha}}^{[\beta,\mathcal{N};v]} = \beta\left(\boldsymbol{v}(\mathbf{r}) - \breve{\boldsymbol{\mu}}^{[\beta,\mathcal{N};v]}\right). \tag{58}$$

We have the following slant asymptotes

$$\tilde{\boldsymbol{w}}^{\left[\beta,\breve{\rho}\left[\beta,\mathcal{N};v\right]\right]} \simeq \beta\left(\boldsymbol{v}(\mathbf{r}) - \boldsymbol{a}^{\left[\mathcal{N};v\right]}\right) - \boldsymbol{b}^{\left[\mathcal{N};v\right]} \tag{59}$$

and

$$\tilde{\boldsymbol{u}}^{[\beta,\breve{\rho}[\beta,\mathcal{N};\boldsymbol{v}]]} \simeq \beta^{-1}\boldsymbol{b}^{[\mathcal{N};\boldsymbol{v}]} + \left(\boldsymbol{a}^{[\mathcal{N};\boldsymbol{v}]} - \boldsymbol{v}(\mathbf{r})\right) \doteq \boldsymbol{a}^{[\mathcal{N};\boldsymbol{v}]} - \boldsymbol{v}(\mathbf{r}). \tag{60}$$

The 2vecs $\boldsymbol{a}^{[\mathcal{N};\boldsymbol{v}]}$ and $\boldsymbol{b}^{[\mathcal{N};\boldsymbol{v}]}$ describing the asymptotic behavior of $\breve{\boldsymbol{\alpha}}[\beta,\mathcal{N};\boldsymbol{v}]$ and $\breve{\boldsymbol{\mu}}[\beta,\mathcal{N};\boldsymbol{v}]$ at low temperature are discussed in Sec. V.B. (see Eqs.(135) and (136)). After inserting these forms into Eq.(27) and taking the $\beta \to \infty$ limit, we have $\lim_{\beta\to\infty} \hat{\Gamma}_{(I)}^{[\beta,\breve{\rho}[\beta,\mathcal{N};\boldsymbol{v}]]} = \lim_{\beta\to\infty} \hat{\Gamma}_{(I)}^{[\beta,\boldsymbol{v}]}$ – the result of Eq.(54) can be used. The asymptotic relations (59) or (60) together with the $\int[\boldsymbol{\rho}] = \mathcal{N}$ condition form the equivalence conditions between $\{\hat{H},\hat{\mathcal{N}}\}$ system and $\{\hat{F}_{\rm int},\hat{\boldsymbol{\rho}}(\mathbf{r})\}$ system at 0K limit.

The macrostate spin Massieu universal functional, Eq.(40), at low temperature can be approximated by the asymptote

$$G_{(I)}\left[\beta,\boldsymbol{\rho}_{(I)}\right] \simeq S_{(I)}^{[\boldsymbol{\rho}_{(I)}]} - \beta F_{{\rm int},(I)}^{[\boldsymbol{\rho}_{(I)}]}, \tag{61}$$

where $F_{{\rm int},(I)}^{[\boldsymbol{\rho}_{(I)}]}$ is the 0K limit of the macrostate spin canonical universal functional, Eq.(44), namely

$$F_{(I)}\left[\beta,\boldsymbol{\rho}_{(I)}\right] = F_{{\rm int},(I)}^{[\beta,\boldsymbol{\rho}_{(I)}]} - \beta^{-1} S_{(I)}^{[\beta,\boldsymbol{\rho}_{(I)}]} \simeq F_{{\rm int},(I)}^{[\boldsymbol{\rho}_{(I)}]} - \beta^{-1} S_{(I)}^{[\boldsymbol{\rho}_{(I)}]} \doteq F_{{\rm int},(I)}^{[\boldsymbol{\rho}_{(I)}]} \;. \tag{62}$$

It can be evaluated next as the 0K limit of Eq.(48)

$$\forall \boldsymbol{\rho} \in \{\boldsymbol{\rho}_{(I)}\},\; F_{(I)}[\beta,\boldsymbol{\rho}] \doteq \underset{\hat{\Gamma}\in\{\hat{\Gamma}_{(I)}\}}{\rm Min} \left\{ {\rm Tr}\,\hat{\Gamma}\hat{F}_{\rm int} \middle| {\rm Tr}\,\hat{\Gamma}\hat{\boldsymbol{\rho}} = \boldsymbol{\rho} \right\} \equiv F_{{\rm L},(I)}[\boldsymbol{\rho}]. \tag{63}$$

Here $F_{{\rm L},(I)}[\boldsymbol{\rho}]$ is the Lieb-Valone universal functional, defined in Appendix of Paper II for application in SDFT. Thus we have an important relation

$$F_{{\rm int},(I)}^{[\boldsymbol{\rho}]} \equiv F_{{\rm L},(I)}[\boldsymbol{\rho}], \qquad \forall \boldsymbol{\rho} \in \{\boldsymbol{\rho}_{(I)}\}. \tag{64}$$

The notation used here, $F_{\mathrm{L},(I)}$, replacing the previous notation $F_{\mathrm{L}}$, Eq.(II.A4), reflects the fact that minimization in Eq.(63) is performed in the subspace (set) of $(I)$ macrostate DMs; this is in agreement with Eq.(II.A5) accompanying the definition of $F_{\mathrm{L}}$.

The combination of Eq.(63) with Eq.(47) and (56) yields the zero-temperature ground-state energy of $(I)$ macrostate as a fnl. of the equilibrium density

$$A_{(I)}^{[\boldsymbol{v}]}=E_{(0,I)}^{[\boldsymbol{v}]}=F_{\mathrm{L},(I)}\left[\boldsymbol{\rho}_{\mathrm{eq},(I)}^{[\boldsymbol{v}]}\right]+\int\left[\boldsymbol{v}\boldsymbol{\rho}_{\mathrm{eq},(I)}^{[\boldsymbol{v}]}\right], \qquad (65)$$

where according to Eq.(49), the $(I)$ macrostate equilibrium density at 0K limit is

$$\boldsymbol{\rho}_{\mathrm{eq},(I)}^{[\boldsymbol{v}]}(\mathbf{r})=\mathrm{Tr}\,\hat{\varGamma}_{(I)}^{[\boldsymbol{v}]}\hat{\boldsymbol{\rho}}(\mathbf{r})\doteq\mathrm{Tr}\,\hat{\varGamma}_{(I)}^{[\beta,\boldsymbol{v}]}\hat{\boldsymbol{\rho}}(\mathbf{r})=\breve{\boldsymbol{\rho}}_{\mathrm{eq},(I)}^{[\beta,\boldsymbol{v}]}(\mathbf{r}), \qquad (66)$$

So, using Eqs.(54), (39),

$$\boldsymbol{\rho}_{\mathrm{eq},(I)}^{[\boldsymbol{v}]}(\mathbf{r})=\left(d_{(0,I)}^{[\boldsymbol{v}]}\right)^{-1}\sum_{K\in(0,I)}\boldsymbol{\rho}_{K}^{[\boldsymbol{v}]}(\mathbf{r}),\ \boldsymbol{\rho}_{K}^{[\boldsymbol{v}]}(\mathbf{r})=\left\langle\Psi_{K}^{[\boldsymbol{v}]}\middle|\hat{\boldsymbol{\rho}}(\mathbf{r})\middle|\Psi_{K}^{[\boldsymbol{v}]}\right\rangle. \qquad (67)$$

It is equal-weights ensemble (equi-ensemble) density of all pure ground state densities in the $\left\{\mathcal{N}_{(I)},\mathcal{S}_{(I)}\right\}$ subset of the Hilbert space). It should be noted that the expression (67) is insensitive to the choice of a basis in the subspace $\left\{\left|\varPsi_{K}^{[\boldsymbol{v}]}\right\rangle,K\in(0,I)\right\}$, and that the symmetry of this density is the same as the symmetry of $\boldsymbol{v}(\mathbf{r})$.

To make contact with the traditional SDFT let us define the energy functional for the $(I)$ macrostate

$$\begin{aligned}\forall\boldsymbol{\rho}\in\left\{\boldsymbol{\rho}_{(I)}\right\}\qquad \tilde{E}_{(I)}[\boldsymbol{\rho},\boldsymbol{v}]&=\underset{\hat{\varGamma}\in\left\{\hat{\varGamma}_{(I)}\right\}}{\mathrm{Min}}\left\{\mathrm{Tr}\,\hat{\varGamma}\hat{H}[\boldsymbol{v}]\middle|\mathrm{Tr}\,\hat{\varGamma}\hat{\boldsymbol{\rho}}=\boldsymbol{\rho}\right\}\\&=F_{\mathrm{L},(I)}[\boldsymbol{\rho}]+\int[\boldsymbol{v}\boldsymbol{\rho}]=\mathrm{Tr}\,\hat{\varGamma}_{(I)}^{[\boldsymbol{\rho},\boldsymbol{v}]}\hat{H}[\boldsymbol{v}],\end{aligned} \qquad (68)$$

see Eqs.(64), (63). Here $\hat{\Gamma}_{(I)}^{[\rho,v]}$ is the minimizer (in fact, for some argument $\boldsymbol{\rho}$ it may be not unique; a set of DM may play the role of minimizers). The ground-state energy of $(I)$ macrostate must coincide with the result of minimization

$$E_{(0,I)}^{[v]} = \underset{\rho\in\{\rho_{(I)}\}}{\text{Min}}\, \tilde{E}_{(I)}[\boldsymbol{\rho},\boldsymbol{v}] = \tilde{E}_{(I)}\left[\boldsymbol{\rho}_{\text{gs},(I)}^{[v]},\boldsymbol{v}\right] = F_{\text{L},(I)}\left[\boldsymbol{\rho}_{\text{gs},(I)}^{[v]}\right] + \int\left[\boldsymbol{v}\boldsymbol{\rho}_{\text{gs},(I)}^{[v]}\right]. \tag{69}$$

It is easy to check that the minimizer must be

$$\boldsymbol{\rho}_{\text{gs},(I)}^{[v]}(\mathbf{r}) = \text{Tr}\,\hat{\Gamma}_{\text{gs},(I)}^{[v]}\hat{\boldsymbol{\rho}}(\mathbf{r}), \tag{70}$$

where

$$\hat{\Gamma}_{\text{gs},(I)}^{[v]} = \left\{ \sum_{K\in(0,I)} p_{\text{gs},K}^{(I)} \left|\Psi_K^{[v]}\right\rangle\left\langle\Psi_K^{[v]}\right| \middle| p_{\text{gs},K}^{(I)} \geq 0,\ \sum_{K\in(0,I)} p_{\text{gs},K}^{(I)} = 1 \right\} \tag{71}$$

(see the comment after Eq.(6) about the range of such set of DMs). So the ground-state density

$$\boldsymbol{\rho}_{\text{gs},(I)}^{[v]}(\mathbf{r}) = \sum_{K\in(0,I)} p_{\text{gs},K}^{(I)}\boldsymbol{\rho}_K^{[v]}(\mathbf{r}), \tag{72}$$

is, in fact, a set of densities $\left\{\boldsymbol{\rho}_{\text{gs},(I)}^{[v]}\right\}$. The form (70) of the minimizer is due to the property of a mesostate, Eq.(17),

$$\hat{H}[\boldsymbol{v}]\left|\Psi_K^{[v]}\right\rangle = E_{(0,I)}^{[v]}\left|\Psi_K^{[v]}\right\rangle \qquad \forall K \in (0,I), \tag{73}$$

from which follows $\text{Tr}\,\hat{\Gamma}_{\text{gs},(I)}^{[v]}\hat{H}[\boldsymbol{v}] = E_{(0,I)}^{[v]}\sum_{K\in(0,I)} p_{\text{gs},K}^{(I)} = E_{(0,I)}^{[v]}$ .The group-theoretical view on Eq.(73) tells, that each $\Psi_{K\in(0,I)}^{[v]}$ belongs to the row of some $d_{(0,I)}^{[v]}$-dimensional irreducible representation of the symmetry group of the external potential energy operator $\hat{V} = \int[\boldsymbol{v}\hat{\boldsymbol{\rho}}]$, therefore the matrix elements of this operator are equal for various $K$

$$\left\langle\Psi_K\left|\int[\boldsymbol{v}\hat{\boldsymbol{\rho}}]\right|\Psi_K\right\rangle \equiv \int\left[\boldsymbol{v}\boldsymbol{\rho}_K^{[v]}\right] = \int\left[\boldsymbol{v}\boldsymbol{\rho}_{K'}^{[v]}\right], \qquad \forall K,K' \in (0,I). \tag{74}$$

From Eq.(73) we have also

$$E_{(0,I)}^{[\mathbf{v}]} = F_{\mathrm{int},K}^{[\mathbf{v}]} + \int \left[ \mathbf{v} \boldsymbol{\rho}_K^{[\mathbf{v}]} \right], \qquad \forall K \in (0, I) \tag{75}$$

where $F_{\mathrm{int},K}^{[\mathbf{v}]} = \left\langle \Psi_K^{[\mathbf{v}]} \middle| \hat{F}_{\mathrm{int}} \middle| \Psi_K^{[\mathbf{v}]} \right\rangle$. When combined with Eq.(74), Eq.(75) results in

$$F_{\mathrm{int},K}^{[\mathbf{v}]} = F_{\mathrm{int},K'}^{[\mathbf{v}]}, \quad \forall K, K' \in (0, I). \tag{76}$$

Therefore

$$F_{\mathrm{int},(I)} \left[ \boldsymbol{\rho}_{\mathrm{gs},(I)}^{[\mathbf{v}]} \right] = \mathrm{Tr}\, \hat{\Gamma}_{\mathrm{gs},(I)}^{[\mathbf{v}]} \hat{F}_{\mathrm{int}} = \sum_{K \in (0,I)} p_{\mathrm{gs},K}^{(I)} F_{\mathrm{int},K}^{[\mathbf{v}]} = F_{\mathrm{int},K'}^{[\mathbf{v}]}, \tag{77}$$

for any $K' \in (0, I)$. For the same reasons we have

$$F_{\mathrm{int},(I)} \left[ \boldsymbol{\rho}_{\mathrm{eq},(I)}^{[\mathbf{v}]} \right] = F_{\mathrm{int},K'}^{[\mathbf{v}]} = F_{\mathrm{int},(I)} \left[ \boldsymbol{\rho}_{\mathrm{gs},(I)}^{[\mathbf{v}]} \right]. \tag{78}$$

This result yields (using Eq.(61))

$$G_{(I)} \left[ \beta, \boldsymbol{\rho}_{\mathrm{eq},(I)}^{[\mathbf{v}]} \right] - G_{(I)} \left[ \beta, \boldsymbol{\rho}_{\mathrm{gs},(I)}^{[\mathbf{v}]} \right] \simeq S_{(I)}^{\left[ \boldsymbol{\rho}_{\mathrm{eq},(I)}^{[\mathbf{v}]} \right]} - S_{(I)}^{\left[ \boldsymbol{\rho}_{\mathrm{gs},(I)}^{[\mathbf{v}]} \right]} \geq 0, \tag{79}$$

since the equilibrium density maximizes the entropy over the set of the ground-state densities and only $\boldsymbol{\rho}_{\mathrm{eq},(I)}^{[\mathbf{v}]}(\mathbf{r})$ is the 0K limit, Eq.(66), of $\boldsymbol{\rho}_{(I)}^{[\beta;\mathbf{v}]}(\mathbf{r})$ – the maximizer in Eq.(46) and the minimizer in Eq.(47).

Obtained results for the $(I)$ macrostate are suitable to discuss the traditional SDFT, formulated in the Hilbert space of $\mathcal{N}$-electron wave functions, being, in addition, the eigenfunctions of $\hat{S}_z$, i.e. characterized by a fixed pair of integers $\boldsymbol{\mathcal{N}}_{(I)} = \left( \mathcal{N}_{(I)}, \mathcal{S}_{(I)} \right)$. As we see, the equilibrium density, which at zero temperature belongs to the set of the ground-state densities, Eq.(72), happens to be the equi-ensemble density, Eq.(67), having the symmetry of the external potential $\mathbf{v}(\mathbf{r})$. It should be stressed that the equilibrium density, $\boldsymbol{\rho}_{\mathrm{eq},(I)}^{[\mathbf{v}]}(\mathbf{r})$, is simultaneously the maximizer of the macrostate Massieu function and the minimizer of the macrostate spin Helmholtz function at low temperature. This is crucial fact, despite vanishing at 0K limit the contribution of the entropy $S_{(I)}^{\left[ \boldsymbol{\rho}_{(I)} \right]}$ to $F_{(I)} \left[ \beta, \boldsymbol{\rho}_{(I)} \right]$, Eq.(44).

To conclude this section, brief comparison with the established methods which are used to treat the multiplet problem seems appropriate.[20,25-29] Eq. (79) justifies Eq.(67) as the answer to the question how symmetry should be treated in DFT, the so called symmetry dilemma — the 2vec equilibrium density, Eq.(67), has to be used instead of some other ground-state density, Eq.(72), to be in agreement with the demand of the minimum energy and the maximum entropy. The generalized Hohenberg-Kohn theorem – the one-to-one mapping between the equilibrium density 2vec (admitting fractional electron-number 2vec, defined in the Fock space) and the external potential 2vec, was demonstrated in Paper II at *finite temperature* (here commented below Eq.(42)). At *the 0K limit* of the thermodynamic approach in the $\left\{\mathcal{N}_{(I)},\mathcal{S}_{(I)}\right\}$ sector of the Hiblert space this theorem reduces to two maps: $\boldsymbol{v}(\mathbf{r})\to\boldsymbol{\rho}_{\text{eq},(I)}(\mathbf{r})$ (one-to-one), and $\boldsymbol{\rho}_{\text{gs},(I)}(\mathbf{r})\to\boldsymbol{v}(\mathbf{r})$ (many-to-one). The last map can be shown as a direct generalization (to ensemble 2vec. ground-state densities) of the map $\rho_{\text{gs}}(\mathbf{r})\to v_{\text{ext}}(\mathbf{r})$ proven in Ref.[24] for densities derived from degenerate ground-state functions. The admissible functional spaces of density and potential 2vecs are described after Eq.(II.35).

# IV. THE SPIN-GRAND-CANONICAL ENSEMBLE AT 0K LIMIT

## A. Entropy representation

At finite temperature, the equilibrium state under the spin-grand-canonical conditions can be equivalently described by any one of the state functions – the vertices on the KΩBX surface of the mnemonic cube presented on Fig. 1. To analyze the 0K limit, we start with the entropy representation of the eq-DM operator $\hat{\Gamma}_{\text{eq}}^{[\hat{O}]}$ for $\hat{O}=\hat{O}[\beta,\boldsymbol{\alpha};\boldsymbol{v}]=-\beta\hat{H}[\boldsymbol{v}]-\boldsymbol{\alpha}\boldsymbol{\mathcal{N}}$, which can be rewritten in terms of macrostates, using Eqs.(18) – (20) as

$$\hat{\Gamma}_{\mathrm{eq}}^{[\beta,\alpha;\mathbf{v}]} = \frac{\exp\left(-\beta\hat{H}[\mathbf{v}] - \boldsymbol{\alpha}\hat{\boldsymbol{\mathcal{N}}}\right)}{\mathrm{Tr}\exp\left(-\beta\hat{H}[\mathbf{v}] - \boldsymbol{\alpha}\hat{\boldsymbol{\mathcal{N}}}\right)} = \sum_{I}\omega_{(I)}^{[\beta,\alpha;\mathbf{v}]}\hat{\Gamma}_{(I),\mathrm{eq}}^{[\beta,\mathbf{v}]}, \tag{80}$$

where the weight of the macrostate, Eq.(22), is evaluated in Appendix as

$$\omega_{(I)}^{[\beta,\alpha;\mathbf{v}]} \equiv \frac{\mathrm{Tr}\,\hat{g}_{(I)}\exp\left(-\beta\hat{H}[\mathbf{v}] - \boldsymbol{\alpha}\hat{\boldsymbol{\mathcal{N}}}\right)}{\mathrm{Tr}\exp\left(-\beta\hat{H}[\mathbf{v}] - \boldsymbol{\alpha}\hat{\boldsymbol{\mathcal{N}}}\right)} = \frac{\exp\left(Y_{(I)}^{[\beta;\mathbf{v}]} - \boldsymbol{\alpha}\boldsymbol{\mathcal{N}}_{(I)}\right)}{\sum_{J}\exp\left(Y_{(J)}^{[\beta;\mathbf{v}]} - \boldsymbol{\alpha}\boldsymbol{\mathcal{N}}_{(J)}\right)}. \tag{81}$$

It proves convenient to sort the macrostates (renumber them) in a lexicographical order which is based on $\left(E_{(0,I)}, \mathcal{N}_{(I)}, \mathcal{S}_{(I)}\right)$ (if $I < J$ than $E_{(0,I)}^{[\mathbf{v}]} \le E_{(0,J)}^{[\mathbf{v}]}$, if $E_{(0,I)}^{[\mathbf{v}]} = E_{(0,J)}^{[\mathbf{v}]}$ and $\mathcal{N}_{(I)} = \mathcal{N}_{(J)}$ than $\mathcal{S}_{(I)} < \mathcal{S}_{(J)}$). After dividing the numerator and denominator in Eq.(81) by $\left(Y_{(0)}^{[\beta;\mathbf{v}]} - \boldsymbol{\alpha}\boldsymbol{\mathcal{N}}_{(0)}\right)$ and using the asymptotic form of $Y_{(I)}$, Eq.(57), we have

$$\begin{aligned}\omega_{(I)}^{[\beta,\alpha;\mathbf{v}]} &= \frac{\exp\left(\Delta Y_{(I)}^{[\beta;\mathbf{v}]} - \boldsymbol{\alpha}\Delta\boldsymbol{\mathcal{N}}_{(I)}\right)}{\sum_{J}\exp\left(\Delta Y_{(J)}^{[\beta;\mathbf{v}]} - \boldsymbol{\alpha}\Delta\boldsymbol{\mathcal{N}}_{(J)}\right)} \\ &\simeq \frac{\exp\left(\Delta S_{(0,I)}^{[\mathbf{v}]} - \beta\Delta E_{(0,I)}^{[\mathbf{v}]} - \boldsymbol{\alpha}\Delta\boldsymbol{\mathcal{N}}_{(I)}\right)}{\sum_{J}\exp\left(\Delta S_{(0,J)}^{[\mathbf{v}]} - \beta\Delta E_{(0,J)}^{[\mathbf{v}]} - \boldsymbol{\alpha}\boldsymbol{\mathcal{N}}_{(J)}\right)}\end{aligned} \tag{82}$$

where $\Delta o_{(I)}$ is the difference between the value of the quantity $o$ in the $(I)$ macrostate and the $(0)$ macrostate, e.g., $\Delta\boldsymbol{\mathcal{N}}_{(I)} = \boldsymbol{\mathcal{N}}_{(I)} - \boldsymbol{\mathcal{N}}_{(0)}$, $\Delta E_{(0,I)} = E_{(0,I)} - E_{(0,0)}$.

Depending on the degeneracy of the lowest eigenenergy, two cases should be analyzed:

*Case A*. A non-degenerate case, for which $E_{(0,0)}^{[\mathbf{v}]} < E_{(0,1)}^{[\mathbf{v}]}$.

It means that $\Delta E_{(0,I)}^{[\mathbf{v}]} > 0$ for all macrostates different than the $(0)$ macrostate. Therefore, at 0K limit,

$$\omega_{(0)}^{[\alpha;\mathbf{v}]} = 1; \qquad \omega_{(I)}^{[\alpha;\mathbf{v}]} = 0 \quad \text{for} \quad I = 1, 2, \ldots. \tag{83}$$

Using the macrostate eq-DM from Eq.(54) we find

$$\left(\langle \hat{H} \rangle, \langle \hat{\mathcal{N}} \rangle, \langle \hat{\mathcal{S}} \rangle\right) \doteq \left(E^{[v]}_{(0,0)}, \mathcal{N}_{(0)}, \mathcal{S}_{(0)}\right) \tag{84}$$

These results are independent of $\boldsymbol{\alpha}$. When the magnetic field is absent, spin singlet state is nondegenerate, $\mathcal{S}_{(0)}$ is equal zero. In the presence of the magnetic field, $\mathcal{S}_{(0)}$ may be one of integer numbers admissible for the system with $\mathcal{N}_{(0)}$ electrons. Its value depends on the characteristic of the magnetic field.

*Case B*. A degenerate case. Here $E^{[v]}_{(0,0)} = E^{[v]}_{(0,1)} = ... = E^{[v]}_{(0,Q-1)} < E^{[v]}_{(0,Q)}$.

Excluding accidental degeneracy, in the *absence* of the magnetic field such situation can occur as the spin *Q*-plet, where, for $I < Q$, we observe $\mathcal{N}_{(I)} = \mathcal{N}_{(0)}$, $\mathcal{S}_{(I)} = \mathcal{S}_{(0)} + 2I$ with $\mathcal{S}_{(0)} = -Q+1$, and $S^{[v]}_{(0,I)} = S^{[v]}_{(0,0)}$ due to the fact that the spin *Q*-plet components belong to the same irreducible representation of the external potential symmetry group, thus the macrostate entropy is the same (e.g., in atomic case, the orbital multiplicity $(2L+1)$ is the same for all components of the electronic term). From Eq.(82) we find

$$\omega^{[\alpha;v]}_{(I)} = \frac{\exp(-2I\alpha_{\mathrm{S}})}{\sum_{K=0}^{Q-1} \exp(-2K\alpha_{\mathrm{S}})} \quad \text{for } 0 \le I \le (Q-1); \qquad \omega^{[\alpha;v]}_{(I)} = 0 \quad \text{for } I \ge Q. \tag{85}$$

Using the macrostate eq-DM from Eq.(54), we find the 0K limit of the average values:

$$\left(\langle \hat{H} \rangle, \langle \hat{\mathcal{N}} \rangle, \langle \hat{\mathcal{S}} \rangle\right) \doteq \left(E^{[v]}_{(0,0)}, \mathcal{N}_{(0)}, \bar{\mathcal{S}}[\alpha_{\mathrm{S}}]\right), \qquad \bar{\mathcal{S}} \in \left[\mathcal{S}_{(0)}, -\mathcal{S}_{(0)}\right]. \tag{86}$$

This result is independent of $\alpha_{\mathrm{N}}$ but it dependens on $\alpha_{\mathrm{S}}$. We have the following limits:

$$\lim_{\alpha_{\mathrm{S}} \to \pm\infty} \bar{\mathcal{S}}[\alpha_{\mathrm{S}}] = \mp \mathcal{S}_{(0)} = \pm (Q-1). \tag{87}$$

In the case of the carbon-like atom, an illustrative example, at 0K limit the lowest energy is for the anion, $\mathcal{N}_{(0)} = 7$, so $\left(\langle \hat{\mathcal{N}} \rangle, \langle \hat{\mathcal{S}} \rangle\right)$ are represented by points lying on the line between $(7,-3)$ and $(7,3)$, i.e., ${}^4\mathrm{S}^{\circ}_{0,-3/2}$ and ${}^4\mathrm{S}^{\circ}_{0,3/2}$ microstates of the spin quadruplet, the odd parity is indicated by degree symbol. In the *presence* of the magnetic field, the macrostate Helmholtz fn. is no longer an even function of the spin number; at 0K limit the ensemble consists only of one macrostate with the lowest ground-state energy (for the neutral carbon atom this is the macrostate evolved from ${}^3\mathrm{P}_{0,-1}$ at $B=0$). In concluding, for entropy representation, the spin-grand-canonical ensemble with given $\boldsymbol{\alpha}=(\alpha_\mathrm{N},\alpha_\mathrm{S})$ at 0K limit represents the ensemble of the macrostates belonging to the same electronic term (with the same total spin momentum) when the magnetic field is absent, or to the energetically lowest macrostate when the magnetic field is present.

## B. Energy representation

In the energy representation, the situation is more complicated. Using the Massieu-Planck transformation for changing variables, $\boldsymbol{\alpha}=-\beta\boldsymbol{\mu}$, the eq-DM operator has the form:

$$\hat{\Gamma}_{\mathrm{eq}}^{[\beta,\mu;v]}=\frac{\exp\left(-\beta\left(\hat{H}[\boldsymbol{v}]-\boldsymbol{\mu}\hat{\boldsymbol{\mathcal{N}}}\right)\right)}{\mathrm{Tr}\exp\left(-\beta\left(\hat{H}[\boldsymbol{v}]-\boldsymbol{\mu}\hat{\boldsymbol{\mathcal{N}}}\right)\right)}=\sum_I \omega_{(I)}^{[\beta,\mu;v]}\hat{\Gamma}_{(I)}^{[\beta,v]} \tag{88}$$

where the weight of the $(I)$ macrostate obtained as in Eq.(81) (see Appendix), is

$$\omega_{(I)}^{[\beta,\mu;v]}\equiv\frac{\mathrm{Tr}\,\hat{g}_{(I)}\exp\left(-\beta\left(\hat{H}[\boldsymbol{v}]-\boldsymbol{\mu}\hat{\boldsymbol{\mathcal{N}}}\right)\right)}{\mathrm{Tr}\exp\left(-\beta\left(\hat{H}[\boldsymbol{v}]-\boldsymbol{\mu}\hat{\boldsymbol{\mathcal{N}}}\right)\right)}=\frac{\exp\left(-\beta\left(E_{(I)}^{[\beta;v]}-\boldsymbol{\mu}\boldsymbol{\mathcal{N}}_{(I)}\right)+S_{(I)}^{[\beta;v]}\right)}{\sum_J\exp\left(-\beta\left(E_{(J)}^{[\beta;v]}-\boldsymbol{\mu}\boldsymbol{\mathcal{N}}_{(J)}\right)+S_{(J)}^{[\beta;v]}\right)}. \tag{89}$$

At fixed $\boldsymbol{\mu}$, the macrostates are renumbered in a new lexicographical order in which the energy of the $(0,I)$ mesostate is replaced by the combination $\left(E_{(0,I)}^{[v]}-\boldsymbol{\mu}\boldsymbol{\mathcal{N}}_{(I)}\right)$, namely

$$E_{(0,0)}^{[v]}-\boldsymbol{\mu}\boldsymbol{\mathcal{N}}_{(0)}\le E_{(0,1)}^{[v]}-\boldsymbol{\mu}\boldsymbol{\mathcal{N}}_{(1)}\le E_{(0,2)}^{[v]}-\boldsymbol{\mu}\boldsymbol{\mathcal{N}}_{(2)}\le E_{(0,3)}^{[v]}-\boldsymbol{\mu}\boldsymbol{\mathcal{N}}_{(3)}\le\ldots \tag{90}$$

This ordering depends strongly on the value of the 2vec source $\boldsymbol{\mu}=(\mu_\mathrm{N},\mu_\mathrm{S})$. After dividing the numerator and denominator in Eq.(89) by $\left(E_{(0)}^{[\beta,v]}-\boldsymbol{\mu}\boldsymbol{\mathcal{N}}_{(0)}\right)$, we have

$$\omega_{(I)}^{[\beta,\mu;v]} \simeq \frac{\exp\left(-\beta\left(\Delta E_{(I)}^{[\beta;v]} - \boldsymbol{\mu}\Delta\boldsymbol{\mathcal{N}}_{(I)}\right) + S_{(I)}^{[\beta;v]}\right)}{\exp\left(S_{(0)}^{[\beta;v]}\right) + \sum_{J\geq 1}\exp\left(-\beta\left(\Delta E_{(J)}^{[\beta;v]} - \boldsymbol{\mu}\Delta\boldsymbol{\mathcal{N}}_{(J)}\right) + S_{(J)}^{[\beta;v]}\right)}, \tag{91}$$

where $\Delta E_{(I)}^{[\beta;v]} = E_{(I)}^{[\beta;v]} - E_{(0)}^{[\beta;v]}$, $\Delta\boldsymbol{\mathcal{N}}_{(I)} = \boldsymbol{\mathcal{N}}_{(I)} - \boldsymbol{\mathcal{N}}_{(0)}$.

We introduce *the degeneracy indicator q of the ensemble ground state* (EGS) at 0K limit as the position of the first sharp inequality in Eq.(90), namely when

$$E_{(0,0)}^{[v]} - \boldsymbol{\mu}\boldsymbol{\mathcal{N}}_{(0)} = ... = E_{(0,\text{q-1})}^{[v]} - \boldsymbol{\mu}\boldsymbol{\mathcal{N}}_{(\text{q-1})} < E_{(0,\text{q})}^{[v]} - \boldsymbol{\mu}\boldsymbol{\mathcal{N}}_{(\text{q})} \leq ... \tag{92}$$

is satisfied. At 0K limit, we distinguish two types of EGS: the *isoelectronic* EGS (i-EGS), built of macrostates with the same number of electrons, and the *unisoelectronic* EGS (ui-EGS), when the macrostates with different number of electrons occur. The mapping of $(\mu_\text{N}, \mu_\text{S})$ into $(\mathcal{N}, \mathcal{S})$ at $\beta = 1/300\text{K}$ is shown on Fig. 3 as $\mathcal{N} = \mathcal{N}[\beta, \mu_\text{N}, \mu_\text{S}]$ (left panel) and $\mathcal{S} = \mathcal{S}[\beta, \mu_\text{N}, \mu_\text{S}]$ (right panel). These two functions are continuous and smooth. What is more, the points $(\mu_\text{N}, \mu_\text{S})$ are in one-to-one correspondence with the points $(\mathcal{N}, \mathcal{S})$, as it was discussed in general in Paper II. The 0K limit of these mappings illustrated on Fig. 4 and 5 shows strikingly different behavior: a two-dimensional set of $(\mu_\text{N}, \mu_\text{S})$ points is mapped into one $(\mathcal{N}, \mathcal{S})$ point, or one-dimensional set into one point, and so on. To discuss this in detail, a few cases are to be distinguished.

*Case A.* Absent degeneracy, $q = 1$. At 0K limit, we have

$$\omega_{(0)}^{[\mu;v]} = 1; \quad \omega_{(I)}^{[\mu;v]} = 0 \text{ for } I > 0 \tag{93}$$

and the expectation values of the operators $\left(\hat{H}, \hat{\mathcal{N}}, \hat{\mathcal{S}}\right)$ are

$$\left(\left\langle\hat{H}\right\rangle, \left\langle\hat{\mathcal{N}}\right\rangle, \left\langle\hat{\mathcal{S}}\right\rangle\right) \doteq \left(E_{(0,0)}^{[v]}, \mathcal{N}_{(0)}, \mathcal{S}_{(0)}\right). \tag{94}$$

When the condition of absent degeneracy is satisfied for some chosen point $(\mu_{\mathrm{N}}, \mu_{\mathrm{S}})$, it is satisfied also in the vicinity of this point. The whole region is defined by a set of inequalities:

$$
\begin{aligned}
&\ldots < -\frac{1}{2}\Delta E_{0,-2}^{[v]} < \mu_{\mathrm{S}} < \frac{1}{2}\Delta E_{0,+2}^{[v]} < \ldots, \quad \text{for } \Delta\mathcal{N}_{(I)} = 0 \\
&\ldots < -\left(\Delta E_{-1,l}^{[v]} - l\mu_{\mathrm{S}}\right) < \mu_{\mathrm{N}} < \Delta E_{+1,k}^{[v]} - k\mu_{\mathrm{S}} < \ldots, \text{ for } \Delta\mathcal{N}_{(I)} = \pm 1 \\
&\ldots < -\frac{1}{2}\left(\Delta E_{-2,l}^{[v]} - l\mu_{\mathrm{S}}\right) < \mu_{\mathrm{N}} < \frac{1}{2}\left(\Delta E_{+2,k}^{[v]} - k\mu_{\mathrm{S}}\right) < \ldots, \quad \text{for } \Delta\mathcal{N}_{(I)} = \pm 2 \\
&\ldots \\
&\ldots < -\frac{1}{\Delta\mathcal{N}}\left(\Delta E_{-\Delta\mathcal{N},l}^{[v]} - l\mu_{\mathrm{S}}\right) < \mu_{\mathrm{N}} < \frac{1}{\Delta\mathcal{N}}\left(\Delta E_{+\Delta\mathcal{N},k}^{[v]} - k\mu_{\mathrm{S}}\right) < \ldots, \text{ for } \Delta\mathcal{N}_{(I)} = \pm\Delta\mathcal{N}
\end{aligned}
\tag{95}
$$

with the notation $\Delta E_{\Delta\mathcal{N}_{(I)},\Delta\mathcal{S}_{(I)}}^{[v]} = E_{(0,I)}^{[v]} - E_{(0,0)}^{[v]}$ and with $\Delta\mathcal{S}_{(I)}$ represented by allowed integer $k$ or $l$ (i.e., of the same parity as $\Delta\mathcal{N}_{(I)}$). It should be noted that energy shifts in the first line of Eq.(95) for $\Delta\mathcal{S}_{(I)} = 2$ or -2 represent the spin down-flip (*sf*-) and spin up-flip energies (*sf*+), where $E_{(0)}^{sf-,[v]} = \Delta E_{0,-2}^{[v]}$ and $E_{(0)}^{sf+,[v]} = \Delta E_{0,+2}^{[v]}$, respectively [30]. In the second line, the differences of the energy on the left side is the ionization energy and on the right side is the electron affinity energy with respect to the reference $\left(\mathcal{N}_{(0)}, \mathcal{S}_{(0)}\right)$, where $\left(\Delta\mathcal{N}_{(I)}, \Delta\mathcal{S}_{(I)}\right)$ equals to $(-1, l)$ or $(+1, k)$, respectively. We suppose that the fulfillment of the first two inequalities results in the fulfillment of the inequalities for the remaining changes in the electron number. This supposition will be discussed in detail in the spin canonical ensemble.

In summary, at 0K limit, for all points $(\mu_{\mathrm{N}}, \mu_{\mathrm{S}})$ which lie in the interior of a convex polygon defined by inequalities (95) the expectation values of operators are constants (i.e., independent of $(\mu_{\mathrm{N}}, \mu_{\mathrm{S}})$) given in Eq.(95). For the carbon atom, (using notation $(0, I) \to \left[\mathcal{N}_{(I)}, \mathcal{S}_{(I)}\right]$) when $(\mu_{\mathrm{N}}, \mu_{\mathrm{S}}) \in$ $\left\{\left(E_{[6,0]}^{[v]} - E_{[6,2]}^{[v]}\right) = 0 < \mu_{\mathrm{S}} < \frac{1}{2}\left(E_{[6,4]}^{[v]} - E_{[6,2]}^{[v]}\right); -I - \mu_{\mathrm{S}} < \mu_{\mathrm{N}} < -A - \mu_{\mathrm{S}}\right\}$, we have $\left(\langle\hat{H}\rangle,\langle\hat{\mathcal{N}}\rangle,\langle\hat{\mathcal{S}}\rangle\right) \doteq \left(E_{[6,2]}^{[v]},6,2\right)$, and the trapezoid form of the polygon (see Fig. 4, right panel, the region having $(\mathcal{N},\mathcal{S})=(6,2)$). In general, the possible forms of the polygon extend between a trapezoid and a decagon in the case of magnetic field absence. Very rarely in the presence of a magnetic field, for the very high spin multiplets (as quintuplet, sextuplet, and so on), the polygon with more than ten sides can occur. In the $(\mu_S,\mu_N)$ plane, all these polygons have two sides parallel to $\mu_N$-axis. On the right panel of Fig. 4 and Fig. 5, the interiors of the polygons with integer $\mathcal{N}$ and $\mathcal{S}$ are presented. The first derivative of the spin Helmholtz free energy with respect to $\mu_S$ (i.e., the average spin number) is discontinuous on the boundaries of each polygon (see Fig. 4, right panel). The derivative with respect to $\mu_N$ (i.e., the average electron number) is discontinuous on the boundaries, but excluding the sides parallel to $\mu_N$-axis (see Fig. 4, left panel, Fig. 5, right panel).

*Case B.* The degeneracy is present, $q>1$. From Eq.(92), using $\exp\left(S_{(I)}^{[v]}\right)=d_{(0,I)}^{[v]}$ due to Eq.(55), we have at 0K limit

$$\omega_{(I)}^{[\mu;v]}=\frac{d_{(0,I)}^{[v]}}{d_{\text{sum}}} \text{ for } I<q;\ d_{\text{sum}}=\sum_{I=0}^{q-1} d_{(0,I)}^{[v]};\qquad \omega_{(I)}^{[\mu;v]}=0 \text{ for } I\geq q\,. \tag{96}$$

Using Eq.(54) for $\hat{\Gamma}_{(I)}^{[\beta,v]}$ and above equation on the 0K limit of Eq.(88) we find

$$\hat{\Gamma}_{\text{eq}}^{[\mu;v]}=\frac{1}{d_{\text{sum}}}\sum_{I=0}^{q-1} d_{(0,I)}^{[v]}\hat{\Gamma}_{(0,I)}^{[v]}\,. \tag{97}$$

So average values at 0K limit are

$$\left(\langle\hat{H}\rangle,\langle\hat{\mathcal{N}}\rangle,\langle\hat{\mathcal{S}}\rangle\right) \doteq \frac{1}{d_{\text{sum}}}\sum_{I=0}^{q-1} d_{(0,I)}^{[v]}\left(E_{(0,I)}^{[v]},\mathcal{N}_{(I)},\mathcal{S}_{(I)}\right). \tag{98}$$

These expressions can be simplified for special subcases. Absence of the magnetic field is assumed for them.

*Case B.1.* All macrostates are members of the spin *q*-plet. Therefore,

$$d_{(0,I)}^{[v]}=d_{(0,0)}^{[v]},\ E_{(0,I)}^{[v]}=E_{(0,0)}^{[v]},\ \mathcal{N}_{(I)}=\mathcal{N}_{(0)} \text{ and } \mathcal{S}_{(I)}=-(q-1)+2I \quad \text{for } I<q. \tag{99}$$

From Eq.(96), the weights are equal $1/q$ and from Eqs.(98) and (99),

$$\left(\langle\hat{H}\rangle,\langle\hat{\mathcal{N}}\rangle,\langle\hat{\mathcal{S}}\rangle\right)\doteq\left(E_{(0,0)}^{[v]},\mathcal{N}_{(0)},0\right). \tag{100}$$

From Eq.(99) used in Eq.(92) we find $\mu_{\mathrm{S}}=0$, while $\mu_{\mathrm{N}}$ is limited by the sharp inequality in the second line of Eq.(95). The $(\mu_{\mathrm{N}},\mu_{\mathrm{S}})$ points lie on the open red segments in Fig. 4, right panel. On example of the carbon-like atom, the red segment for $\mathcal{N}=5$ is due to the cation doublet, that for $\mathcal{N}=6$ – the neutral triplet, that for $\mathcal{N}=7$ – the anion quadruplet.

*Case B.2.* Here $q=2$, but two macrostates are not members of the spin doublet. The weights at 0K limit are

$$\omega_{(I)}^{[\mu;v]}=\frac{d_{(0,I)}^{[v]}}{d_{(0,0)}^{[v]}+d_{(0,1)}^{[v]}} \quad \text{for } I<2; \qquad \omega_{(I)}^{[\mu;v]}=0 \quad \text{for } I\ge 2; \tag{101}$$

For the i-EGS, $\mathcal{N}_{(0)}=\mathcal{N}_{(1)}=\mathcal{Z}$, the macrostates $(0)$ and $(1)$ are members of different spin multiplets. Then $E_{(0,1)}^{[v]}\neq E_{(0,0)}^{[v]}$ and it is possible to have $d_{(0,1)}^{[v]}\neq d_{(0,0)}^{[v]}$. So, average values at 0K limit are

$$\left(\langle\hat{H}\rangle,\langle\hat{\mathcal{N}}\rangle,\langle\hat{\mathcal{S}}\rangle\right)\doteq\left(\frac{\left(d_{(0,0)}^{[v]}E_{(0,0)}^{[v]}+d_{(0,1)}^{[v]}E_{(0,0)}^{[v]}\right)}{d_{(0,0)}^{[v]}+d_{(0,1)}^{[v]}},\mathcal{N}_{(0)},\frac{\left(d_{(0,0)}^{[v]}\mathcal{S}_{(0)}+d_{(0,1)}^{[v]}\mathcal{S}_{(1)}\right)}{d_{(0,0)}^{[v]}+d_{(0,1)}^{[v]}}\right). \tag{102}$$

It should be noted that $\langle\hat{\mathcal{S}}\rangle$ is no longer integer (contrary to $\langle\hat{\mathcal{N}}\rangle$). From the degeneracy condition, $E_{(0,0)}^{[v]}-\boldsymbol{\mu}\boldsymbol{\mathcal{N}}_{(0)}=E_{(0,1)}^{[v]}-\boldsymbol{\mu}\boldsymbol{\mathcal{N}}_{(1)}$, the spin source is determined

$$\mu_{\mathrm{S}}=\left(E_{(0,1)}^{[v]}-E_{(0,0)}^{[v]}\right)/\left(\mathcal{S}_{(1)}-\mathcal{S}_{(0)}\right), \tag{103}$$

while $\mu_{\mathrm{N}}$ is limited by the sharp inequality in the second line of Eq.(95). The $(\mu_{\mathrm{N}},\mu_{\mathrm{S}})$ points lie on open segments – common sites of two polygons, which are parallel to $\mu_{\mathrm{N}}$-axis (but different than a red segment) (see Fig. 5, left panel, open line segments parallel to $\mu_{\mathrm{N}}$-axis).

For the ui-EGS, $\mathcal{N}_{(0)}=\mathcal{N}_{(1)}\pm 1$, $\mu_{\mathrm{S}}$ is limited by the sharp inequality in the first line of Eq.(95). From equality in Eq.(92), we have

$$\mu_{\mathrm{N}}=\pm\left(E_{(0,1)}^{[v]}-E_{(0,0)}^{[v]}-\left(\mathcal{S}_{(1)}-\mathcal{S}_{(0)}\right)\mu_{\mathrm{S}}\right). \tag{104}$$

The set of points $(\mu_{\mathrm{N}},\mu_{\mathrm{S}})$ corresponding to this case is an open segment – a common site of two polygons, not parallel to $\mu_{\mathrm{N}}$-axis. $\langle\hat{\mathcal{N}}\rangle$ and $\langle\hat{\mathcal{S}}\rangle$ are no longer integer.

*Case B.3.* We consider $q=3$, so $E_{(0,0)}^{[v]}-\boldsymbol{\mu}\boldsymbol{\mathcal{N}}_{(0)}=E_{(0,1)}^{[v]}-\boldsymbol{\mu}\boldsymbol{\mathcal{N}}_{(1)}=E_{(0,2)}^{[v]}-\boldsymbol{\mu}\boldsymbol{\mathcal{N}}_{(2)}<\ldots.$. In addition, it is assumed that the three macrostates are members of three different spin multiplets, $E_{(0,0)}^{[v]}$, $E_{(0,1)}^{[v]}$, $E_{(0,2)}^{[v]}$ are different. For the ui-EGS, we can distinguish first the case $\mathcal{N}_{(0)}=\mathcal{N}_{(1)}=\mathcal{Z}$ and $\mathcal{N}_{(2)}=\mathcal{Z}\pm 1$. Only one point $(\mu_{\mathrm{N}},\mu_{\mathrm{S}})$ – a common vertex of three polygons, $\mu_{\mathrm{S}}=\left(E_{(0,1)}^{[v]}-E_{(0,0)}^{[v]}\right)\Big/\left(\mathcal{S}_{(1)}-\mathcal{S}_{(0)}\right)$ and $\mu_{\mathrm{N}}=\left(E_{(0,2)}^{[v]}-E_{(0,0)}^{[v]}-\left(\mathcal{S}_{(2)}-\mathcal{S}_{(0)}\right)\mu_{\mathrm{S}}\right)\Big/\left(\mathcal{N}_{(2)}-\mathcal{N}_{(0)}\right)$ satisfies the degeneracy conditions (see Fig. 4., right panel, black points on the non-red lines).

According to the supposition formulated after Eq.(95), the difference in the electron numbers between the macrostates may be equal $\pm 1$ or 0. However, the case when $\mathcal{N}_{(0)}=\mathcal{Z}$, $\mathcal{N}_{(1)}=\mathcal{Z}\pm 1$ and $\mathcal{N}_{(2)}=\mathcal{Z}\mp 1$, is not observed on Fig. 4 and 5. So for atomic or molecular systems with $q=3$, the ensemble consists of three macrostates, two of which have the same electron number.

*Case B.4.* We consider $E_{(0,0)}^{[v]}-\boldsymbol{\mu}\boldsymbol{\mathcal{N}}_{(0)}=\ldots=E_{(0,q-1)}^{[v]}-\boldsymbol{\mu}\boldsymbol{\mathcal{N}}_{(q-1)}<E_{(0,q)}^{[v]}-\boldsymbol{\mu}\boldsymbol{\mathcal{N}}_{(q)}$ and $q\geq 3$.

For ui-EGS, the ensemble of $q$ macrostates is assumed to be a mix of components belonging to the $Q$-plet and $(Q\pm 1)$-plet, where $Q=(q\mp 1)/2$. Their electron numbers differ by one. Only one point $(\mu_{\mathrm{N}},\mu_{\mathrm{S}})=\left(\left(E_{(0,Q)}^{[v]}-E_{(0,0)}^{[v]}\right)/\left(\mathcal{N}_{(Q)}-\mathcal{N}_{(0)}\right),0\right)$ corresponds to this case (see Fig. 4, right panel, black dots ending red segments). For the carbon atom example, when $q=3$, $\boldsymbol{\mathcal{N}}_{(0)}=(4,0)$, $\boldsymbol{\mathcal{N}}_{(1)}=(5,-1)$, $\boldsymbol{\mathcal{N}}_{(2)}=(5,1)$ then $(\mu_{\mathrm{N}},\mu_{\mathrm{S}})$ is the black dot between the red segment of $\mathcal{N}=5$ and the red polygon of $\mathcal{N}=4$; when $q=5$, the ensemble is a mix of doublet and triplet components (the cation and neutral form of carbon), then $(\mu_{\mathrm{N}},\mu_{\mathrm{S}})$ is the black dot between the red segments of $\mathcal{N}=6$ and $\mathcal{N}=5$.

For the i-EGS situation, we assume that the $q$ macrostates are members of one spin multiplet. We can distinguish two subcases. For the first, all $q$ macrostates have the same energy, $E_{(0,0)}^{[v]}=E_{(0,1)}^{[v]}=\ldots=E_{(0,q)}^{[v]}$. This was discussed in Case B.1. The second subcase, the situation $E_{(0,0)}^{[v]}< E_{(0,1)}^{[v]}<\ldots<E_{(0,q-1)}^{[v]}$ is considered. From the degeneracy condition and i-EGS situation we find

$$\mu_{\mathrm{S}}=\frac{\left(E_{(0,I)}^{[v]}-E_{(0,J)}^{[v]}\right)}{\left(\mathcal{S}_{(I)}-\mathcal{S}_{(J)}\right)}\quad \text{for } I,J<q. \tag{105}$$

Such situation is possible only in the presence of a uniform, linear (along *z*-direction) magnetic field. The weights and average values are as in Case B.1.

The presence of a nonuniform magnetic field may lower the value of the degeneracy indicator of the EGS so e.g., the red line on Fig. 4 can be shifted to values $\mu_{\mathrm{S}}\neq 0$. But the discussion presented in this section will be still valid.

# V. THE SPIN-CANONICAL ENSEMBLE AT 0K LIMIT

## A. Energy surface at 0K limit

At finite temperature, the equilibrium state under the spin-canonical conditions can be equivalently described by any one of the state functions – the vertices on the YAFG surface of the mnemonic cube presented on Fig. 1. To analyze the 0K limit, we start with the eq-DM operator, which can be rewritten in terms of macrostate contributions using Eqs.(18)-(20) as

$$\hat{\varGamma}_{\mathrm{eq}}^{[\beta,\boldsymbol{\mathcal{N}};\boldsymbol{v}]}=\frac{\exp\left(-\beta\hat{H}[\boldsymbol{v}]-\boldsymbol{\alpha}^{[\beta,\boldsymbol{\mathcal{N}};\boldsymbol{v}]}\left(\hat{\boldsymbol{\mathcal{N}}}-\boldsymbol{\mathcal{N}}\right)\right)}{\operatorname{Tr}\exp\left(-\beta\hat{H}[\boldsymbol{v}]-\boldsymbol{\alpha}^{[\beta,\boldsymbol{\mathcal{N}};\boldsymbol{v}]}\left(\hat{\boldsymbol{\mathcal{N}}}-\boldsymbol{\mathcal{N}}\right)\right)}=\sum_{I}\omega_{(I)}^{[\beta,\boldsymbol{\mathcal{N}};\boldsymbol{v}]}\hat{\varGamma}_{(I)}^{[\beta,\boldsymbol{v}]}, \tag{106}$$

where the macrostate weight is (see Eq.(23))

$$\begin{aligned}\omega_{(I)}^{[\beta,\boldsymbol{\mathcal{N}};\boldsymbol{v}]}&=\operatorname{Tr}\hat{g}_{(I)}\exp\left(-\beta\hat{H}[\boldsymbol{v}]-\breve{\boldsymbol{\alpha}}^{[\beta,\boldsymbol{\mathcal{N}};\boldsymbol{v}]}\left(\hat{\boldsymbol{\mathcal{N}}}-\boldsymbol{\mathcal{N}}\right)\right)\Big/\Xi^{[\beta,\boldsymbol{\mathcal{N}};\boldsymbol{v}]}\\&=\operatorname{Tr}\hat{g}_{(I)}\exp\left(-\beta\left(\hat{H}[\boldsymbol{v}]-\breve{\boldsymbol{\mu}}^{[\beta,\boldsymbol{\mathcal{N}};\boldsymbol{v}]}\left(\hat{\boldsymbol{\mathcal{N}}}-\boldsymbol{\mathcal{N}}\right)\right)\right)\Big/\Xi^{[\beta,\boldsymbol{\mathcal{N}};\boldsymbol{v}]}.\end{aligned} \tag{107}$$

These two forms are related through the Massieu-Planck transformation: $\breve{\boldsymbol{\alpha}}[\beta,\boldsymbol{\mathcal{N}};\boldsymbol{v}]=-\beta\breve{\boldsymbol{\mu}}[\beta,\boldsymbol{\mathcal{N}};\boldsymbol{v}]$.

The macrostate weights are involved in various sum rules, e.g. the normalization condition

$$\sum_{I}\omega_{(I)}^{[\beta,\boldsymbol{\mathcal{N}};\boldsymbol{v}]}=1, \tag{108}$$

the spin-canonical conditions 2vec

$$\sum_{I}\omega_{(I)}^{[\beta,\boldsymbol{\mathcal{N}};\boldsymbol{v}]}\boldsymbol{\mathcal{N}}_{(I)}=\boldsymbol{\mathcal{N}}\ , \tag{109}$$

the evaluation of $\langle\hat{H}\rangle$ and $\langle-\ln\hat{\varGamma}\rangle$ – the energy and the entropy of whole system in the macrostate resolution (see Eq.(14))

$$\sum_{I}\omega_{(I)}^{[\beta,\boldsymbol{\mathcal{N}};\boldsymbol{v}]}E_{(I)}^{[\beta;\boldsymbol{v}]}=E^{[\beta,\boldsymbol{\mathcal{N}};\boldsymbol{v}]}, \tag{110}$$

$$\sum_I \omega_{(I)}^{[\beta,\mathcal{N};v]}\left(S_{(I)}^{[\beta;v]}-\ln\omega_{(I)}^{[\beta,\mathcal{N};v]}\right)=S^{[\beta,\mathcal{N};v]}, \tag{111}$$

the components of the spin Helmholtz fn., $A^{[\beta,\mathcal{N};v]}=E^{[\beta,\mathcal{N};v]}-\beta^{-1}S^{[\beta,\mathcal{N};v]}$, Eq.(33).

As discussed in Paper I and II, $A[\beta,\boldsymbol{\mathcal{N}};\boldsymbol{v}]$ is an analytic, strictly convex function of $\boldsymbol{\mathcal{N}}$ at finite $\beta$. All admissible macrostate terms contribute to the average energy and the entropy in Eqs.(110) and (111). At 0K limit, the entropy contribution to the spin Helmholtz fn. vanishes due to $\beta^{-1}\to 0$, while the energy term, $E^{[\mathcal{N};v]}\doteq E^{[\beta,\mathcal{N};v]}$ is continuous and still a convex function of $\boldsymbol{\mathcal{N}}$ (but not strictly convex). As will be demonstrated, the energy surface $E^{[\mathcal{N};v]}=\sum_I \omega_{(I)}^{[\mathcal{N};v]}E_{(0,I)}^{[v]}$ (the 0K limit of Eq.(110)) is a kind of a “bowl”, a truncated polyhedron, a collection of polygons joined at their edges. Since $E_{(0,I)}^{[v]}$ are independent of $\boldsymbol{\mathcal{N}}$, form continuity of $E^{[\mathcal{N};v]}$ follows continuity of $\omega_{(I)}^{[\mathcal{N};v]}$ for each $(I)$. Fig. 2 shows, as an example, a part of the energy surface (left panel), and a projection of this surface on the $(\mathcal{N},\mathcal{S})$ plane (right panel), calculated at $B_z=0$ for the carbon-like atom. The polygons seen there are triangles and trapezoids (for non-coulombic systems, other polygons may be possible). A common side of two polygons is a place of discontinuity of the energy derivative w.r.t. $\mathcal{N}$ and/or $\mathcal{S}$. A set of these segments will be named the *discontinuity pattern* on the $(\mathcal{N},\mathcal{S})$ plane. To prove this important property of a polygon side, let us consider the energy surface as a function of $\boldsymbol{\mathcal{N}}$. From Eq.(134) (taken at 0K limit) follows the equivalence of the energy gradient and the external source 2vec, $\boldsymbol{\mu}[\boldsymbol{\mathcal{N}};\boldsymbol{v}]=\left(\partial E^{[\mathcal{N};v]}/\partial\boldsymbol{\mathcal{N}}\right)_{\boldsymbol{v}}$ $=\sum_I E_{(0,I)}^{[v]}\left(\partial\omega_{(I)}^{[\mathcal{N};v]}/\partial\boldsymbol{\mathcal{N}}\right)_{\boldsymbol{v}}$. As seen from Fig. 6 (confronted with the polygon set of Fig. 2), the components $\mu_{\mathrm{N}}^{[\mathcal{N};v]}$ and $\mu_{\mathrm{S}}^{[\mathcal{N};v]}$ of the source are constants for $\boldsymbol{\mathcal{N}}$ within each polygon, but one of the components or both change their value by leap when $\boldsymbol{\mathcal{N}}$ moves to the neighboring polygon (i.e., is crossing the polygon side). Since $E_{(0,I)}^{[v]}$ is independent of $\boldsymbol{\mathcal{N}}$ this leap is due to leaps of $\left(\partial\omega_{(I)}^{[\mathcal{N};v]}\big/\partial\boldsymbol{\mathcal{N}}\right)_{\mathbf{v}}$.

### B. Discontinuity pattern.

As will be shown, in general at 0K limit, for $(\mathcal{N},\mathcal{S})$ belonging to a particular open projected polygon, only macrostates having $\left(\mathcal{N}_{(I)},\mathcal{S}_{(I)}\right)$ lying on the sides of this polygon contribute to the ensemble with non-vanishing weight, $\omega_{(I)}^{[\mathcal{N};v]}\neq 0$. For $(\mathcal{N},\mathcal{S})$ belonging to an open segment — a projection of the particular edge joining two polygons of the energy surface, non-vanishing weights correspond to macrostates with $\left(\mathcal{N}_{(I)},\mathcal{S}_{(I)}\right)$ lying on the ends of this segment (and also macrostates inside it in the case of multiplets). The collected above properties of $\omega_{(I)}^{[\mathcal{N};v]}$ discussed at fixed $\boldsymbol{\mathcal{N}}$ and for various $(I)$ can be also viewed from another perspective: at chosen (fixed) macrostate $(I)$ and various $\boldsymbol{\mathcal{N}}$.

We define $\mathcal{C}_{(I)}^{[v]}$ – the *contributing region* of the macrostate $(I)$:

$$\mathcal{C}_{(I)}^{[v]}\equiv\left\{\boldsymbol{\mathcal{N}}\left|\omega_{(I)}^{[\mathcal{N};v]}>0\right.\right\},\tag{112}$$

i.e., the set of argument points $\boldsymbol{\mathcal{N}}$ of $\omega_{(I)}^{[\mathcal{N};v]}$ for which the contribution of $(I)$ macrostate to mean values at 0K limit in the sums like Eq.(110) and (111) is non-zero. This $\mathcal{C}_{(I)}^{[v]}$ is an open set, an interior of some polygon, which is a union of all projected energy-surface polygons having $(I)$ on their sides. The weight $\omega_{(I)}^{[\mathcal{N};v]}$ is zero for $\boldsymbol{\mathcal{N}}$ belonging to the polygon sides of $\mathcal{C}_{(I)}^{[v]}$ and outside this polygon. These polygon sides are the discontinuity lines of the energy derivatives, as they are sides of the mentioned projected polygons.

It proves convenient to define for two macrostates $(I)$ and $(J)$ the property of *non-coexistence*, as satisfaction of the condition

$$\mathcal{C}_{(I)}^{[v]} \cap \mathcal{C}_{(J)}^{[v]} = \varnothing , \tag{113}$$

equivalent to the condition

$$\forall \mathcal{N}, \quad \omega_{(I)}^{[\mathcal{N};v]} \omega_{(J)}^{[\mathcal{N};v]} = 0 . \tag{114}$$

The contributing regions of non-coexisting macrostates do not overlap. These regions may touch each other along some side, but this side does not belong to them because they are defined to be open sets. The negation of Eq.(113), namely the condition

$$\mathcal{C}_{(I)}^{[v]} \cap \mathcal{C}_{(J)}^{[v]} \neq \varnothing , \tag{115}$$

equivalent to

$$\exists \mathcal{N}, \quad \omega_{(I)}^{[\mathcal{N};v]} \omega_{(J)}^{[\mathcal{N};v]} > 0 , \tag{116}$$

defines for the macrostates $(I)$ and $(J)$ the property of *coexistence*. For two coexistent macrostates there exists such $\mathcal{N}$ region that the average-value summations (like for $E^{[\mathcal{N};v]} \doteq E^{[\beta,\mathcal{N};v]}$, Eq.(110)) include non-zero contributions of both macrostates. But in other $\mathcal{N}$ regions only one of macrostates may contribute or none.

The $(I)$ macrostate is defined to be *excluded* from ensembles at zero-temperature limit when

$$\mathcal{C}_{(I)}^{[v]} = \varnothing , \tag{117}$$

equivalently, when

$$\forall \mathcal{N}, \quad \omega_{(I)}^{[\mathcal{N};v]} = 0 . \tag{118}$$

The excluded $(I)$ does not enter any average-value summation at 0K limit (although it does contribute at finite temperature).

These properties of $\mathcal{C}_{(I)}^{[v]}$ can be investigated with the help of the following expression for the weight (see Appendix, Eq.(A3))

$$\omega_{(I)}^{[\beta,\mathcal{N};v]} = \exp\left(-\beta\left(\left(A_{(I)}^{[\beta;v]} - A^{[\beta,\mathcal{N};v]}\right) - \boldsymbol{\mu}^{[\beta,\mathcal{N};v]}\left(\boldsymbol{\mathcal{N}}_{(I)} - \boldsymbol{\mathcal{N}}\right)\right)\right), \qquad \forall I . \tag{119}$$

To calculate the weights at 0K limit, we have to know the asymptotic behavior of $\boldsymbol{\mu}^{[\beta,\mathcal{N};v]}$ for large $\beta$ (see Eq.(136)), but to find this, we have to know the same weights for all $(I)$, Eq.(134). To overcome this 'self-consistency' problem, we make such use of Eq.(119) that the dependence on $A^{[\beta,\mathcal{N};v]}$ and $\boldsymbol{\mu}^{[\beta,\mathcal{N};v]}$ is eliminated.

Relations between weights of four chosen macrostates $\{k\} \equiv (I_k)$, $k = 1,2,3,4$, will be deduced from the combination

$$p_{\{1,2,3,4\}}^{[\beta,\mathcal{N};v]} \equiv \frac{\omega_{\{1\}}^{[\beta,\mathcal{N};v]}\omega_{\{3\}}^{[\beta,\mathcal{N};v]}}{\omega_{\{2\}}^{[\beta,\mathcal{N};v]}\omega_{\{4\}}^{[\beta,\mathcal{N};v]}} = \exp\left(-\beta\left(\Delta A_{\{1,2,3,4\}}^{[\beta,\mathcal{N};v]} - \boldsymbol{\mu}^{[\beta,\mathcal{N};v]}\Delta\boldsymbol{\mathcal{N}}_{\{1,2,3,4\}}\right)\right), \tag{120}$$

where the dependence on $A^{[\beta,\mathcal{N};v]}$ is already eliminated. Here the notation $\Delta O_{\{1,2,3,4\}} \equiv O_{\{1\}} - O_{\{2\}} + O_{\{3\}} - O_{\{4\}}$ is introduced, where $O_{\{k\}}$ is the macrostate counterpart of $O$ defined for whole system, e.g., $A_{(I)}^{[\beta,\mathcal{N};v]}$, $S_{(I)}^{[\beta;v]}$, $E_{(I)}^{[\beta;v]}$, $\boldsymbol{\mathcal{N}}_{(I)}$ for $I = I_k$. The permutational symmetry $p_{\{1,2,3,4\}} = p_{\{3,2,1,4\}} = p_{\{3,4,1,2\}} = p_{\{1,4,3,2\}}$ is to be noted. For further considerations, four macrostates representing consecutive vertices of some *parallelogram* are chosen, because this choice results in $\Delta\boldsymbol{\mathcal{N}}_{\{1,2,3,4\}} = 0$. The dependence on $\boldsymbol{\mu}^{[\beta,\mathcal{N};v]}$ in Eq.(120) disappears, and the asymptotic form of $\ln p_{\{1,2,3,4\}}^{[\beta,\mathcal{N};v]}$ at large $\beta$ (see Eqs.(A6) and (A7)) is especially simple

$$\ln\left(p_{\{1,2,3,4\}}^{[\beta,\mathcal{N};v]}\right) = -\beta A_{\{1,2,3,4\}}^{[\beta,\mathcal{N};v]} \simeq -\beta\Delta E_{\{1,2,3,4\}}^{[v]} + \Delta S_{\{1,2,3,4\}}^{[v]} . \tag{121}$$

From possible limits of $p_{\{1,2,3,4\}}^{[\beta,\mathcal{N};v]}$ for $\beta \to \infty$, some conclusions about weights at 0K limit can be drown if $\Delta E^{[v]}_{\{1,2,3,4\}} > 0$ then $\quad p^{[\beta,\mathcal{N};v]}_{\{1,2,3,4\}} \doteq p^{[\mathcal{N};v]}_{\{1,2,3,4\}} = 0$, (122) if $\Delta E^{[v]}_{\{1,2,3,4\}} < 0$ then $\quad p^{[\beta,\mathcal{N};v]}_{\{1,2,3,4\}} \doteq +\infty$, (123) if $\Delta E^{[v]}_{\{1,2,3,4\}} = 0$ then $\quad p^{[\beta,\mathcal{N};v]}_{\{1,2,3,4\}} \doteq \exp\left(\Delta S^{[v]}_{\{1,2,3,4\}}\right)$. (124)

It should be noted that these 0K limits are independent of $\mathcal{N}$.

If $\Delta E^{[v]}_{\{1,2,3,4\}} > 0$ then either both the denominator and the numerator of $p^{[\beta,\mathcal{N};v]}_{\{1,2,3,4\}}$ approach zero but the denominator more slowly than the numerator or $\omega^{[\mathcal{N};v]}_{\{1\}}\omega^{[\mathcal{N};v]}_{\{3\}} = 0$ and $\omega^{[\mathcal{N};v]}_{\{2\}}\omega^{[\mathcal{N};v]}_{\{4\}} > 0$. A common feature of these cases can be written as $\omega^{[v]}_{\{1\}}\omega^{[v]}_{\{3\}} = 0$, showing the property that the $\{1\}$ macrostate and the $\{3\}$ macrostate are non-coexisting (see Eqs.(113) and (114)). When $\Delta E^{[v]}_{\{1,2,3,4\}} < 0$, then the $\{2\}$ macrostate and the $\{4\}$ macrostate are non-coexisting, as it follows from the previous conclusion because $p_{\{2,1,4,3\}} = 1/p_{\{1,2,3,4\}}$.

The analysis of Eq.(124) is similar but more extensive. The non-accidental $\Delta E^{[v]}_{\{1,2,3,4\}} = 0$ can happen when the macrostates $\{1\}$ and $\{2\}$ belong to one multiplet, while the macrostates $\{3\}$ and $\{4\}$ belong also to one multiplet (the same or a different one). Note $\Delta S^{[v]}_{\{1,2,3,4\}} = 0$, because the entropies $S^{[v]}_{\{k\}} = \ln d^{[v]}_{(0,\{k\})}$ are the same for all members of a multiplet, so $p^{[\mathcal{N};v]}_{\{1,2,3,4\}} = 1$. This result can be realized in various $\mathcal{N}$ regions as : (i) $\omega^{[\mathcal{N};v]}_{\{k\}} > 0$ for $k = 1,2,3,4$; (ii) one nonvanishing and three vanishing weights, e.g., $\omega^{[\mathcal{N};v]}_{\{1\}} > 1$, $\omega^{[\mathcal{N};v]}_{\{k\}} = 0$ for $k = 2,3,4$, but $\left(\omega^{[\beta,\mathcal{N};v]}_{\{2\}}\omega^{[\beta,\mathcal{N};v]}_{\{4\}}\right) \Big/ \omega^{[\beta,\mathcal{N};v]}_{\{3\}} \doteq \omega^{[\mathcal{N};v]}_{\{1\}}$; (iii) two nonvanishing and two vanishing weights, e.g., $\omega^{[\mathcal{N};v]}_{\{1\}} = \omega^{[\mathcal{N};v]}_{\{2\}} = 0$ but $\omega^{[\beta,\mathcal{N};v]}_{\{1\}} \Big/ \omega^{[\beta,\mathcal{N};v]}_{\{2\}} \doteq \omega^{[\mathcal{N};v]}_{\{4\}} \Big/ \omega^{[\mathcal{N};v]}_{\{3\}} > 0$; (iv) $\omega^{[\mathcal{N};v]}_{\{k\}} = 0$ for $k = 1,2,3,4$, but $p^{[\beta,\mathcal{N};v]}_{\{1,2,3,4\}} \doteq 1$.

It is obvious that for the case (i) we have $\bigcap_k \mathcal{C}_{\{k\}}^{[v]} \neq \varnothing$, thus meaning the coexistence of four macrostates at 0K limit. This property is true also for the case (ii) occurring for $\boldsymbol{\mathcal{N}}$ belonging to some triangle having $\{1\}$ as vertex, but $\{1\}$ being also a vertex of a trapezoid with $\{k\}$, $k = 2,3,4$ representing the vertices of this trapezoid that are not the triangle vertices. However, we can say nothing about connection of these points by a discontinuity segments (information about particular weights at 0K limit is needed). The case (iii) means that the macrostates with nonvanishing weights are coexisting and connected by discontinuity segment. They can belong either to one multiplet, or to two multiplets which differ by one electron. For the case (iv) $\bigcap_k \mathcal{C}_{\{k\}}^{[v]} = \varnothing$, what means that macrostates belonging to two multiplets: one with $\mathcal{N}_{\{1\}} = \mathcal{N}_{\{2\}} = \mathcal{Z}$, the second with $\mathcal{N}_{\{3\}} = \mathcal{N}_{\{4\}} = \mathcal{Z} + k,\ k \geq 2$, are non-coexisting.

Having these relations, it seems convenient to identify first the excluded macrostates and the discontinuity pattern for the whole set of admissible macrostates entering ensemble, and then to analyze the discontinuity pattern in the vicinity of a given point $\boldsymbol{\mathcal{N}} = (\mathcal{N}, \mathcal{S})$. The first step can be done by investigation of parallelograms of various shapes.

Initially, we suppose that all admissible macrostate points are joined by discontinuity segments, i.e., that each pair of macrostates coexist at 0K limit. The macrostate will be characterized by the label $\left[\mathcal{N}_{(I_k)}, \mathcal{S}_{(I_k)}\right] \equiv \{k\} = (I_k)$ showing its position on the $(\mathcal{N}, \mathcal{S})$ plane. We recall conclusion from Sec.III.D that the mesostate description and macrostate description provide the same information about the equilibrium state at 0K limit. This means that (at this limit) the average macrostate energy is equal to the lowest energy of the mesostates belonging to this macrostate.

*Shape 1*. Let us consider $\boldsymbol{\mathcal{N}}_{\{2\}} = \boldsymbol{\mathcal{N}}_{\{4\}} = (\mathcal{Z}, \Sigma)$, a parallelogram degenerated to a linear object – a segment with $\boldsymbol{\mathcal{N}}_{\{1\}}$ and $\boldsymbol{\mathcal{N}}_{\{3\}}$ on its ends and $\boldsymbol{\mathcal{N}}_{\{2\}}$ in the middle of this segment. Now

$$\Delta E^{[v]}_{\{1,2,3,2\}} = \left( E^{[v]}_{\{1\}} + E^{[v]}_{\{3\}} \right) - 2E^{[v]}_{\{2\}} . \tag{125}$$

If $\Delta E^{[v]}_{\{1,2,3,2\}} > 0$, from the analysis of Eq.(122) follows non-coexistence of the $\{1\}$ macrostate and the $\{3\}$ macrostate — they cannot be connected. This means that we need to remove the (supposed) energy discontinuity segment between $\{1\}$ and $\{3\}$. If $\Delta E^{[v]}_{\{1,2,3,2\}} < 0$, from the analysis of Eq.(123) follows $\mathcal{C}^{[v]}_{\{2\}} = \varnothing$, i.e. the exclusion of $\{2\}$. The $\left( \mathcal{N}_{\{2\}}, \mathcal{S}_{\{2\}}, E^{[v]}_{(0,\{2\})} \right)$ point does not belong to the energy surface $E\left[\boldsymbol{\mathcal{N}}; \boldsymbol{v}\right]$ vs. $\boldsymbol{\mathcal{N}}$ – the convex hull at 0K limit. This point is situated above this surface, so it cannot be the vertex of any polygon of the energy surface. The $\Delta E^{[v]}_{\{1,2,3,2\}} = 0$ case is realized only by three macrostates belonging to one spin multiplet and these macrostates coexist (Shape 1.A case).

*Shape 1.A.* An isoelectronic case, $\mathcal{N}_{\{k\}} = \mathcal{Z}$, $k \in \{1,2,3,2\}$ . Only macrostates with $\mathcal{S}_{\{1\}} = \Sigma - 2l$ and $\mathcal{S}_{\{3\}} = \Sigma + 2l$, $l \in \{1,2,...\}$, if they are admissible, can fulfill $\Delta\boldsymbol{\mathcal{N}}_{\{1,2,3,2\}} = 0$.

The sign of $\Delta E^{[v]}_{\{1,2,3,2\}}$ determines the property of convexity (or lack of it) for the ground-state energy $E^{[v]}_{(0,\{k\})}$ as a function of the macrostate spin number $\mathcal{S}_{\{k\}}$ at fixed electron number. At $B_{\mathrm{z}} = 0$, the (non-strict) convexity of the straight-line connection can be expected, but not guaranteed. This will be discussed on the example of a system with odd $\mathcal{Z} \geq 5$, for which the energy of the spin doublet is the lowest eigenvalue of the Hamiltonian, $E^{[v]}_{(0,[\mathcal{Z},-1])} = E^{[v]}_{(0,[\mathcal{Z},+1])}$. The mesostate $\left( E^{[v]}_{(0,[\mathcal{Z},-3])}, \mathcal{Z}, -3 \right)$ needs to be the member of the next higher multiplet – the quadruplet, and the mesostate $\left( E^{[v]}_{(0,[\mathcal{Z},-3])}, \mathcal{Z}, -5 \right)$ – the sextuplet. So,

$$E^{[v]}_{(0,[\mathcal{Z},-5])} > E^{[v]}_{(0,[\mathcal{Z},-3])} > E^{[v]}_{(0,[\mathcal{Z},-1])} = E^{[v]}_{(0,[\mathcal{Z},1])} < E^{[v]}_{(0,[\mathcal{Z},3])} < E^{[v]}_{(0,[\mathcal{Z},5])} , \tag{126}$$

are satisfied. The convexity of $E^{[v]}_{(0,[\mathcal{Z},\Sigma])}$ vs. $\Sigma$ for $\Sigma=-3,-1,1,3$ is obvious from Eq.(126), and $\Delta E^{[v]}_{\{1,2,3,2\}}=E^{[v]}_{(0,[\mathcal{Z},-3])}-E^{[v]}_{(0,[\mathcal{Z},-1])}>0$ for $\mathcal{S}_{\{2\}}=\Sigma=-1$. This statement cannot be extended for $|\Sigma|>1$. If $E^{[v]}_{(0,[\mathcal{Z},-5])}-E^{[v]}_{(0,[\mathcal{Z},-3])}>E^{[v]}_{(0,[\mathcal{Z},-3])}-E^{[v]}_{(0,[\mathcal{Z},-1])}$, then the convexity is true for $\Sigma=-5,-3,-1$, else the $\left(E^{[v]}_{(0,[\mathcal{Z},-3])},\mathcal{Z},-3\right)$ mesostate have to be excluded, $\Delta E^{[v]}_{\{1,2,3,2\}}<0$. For even $\mathcal{Z}$, the discussion of the convexity can start from a singlet, go via triplet, quintuplet, and so on. In general, we have

$$E^{[v]}_{(0,[\mathcal{Z},\pm\mathcal{Z}])}\geq E^{[v]}_{(0,[\mathcal{Z},\pm(\mathcal{Z}-2)])}\geq\ldots, \tag{127}$$

but this does not guaranteed the convexity of $E^{[v]}_{(0,[\mathcal{Z},\Sigma])}$ vs. $\Sigma$ (but the convexity of $E[\mathcal{N},\mathcal{S};\boldsymbol{v}]$ vs. $\mathcal{S}$ is guaranteed by the property of the state function, $A[\beta,\boldsymbol{\mathcal{N}};\boldsymbol{v}]$)

In some cases, the occurrence of $\Delta E^{[v]}_{\{1,2,3,2\}}<0$ does not mean necessarily that the convexity is violated, but only that the data used for its evaluation were incorrectly classified. As an example, let us consider the even $\mathcal{Z}$-electron system showing the triplet ground state when its energy is calculated using a single-determinant method (the Hartree-Fock or the Kohn-Sham approach). Let $\Phi_{[\mathcal{Z},\Sigma]}$ be the Slater determinant for the system characterized by $\boldsymbol{\mathcal{N}}=(\mathcal{Z},\Sigma)$ and $E\left[\Phi_{[\mathcal{Z},\Sigma]}\right]$ denotes the calculated energy. The energy $E\left[\Phi_{[\mathcal{Z},0]}\right]$ is found higher than the energy $E\left[\Phi_{[\mathcal{Z},+2]}\right]$. It should not be classified as $E^{[v]}_{(0,[\mathcal{Z},0])}$, but as $E^{[v]}_{(1,[\mathcal{Z},0])}$. For the correctly classified mesostate energies, $E^{[v]}_{(0,[\mathcal{Z},-2])}=E^{[v]}_{(0,[\mathcal{Z},0])}=E^{[v]}_{(0,[\mathcal{Z},2])}=E\left[\Phi_{[\mathcal{Z},+2]}\right]$ is the triplet state energy.

As an illustration of $\Delta E^{[v]}_{\{1,2,3,2\}}>0$ case, we take the macrostates $\{\{1\},\{2\},\{3\}\}=\{[6,-4],[6,-2],[6,0]\}$ (first three red points on $\mathcal{Z}=6$ line, Fig. 2) (only two of them are shown). $\mathcal{C}^{[v]}_{[6,-4]}$ is a union of all triangles with a common vertex $[6,-4]$. $\mathcal{C}^{[v]}_{[6,0]}$ is a union of two trapezoids having a common triplet side at $\mathcal{Z}=6$. These two contributing regions do not overlap (they are defined to be open regions), so the $[6,-4]$ and $[6,0]$macrostates do not coexist at 0K limit, $\omega^{[\mathcal{N};v]}_{[6,-4]}\omega^{[\mathcal{N};v]}_{[6,0]}=0$.

The case $\Delta E^{[v]}_{\{1,2,3,2\}}=0$ is realized (at $B_z=0$) by macrostates belonging to one spin multiplet: for even $\mathcal{Z}$ it may be a triplet with the spin numbers $-2,0,2$ (see $\mathcal{Z}=6$ or $\mathcal{Z}=8$ line on Fig. 2), or a quintuplet with spin numbers $-4,-2,0$ or $-2,0,2$, or $0,2,4$, or $-4,0,4$; for odd $\mathcal{Z}$ — a quadruplet with spin numbers $-3,-1,1$ or $-1,1,3$ (see $\mathcal{Z}=7$ line on Fig. 2). Eq.(124) can be rewritten as $\omega^{[\beta,\mathcal{N};v]}_{\{1\}}\omega^{[\beta,\mathcal{N};v]}_{\{3\}}\Big/\left(\omega^{[\beta,\mathcal{N};v]}_{\{2\}}\right)^2\doteq 1$. As was discussed, this case can be realized either by all non-zero weights or through one nonvanishing and three vanishing weights. The first realization happens when the point $(\mathcal{N},\mathcal{S})$ is lying on the open segment joining all macrostates belonging to one spin multiplet, all these macrostates contribute to the ensemble characterized by average values equal this $(\mathcal{N},\mathcal{S})$, e.g., the point lying on the "quadruplet" segment (see $\mathcal{Z}=7$ line on Fig. 2). The second realization may happen only when the weight of the vertex macrostate from numerator of $p$ is non-zero, e.g. $\omega^{[\mathcal{N};v]}_{\{1\}}>0$ for $\{1\}=[7,-3]$, and the point $(\mathcal{N},\mathcal{S})$ is lying on the common side, e.g., the segment joining $[7,-3]$ with $[6,-2]$ of the trapezoid $\{[7,-3],[7,3],[6,2],[6,-2]\}$ and the triangle $\{[7,-3],[6,-2],[6,-4]\}$. Because this segment is a side of $\mathcal{C}^{[v]}_{\{2\}}$, $\{2\}=[7,-1]$, it is the discontinuity segment.

It should be mentioned that equalities $\Delta E^{[v]}_{\{1,2,3,2\}}=0$ and $\Delta S^{[v]}_{\{1,2,3,4\}}=0$ for the members of a multiplet remain true when a homogenous collinear magnetic field is present.

*Shape 1.B.* The considered parallelogram is also degenerate, but with differing electron numbers of participating macrostates. The $\Delta\mathcal{N}_{\{1,2,3,2\}} = 0$ condition is fulfilled by $\{\{1,\},\{2\},\{3\}\} = \{[\mathcal{Z} - \Delta\mathcal{N}, \Sigma - \Delta\mathcal{S}],[\mathcal{Z},\Sigma],[\mathcal{Z} + \Delta\mathcal{N}, \Sigma + \Delta\mathcal{S}]\}$. For simplicity, let us assume $\Delta\mathcal{N} = \Delta\mathcal{S} = 1$. Now

$$\Delta E^{[v]}_{\{1,2,3,2\}} = \left(E^{[v]}_{(0,[\mathcal{Z}+1,\Sigma+1])} + E^{[v]}_{(0,[\mathcal{Z}-1,\Sigma-1])}\right) - 2E^{[v]}_{(0,[\mathcal{Z},\Sigma])}. \tag{128}$$

The case $\Delta E^{[v]}_{\{1,2,3,2\}} = 0$ is uninteresting, as it may happen only accidentally. The situation of $\Delta E^{[v]}_{\{1,2,3,2\}} > 0$ means convexity of the ground-state energy as function of the macrostate position points $\mathcal{N}_{\{k\}}$ lying on a straight line. The $\{1\}$ macrostate and the $\{3\}$ macrostate do not coexist for any value of $\mathcal{N}$. When $\Delta E^{[v]}_{\{1,2,3,2\}} < 0$, the convexity is violated, the $\{2\}$ macrostate is excluded from the ensembles formed at 0K limit. This means that the ensemble with $\mathcal{N} = (\mathcal{Z},\Sigma)$ is realized at 0K limit by the mixture of the $[\mathcal{Z}+1,\Sigma+1]$ and $[\mathcal{Z}-1,\Sigma-1]$ macrostates. This is a rare situation, not occurring for coulombic systems to our best knowledge (see, in particular, Ref.[31]). For example, there are no excluded macrostate points on Fig. 2.

In concluding, the presence of the excluded macrostates can be detected from the Shape 1.A and Shape 1.B analysis performed for all admissible macrostates. For the analysis of the other shapes (given below), it is assumed that all involved macrostates are not excluded and that they are vertices of polygons which built the convex hull at 0K limit (or are members of multiplets).

*Shape 2.* A special parallelogram – a square – is considered, with vertices $\{\{1,\},\{2\},\{3\},\{4\}\} = \{[\mathcal{Z}+1,\Sigma+1],[\mathcal{Z},\Sigma+2],[\mathcal{Z}-1,\Sigma+1],[\mathcal{Z},\Sigma]\}$. For it

$$\Delta E^{[v]}_{\{1,2,3,4\}} = \left(E^{[v]}_{(0,[\mathcal{Z}+1,\Sigma+1])} + E^{[v]}_{(0,[\mathcal{Z}-1,\Sigma+1])}\right) - \left(E^{[v]}_{(0,[\mathcal{Z},\Sigma])} + E^{[v]}_{(0,[\mathcal{Z},\Sigma+2])}\right). \tag{129}$$

Again, the case $\Delta E^{[v]}_{\{1,2,3,4\}} = 0$ is uninteresting as accidental. Satisfaction of $\Delta E^{[v]}_{\{1,2,3,4\}} > 0$, means $\mathcal{C}^{[v]}_{\{1\}} \cap \mathcal{C}^{[v]}_{\{3\}} = \varnothing$, the supposed earlier discontinuity line joining the $[\mathcal{Z}+1, \Sigma+1]$ and $[\mathcal{Z}-1, \Sigma+1]$ macrostates should be removed. For example, in the square with vertices $\{\{1\},\{2\},\{3\},\{4\}\} = \{[6,2],[5,3],[4,2],[5,1]\}$, the $[6,2]$ and $[4,2]$ points are seen disconnected on Fig. 2.

The satisfaction of $\Delta E^{[v]}_{\{1,2,3,4\}} < 0$ means that the $[\mathcal{Z}, \Sigma]$ point and the $[\mathcal{Z}, \Sigma+2]$ point are disconnected (no discontinuity line joining them). It is believed that this case does not occur for coulombic systems: all isoelectronic points are connected on Fig. 2. In general, we observe that $\mathcal{C}^{[v]}_{[\mathcal{Z},\Sigma]} \cap \mathcal{C}^{[v]}_{[\mathcal{Z}\pm 2,\Sigma']} = \varnothing$ and that all macrostate points with the same (integer) electron number belong to the discontinuity line.

An interpretation of the condition $\Delta E^{[v]}_{\{1,2,3,4\}} > 0$ can be done in the case of $(\mathcal{Z}, \Sigma) \equiv (\mathcal{Z}, 0)$ macrostate with the $(0,[\mathcal{Z},0])$ mesostate representing the ground-state of the closed-shell system of electrons (spin singlet). Eq.(129) with $\Sigma = 0$ can be rewritten in terms of the hardness $\eta^{[v]}$ — the difference between the ionization energy and the electron affinity, $\eta^{[v]} = E^{[v]}_{(0,[\mathcal{Z}+1,1])} + E^{[v]}_{(0,[\mathcal{Z}-1,1])} - 2E^{[v]}_{(0,[\mathcal{Z},0])}$, and in terms of the singlet-triplet excitation energy, $E^{[v]}_{\mathrm{ST}} = E^{[v]}_{(0,[\mathcal{Z},2])} - E^{[v]}_{(0,[\mathcal{Z},0])} = E^{[v]}_{(0,[\mathcal{Z},-2])} - E^{[v]}_{(0,[\mathcal{Z},0])}$, namely as $\Delta E^{[v]}_{\{1,2,3,4\}} = \eta^{[v]} - E^{[v]}_{\mathrm{ST}}$. Now, $\Delta E^{[v]}_{\{1,2,3,4\}} > 0$ means that the hardness is larger than the singlet-triplet excitation energy. Validity of such inequality for coulombic systems is confirmed by the experimental and calculated data data,[32] and theoretical study.[33] This means also the convexity of the energy (averaged over $\mathcal{S}_{(I)}$ to $\mathcal{S} = 1$) $E^{[\mathcal{N},1;v]}$ as a function of $\mathcal{N}$, satisfied for three values of $\mathcal{N} = \mathcal{Z}-1, \mathcal{Z}, \mathcal{Z}+1$, namely $E^{[\mathcal{Z},1;v]} = \frac{1}{2}\left(E^{[v]}_{(0,[\mathcal{Z},2])} + E^{[v]}_{(0,[\mathcal{Z},0])}\right) < \frac{1}{2}\left(E^{[\mathcal{Z}+1,1;v]} + E^{[\mathcal{Z}-1,1;v]}\right)$

$= \frac{1}{2}\left(E^{[v]}_{(0,[\mathcal{Z}+1,1])} + E^{[v]}_{(0,[\mathcal{Z}-1,1])}\right)$.

*Shape 3.* A parallelogram with two horizontal sites, having vertices $\{\{1,\},\{2\},\{3\},\{4\}\} = \{[\mathcal{Z},\Sigma],[\mathcal{Z}+1,\Sigma+1],[\mathcal{Z}+1,\Sigma+3],[\mathcal{Z},\Sigma+2]\}$ is considered. For it, $\Delta E^{[v]}_{\{1,2,3,2\}}$ can be rewritten as

$$\Delta E^{[v]}_{\{1,2,3,4\}} = \left(E^{[v]}_{(0,[\mathcal{Z}+1,\Sigma+3])} - E^{[v]}_{(0,[\mathcal{Z}+1,\Sigma+1])}\right) - \left(E^{[v]}_{(0,[\mathcal{Z},\Sigma+2])} - E^{[v]}_{(0,[\mathcal{Z},\Sigma])}\right). \tag{130}$$

If $\Delta E^{[v]}_{\{1,2,3,4\}} > 0$, the $[\mathcal{Z}+1,\Sigma+3]$ and $[\mathcal{Z},\Sigma]$ macrostates are disconnected. This occurs on Fig. 2 for the parallelogram with vertices $\{\{1,\},\{2\},\{3\},\{4\}\} = \{[6,2],[7,3],[7,5],[6,4]\}$: the $[6,2]$ and $[7,5]$ macrostates are not joined by discontinuity line. If $\Delta E^{[v]}_{\{1,2,3,4\}} < 0$ ($\Delta E^{[v]}_{\{2,1,4,3\}} > 0$ ), the $[\mathcal{Z}+1,\Sigma+1]$ macrostate and the $[\mathcal{Z},\Sigma+2]$ macrostate are disconnected. This occurs on Fig. 2 for the parallelogram with vertices $\{\{1,\},\{2\},\{3\},\{4\}\} = \{[5,1],[6,2],[6,4],[5,3]\}$: the $[6,2]$ and $[5,3]$ macrostates do not coexist.

The case $\Delta E^{[v]}_{\{1,2,3,4\}} = 0$ means (see Eq.(130)) that

$$\left(E^{[v]}_{(0,[\mathcal{Z}+1,\Sigma+3])} - E^{[v]}_{(0,[\mathcal{Z}+1,\Sigma+1])}\right) = \left(E^{[v]}_{(0,[\mathcal{Z},\Sigma+2])} - E^{[v]}_{(0,[\mathcal{Z},\Sigma])}\right). \tag{131}$$

Such case is not accidental only if the macrostates $[\mathcal{Z},\Sigma]$ and $[\mathcal{Z},\Sigma+2]$ are members of one multiplet, while the macrostates $[\mathcal{Z}+1,\Sigma+1]$ and $[\mathcal{Z}+1,\Sigma+3]$ — of another multiplet, e.g., the members of triplet and quintuplet. As noted at Shape 1.A, entropies of all members of one multiplet are equal, so now $S^{[v]}_{[\mathcal{Z},\Sigma+2]} = S^{[v]}_{[\mathcal{Z},\Sigma]}$, $S^{[v]}_{[\mathcal{Z}+1,\Sigma+3]} = S^{[v]}_{[\mathcal{Z}+1,\Sigma+1]}$, therefore $\Delta S^{[v]}_{\{1,2,3,2\}}$=0, yielding

$$\frac{\omega^{[\mathcal{N};v]}_{[\mathcal{Z},\Sigma]}}{\omega^{[\mathcal{N};v]}_{[\mathcal{Z},\Sigma+2]}} = \frac{\omega^{[\mathcal{N};v]}_{[\mathcal{Z}+1,\Sigma+3]}}{\omega^{[\mathcal{N};v]}_{[\mathcal{Z}+1,\Sigma+1]}}. \tag{132}$$

Because this relation holds also for other parallelograms with different choice of multiplet members, as long as $\Delta\mathcal{N}_{\{1,2,3,2\}} = 0$, we can conclude that there is no energy derivative discontinuity segments inside parallelograms and the union of these parallelograms is a trapezoid. Other parallelograms with two horizontal sites, but with remaining sides inclined in the opposite direction, e.g. $\{\{1,\},\{2\},\{3\},\{4\}\} =$ $\{[\mathcal{Z},\Sigma],[\mathcal{Z}+1,\Sigma-1],[\mathcal{Z}+1,\Sigma+1],$ $[\mathcal{Z},\Sigma+2]\}$, can be analyzed similarly. Knowing the discontinuity pattern (for given external potential $\boldsymbol{v}$), we can calculate the weights and the external sources conjugate to the $(\mathcal{N},\mathcal{S})$ variables at 0K limit.

## C. The zero-temperature limit of macrostate external sources conjugate to $(\mathcal{N},\mathcal{S})$ variables, and of macrostate weights

The 2vec external source conjugated to the $(\mathcal{N},\mathcal{S})$ variables (the Lagrange multiplier 2vec at the operator $\hat{\mathcal{N}}$) can be calculated according to Eq.(II.18) (see Eq.(31) for $Y$, Eq.(119) for $\omega_{(I)}$, Eq.(30) for $Y_{(I)}$)

$$\breve{\boldsymbol{\alpha}}[\beta,\mathcal{N};\boldsymbol{v}] = \left(\frac{\partial Y^{[\beta,\mathcal{N};\boldsymbol{v}]}}{\partial\mathcal{N}}\right)_{\beta} \quad . \tag{133}$$

So, the function $\breve{\boldsymbol{\mu}}[\beta,\mathcal{N};\boldsymbol{v}] = -\beta^{-1}\breve{\boldsymbol{\alpha}}[\beta,\mathcal{N};\boldsymbol{v}]$ is

$$\begin{aligned}\breve{\boldsymbol{\mu}}[\beta,\mathcal{N};\boldsymbol{v}] &= \left(\frac{\partial E^{[\beta,\mathcal{N};\boldsymbol{v}]}}{\partial\mathcal{N}}\right)_{\beta} - \beta^{-1}\left(\frac{\partial S^{[\beta,\mathcal{N};\boldsymbol{v}]}}{\partial\mathcal{N}}\right)_{\beta} \\ &= \boldsymbol{a}[\beta,\mathcal{N};\boldsymbol{v}] + \beta^{-1}\boldsymbol{b}[\beta,\mathcal{N};\boldsymbol{v}] \\ &= \sum_I \left(\frac{\partial\omega_{(I)}^{[\beta,\mathcal{N};\boldsymbol{v}]}}{\partial\mathcal{N}}\right)_{\beta} E_{(I)}^{[\beta;\boldsymbol{v}]} + \beta^{-1}\sum_I \left(\frac{\partial\omega_{(I)}^{[\beta,\mathcal{N};\boldsymbol{v}]}}{\partial\mathcal{N}}\right)_{\beta}\left(\ln\omega_{(I)}^{[\beta,\mathcal{N};\boldsymbol{v}]} - S_{(I)}^{[\beta;\boldsymbol{v}]}\right)\end{aligned} \tag{134}$$

(the identity $\sum_I\left(\partial\omega_{(I)}^{[\beta,\mathcal{N};\boldsymbol{v}]}\big/\partial\mathcal{N}\right) = 0$, stemming from Eq.(108), was used).

The definitions of $\boldsymbol{a}$ and $\boldsymbol{b}$ components of $\breve{\boldsymbol{\mu}}$, introduced in first line, are used to obtain the second and third line. As we see, $\boldsymbol{a}$ is the gradient (w.r.t. $\boldsymbol{\mathcal{N}}$) of the average energy, while $\boldsymbol{b}$ is the gradient of the entropy, with the minus sign.

The large-$\beta$ behavior of $\breve{\boldsymbol{\alpha}}\left[\beta,\boldsymbol{\mathcal{N}};\boldsymbol{v}\right]$ and $\breve{\boldsymbol{\mu}}\left[\beta,\boldsymbol{\mathcal{N}};\boldsymbol{v}\right]$ stemming from Eq.(134) can be written as a slant asymptote (see Appendix)

$$\breve{\boldsymbol{\alpha}}\left[\beta,\boldsymbol{\mathcal{N}};\boldsymbol{v}\right]=-\beta\boldsymbol{a}\left[\beta,\boldsymbol{\mathcal{N}},\boldsymbol{v}\right]-\boldsymbol{b}\left[\beta,\boldsymbol{\mathcal{N}},\boldsymbol{v}\right]\simeq-\beta\boldsymbol{a}\left[\boldsymbol{\mathcal{N}},\boldsymbol{v}\right]-\boldsymbol{b}\left[\boldsymbol{\mathcal{N}},\boldsymbol{v}\right], \tag{135}$$

$$\breve{\boldsymbol{\mu}}\left[\beta,\boldsymbol{\mathcal{N}};\boldsymbol{v}\right]=\boldsymbol{a}\left[\beta,\boldsymbol{\mathcal{N}},\boldsymbol{v}\right]+\beta^{-1}\boldsymbol{b}\left[\beta,\boldsymbol{\mathcal{N}},\boldsymbol{v}\right]\simeq\boldsymbol{a}\left[\boldsymbol{\mathcal{N}},\boldsymbol{v}\right]+\beta^{-1}\boldsymbol{b}\left[\boldsymbol{\mathcal{N}},\boldsymbol{v}\right], \tag{136}$$

where the 0K limiting values are $\boldsymbol{a}\left[\beta,\boldsymbol{\mathcal{N}};\boldsymbol{v}\right]\doteq\boldsymbol{a}\left[\boldsymbol{\mathcal{N}},\boldsymbol{v}\right]$, $\boldsymbol{b}\left[\beta,\boldsymbol{\mathcal{N}};\boldsymbol{v}\right]\doteq\boldsymbol{b}\left[\boldsymbol{\mathcal{N}},\boldsymbol{v}\right]$.

The zero-temperature $\boldsymbol{a}$ and $\boldsymbol{b}$ can be calculated from three conditions, Eqs.(108)-(109), applied to the weights, Eq.(119). In particular, using Eqs. (119),(34) and Eq.(136), the large-$\beta$ behavior of the ratio between two macrostate weights is

$$\begin{aligned}\frac{\omega_{(I)}^{[\beta,\boldsymbol{\mathcal{N}};\boldsymbol{v}]}}{\omega_{(J)}^{[\beta,\boldsymbol{\mathcal{N}};\boldsymbol{v}]}}&=\exp\left(-\beta\left(\left(E_{(I)}^{[\beta;\boldsymbol{v}]}-E_{(J)}^{[\beta;\boldsymbol{v}]}\right)-\boldsymbol{a}^{[\beta,\boldsymbol{\mathcal{N}};\boldsymbol{v}]}\left(\boldsymbol{\mathcal{N}}_{(I)}-\boldsymbol{\mathcal{N}}_{(J)}\right)\right)\right)\\&\quad\times\exp\left(\boldsymbol{b}^{[\beta,\boldsymbol{\mathcal{N}};\boldsymbol{v}]}\left(\boldsymbol{\mathcal{N}}_{(I)}-\boldsymbol{\mathcal{N}}_{(J)}\right)+\left(S_{(I)}^{[\beta;\boldsymbol{v}]}-S_{(J)}^{[\beta;\boldsymbol{v}]}\right)\right)\\&\quad\simeq\exp\left(-\beta\left(\left(E_{(I)}^{[\boldsymbol{v}]}-E_{(J)}^{[\boldsymbol{v}]}\right)-\boldsymbol{a}^{[\boldsymbol{\mathcal{N}};\boldsymbol{v}]}\left(\boldsymbol{\mathcal{N}}_{(I)}-\boldsymbol{\mathcal{N}}_{(J)}\right)\right)\right)\\&\quad\times\exp\left(\boldsymbol{b}^{[\boldsymbol{\mathcal{N}};\boldsymbol{v}]}\left(\boldsymbol{\mathcal{N}}_{(I)}-\boldsymbol{\mathcal{N}}_{(J)}\right)+\left(S_{(I)}^{[\boldsymbol{v}]}-S_{(J)}^{[\boldsymbol{v}]}\right)\right)\doteq\frac{\omega_{(I)}^{[\boldsymbol{\mathcal{N}};\boldsymbol{v}]}}{\omega_{(J)}^{[\boldsymbol{\mathcal{N}};\boldsymbol{v}]}}.\end{aligned} \tag{137}$$

In the case of $\boldsymbol{\mathcal{N}}$ lying inside a *triangle* $\{\{1\},\{2\},\{3\}\}$, the three weights can be calculated directly from three components of the conditions, Eqs.(108)-(109). Based on the discussion about the discontinuity segments at Shape 2 , $\mathcal{C}_{[\mathcal{Z},\Sigma]}^{[\boldsymbol{v}]}\cap\mathcal{C}_{[\mathcal{Z}\pm2,\Sigma']}^{[\boldsymbol{v}]}=\varnothing$ implies that the difference in the electron numbers between the vertices of this triangle can be equal $\pm1$ or 0; let us take

$$\{\{1\},\{2\},\{3\}\}=\left\{[\mathcal{Z},\Sigma],[\mathcal{Z},\Sigma+2],[\mathcal{Z}+\Delta\mathcal{N},\Sigma+\Delta\mathcal{S}]\right\},\ \Delta\mathcal{S}\in\{-1,1,3\},\Delta\mathcal{N}\in\{-1,1\}. \tag{138}$$

The zero-temperature weights are

$$\begin{cases} \omega_{\{3\}}^{[\mathcal{N};v]} = \Delta\mathcal{N}\left(\mathcal{N}-\mathcal{Z}\right) \\ \omega_{\{2\}}^{[\mathcal{N};v]} = \frac{1}{2}\left(\mathcal{S}-\Sigma-\Delta\mathcal{S}\omega_{\{3\}}^{[\mathcal{N};v]}\right) = \frac{1}{2}\left(\mathcal{S}-\Sigma-\Delta\mathcal{N}\Delta\mathcal{S}\left(\mathcal{N}-\mathcal{Z}\right)\right) \\ \omega_{\{1\}}^{[\mathcal{N};v]} = 1-\omega_{\{2\}}^{[\mathcal{N};v]}-\omega_{\{3\}}^{[\mathcal{N};v]} = 1-\frac{1}{2}\left(\mathcal{S}-\Sigma-\Delta\mathcal{N}\left(\Delta\mathcal{S}-2\right)\left(\mathcal{N}-\mathcal{Z}\right)\right) \end{cases} \tag{139}$$

Using these weights in Eq.(137) at 0K limit, we have

$$\boldsymbol{a}^{[\mathcal{N};v]}\left(\boldsymbol{\mathcal{N}}_{\{k\}}-\boldsymbol{\mathcal{N}}_{\{l\}}\right) = \left(E_{\{k\}}^{[v]}-E_{\{l\}}^{[v]}\right), \quad k,l \in \{1,2,3\}, \tag{140}$$

$$\boldsymbol{b}^{[\mathcal{N};v]}\left(\boldsymbol{\mathcal{N}}_{\{k\}}-\boldsymbol{\mathcal{N}}_{\{l\}}\right) = \ln\frac{\omega_{\{k\}}^{[\mathcal{N};v]}}{\omega_{\{l\}}^{[\mathcal{N};v]}} - \ln\frac{d_{(0,\{k\})}^{[v]}}{d_{(0,\{l\})}^{[v]}}, \quad k,l \in \{1,2,3\}. \tag{141}$$

These relations can be rewritten as

$$\breve{\mu}_{\mathrm{S}}^{[\mathcal{N};v]} = a_{\mathrm{S}}^{[\mathcal{N};v]} = \frac{1}{2}\left(E_{(0,\{2\})}^{[v]}-E_{(0,\{1\})}^{[v]}\right), \tag{142}$$

$$\breve{\mu}_{\mathrm{N}}^{[\mathcal{N};v]} = a_{\mathrm{N}}^{[\mathcal{N};v]} = \pm\left(E_{(0,\{3\})}^{[v]}-E_{(0,\{1\})}^{[v]}-\mu_{\mathrm{S}}^{[\mathcal{N};v]}\Delta\mathcal{S}\right). \tag{143}$$

$$b_{\mathrm{S}}^{[\mathcal{N};v]} = \frac{1}{2}\ln\frac{\omega_{\{2\}}^{[\mathcal{N};v]}}{\omega_{\{1\}}^{[\mathcal{N};v]}}\frac{d_{(0,\{1\})}^{[v]}}{d_{(0,\{2\})}^{[v]}}, \tag{144}$$

$$b_{\mathrm{N}}^{[\mathcal{N};v]} = \Delta\mathcal{N}\left(\ln\frac{\omega_{\{3\}}^{[\mathcal{N};v]}}{\omega_{\{1\}}^{[\mathcal{N};v]}}\frac{d_{(0,\{1\})}^{[v]}}{d_{(0,\{3\})}^{[v]}}-\Delta\mathcal{S}b_{\mathrm{S}}^{[\mathcal{N};v]}\right). \tag{145}$$

Important conclusion which can be drawn from Eqs. (142) and (143) is that $\breve{\mu}_{\mathrm{N}}$ and $\breve{\mu}_{\mathrm{S}}$ have the same values for each point $\left(\mathcal{N},\mathcal{S}\right)$ inside the triangle.

The flat-plane behavior of the energy as a function of the $\left(\mathcal{N},\mathcal{S}\right)$ points inside the triangle defined in Eq.(138) is

$$E^{[\mathcal{N};v]} = \omega_{[\mathcal{Z},\Sigma]}^{[\mathcal{N};v]}E_{(0,[\mathcal{Z},\Sigma])}^{[v]} + \omega_{[\mathcal{Z},\Sigma+2]}^{[\mathcal{N};v]}E_{(0,[\mathcal{Z},\Sigma+2])}^{[v]} + \omega_{[\mathcal{Z}+\Delta\mathcal{N},\Sigma+\Delta\mathcal{S}]}^{[\mathcal{N};v]}E_{(0,[\mathcal{Z}+\Delta\mathcal{N},\Sigma+\Delta\mathcal{S}])}^{[v]}. \tag{146}$$

with the weights given in Eq.(139), seen to be linear in $\mathcal{N}$ and $\mathcal{S}$. For example (Fig. 2), the case of $\Delta\mathcal{N}=-1$ and $\Delta\mathcal{S}=-1$ is shown in the $\{[6,2],[6,4],[5,1]\}$ triangle, that of $\Delta\mathcal{N}=+1$ and $\Delta\mathcal{S}=+1$ – in the $\{[6,2],[6,4],[7,3]\}$ triangle, and that of $\Delta\mathcal{N}=+1$ and $\Delta\mathcal{S}=+3$ – in the $\{[5,1],[5,3],[6,4]\}$ triangle.

Calculating weights at vertices as limits w.r.t. $\boldsymbol{\mathcal{N}}$ lying inside the triangle, and denoting $\omega_{\{l\}}\left[\boldsymbol{\mathcal{N}}_{\{k\}};\boldsymbol{v}\right]=\lim_{\boldsymbol{\mathcal{N}}\to\boldsymbol{\mathcal{N}}_{\{k\}}}\omega_{\{l\}}\left[\boldsymbol{\mathcal{N}};\boldsymbol{v}\right]$, we find from Eq.(139) $\omega_{\{l\}}\left[\boldsymbol{\mathcal{N}}_{\{k\}};\boldsymbol{v}\right]=\delta_{kl}$, i.e., the weight at the $\{k\}$ vertex is 1, there are no contribution of remaining vertices.

Some generalization for the piecewise-flat-plane energies as a function of (fractional) electron number and spin number was already presented, but either for the hydrogen case[3], or did not cover all cases.[10,12,31]

Consider now the case of $\boldsymbol{\mathcal{N}}$ lying inside a *trapezoid*, for example, the doublet-triplet one, $\{\{1\},\{2\},\{3\},\{4\},\{5\}\}=\{[\mathcal{Z},+1],[\mathcal{Z},-1],[\mathcal{Z}\pm 1,-2],[\mathcal{Z}\pm 1,0],[\mathcal{Z}\pm 1,2]\}$.

Weights (not linear in $\boldsymbol{\mathcal{N}}$) are to be determined numerically for each argument $(\mathcal{N},\mathcal{S})$. With $\omega_{\{k\}}\left[\boldsymbol{\mathcal{N}};\boldsymbol{v}\right]$ abbreviated to $\omega_{\{k\}}$, the spin canonical conditions (109) can be rewritten as follows

$$\omega^{\pm}\equiv\omega_{\{3\}}+\omega_{\{4\}}+\omega_{\{5\}}=\pm\left(\mathcal{N}-\mathcal{Z}\right), \tag{147}$$

$$\left(\omega_{\{1\}}-\omega_{\{2\}}\right)+2\left(\omega_{\{3\}}-\omega_{\{5\}}\right)=\left(1-\omega^{\pm}\right)\mathcal{S}^{0}+\omega^{\pm}\mathcal{S}^{\pm}=\mathcal{S}, \tag{148}$$

where $\mathcal{S}^{0}$ and $\mathcal{S}^{\pm}$ are an average spin numbers for the multiplets with $\mathcal{Z}$ and $(\mathcal{Z}\pm 1)$ electrons, respectively. Due to the energy degeneracy within each multiplet, the energy surface for the trapezoid is a flat-plane one:

$$E^{[\mathcal{N};v]}=\left(1-\omega^{\pm}\right)E_{(0,[\mathcal{Z},1])}+\omega^{\pm}E_{(0,[\mathcal{Z}\pm 1,2])}=E_{(0,[\mathcal{Z},1])}\pm\left(\mathcal{N}-\mathcal{Z}\right)\left(E_{(0,[\mathcal{Z}\pm 1,2])}-E_{(0,[\mathcal{Z},1])}\right), \tag{149}$$

independent of $\mathcal{S}$.

While detailed expressions for $\omega_{\{k\}}\left[\boldsymbol{\mathcal{N}};\boldsymbol{v}\right]$ are not needed for calculation of the mean value of the Hamiltonian, they are necessary for such calculations concerning other operators. Therefore asymptotic form of $\breve{\boldsymbol{\mu}}[\beta,\boldsymbol{\mathcal{N}};\boldsymbol{v}]$ is of interest. Knowing $E[\boldsymbol{\mathcal{N}};\boldsymbol{v}]$, Eq.(149), $\breve{\boldsymbol{\mu}}[\beta,\boldsymbol{\mathcal{N}};\boldsymbol{v}]$ can be calculated from Eq.(134), at 0K limit

$$\breve{\mu}_{\mathrm{S}}^{[\mathcal{N};v]} = a_{\mathrm{S}}^{[\mathcal{N};v]} = \left(\partial E[\boldsymbol{\mathcal{N}};\boldsymbol{v}]/\partial\mathcal{S}\right) = 0\,, \tag{150}$$

$$\breve{\mu}_{\mathrm{N}}^{[\mathcal{N};v]} = a_{\mathrm{N}}^{[\mathcal{N};v]} = \left(\partial E[\boldsymbol{\mathcal{N}};\boldsymbol{v}]/\partial\mathcal{N}\right) = \pm\left(E_{(0,[\mathcal{Z}\pm 1,2])} - E_{(0,[\mathcal{Z},1])}\right). \tag{151}$$

So 2vec $\breve{\boldsymbol{\mu}}[\boldsymbol{\mathcal{N}};\boldsymbol{v}] = \boldsymbol{a}[\boldsymbol{\mathcal{N}};\boldsymbol{v}]$ have the same values at each point $(\mathcal{N},\mathcal{S})$ inside the trapezoid. Lets us note that the average multiplets spin numbers in Eq.(148) are unknown. The average spin number for the spin $Q$-plet is calculated using the conditional probability that the system is in the $(I)$ macrostate provided we know it to be in the spin $Q$-plet. Using the large-$\beta$ behavior of the $(I)$ macrostate weight, Eq.(23), the conditional probability for the $(I)$ macrostate which belonging to the spin $Q$-plet at 0K limit is equal to

$$\begin{aligned}\frac{\omega_{(I)}^{[\beta,\mathcal{N};v]}}{\sum_{(J)\in Q-\mathrm{plet}}\omega_{(J)}^{[\beta,\mathcal{N};v]}} &\simeq \frac{\exp\left(-\beta\left(E_{(0,I)}^{[v]} - \boldsymbol{a}^{[\mathcal{N};v]}\boldsymbol{\mathcal{N}}_{(I)}\right) + \boldsymbol{b}^{[\mathcal{N};v]}\boldsymbol{\mathcal{N}}_{(I)} + S_{(I)}^{[v]}\right)}{\sum_{(J)\in Q-\mathrm{plet}}\exp\left(-\beta\left(E_{(0,J)}^{[v]} - \boldsymbol{a}^{[\mathcal{N};v]}\boldsymbol{\mathcal{N}}_{(J)}\right) + \boldsymbol{b}^{[\mathcal{N};v]}\boldsymbol{\mathcal{N}}_{(J)}S_{(J)}^{[v]}\right)} \\ &\doteq \frac{\exp\left(b_{\mathrm{S}}^{[\mathcal{N};v]}\mathcal{S}_{(I)}\right)}{\sum_{(J)\in Q-\mathrm{plet}}\exp\left(b_{\mathrm{S}}^{[\mathcal{N};v]}\mathcal{S}_{(J)}\right)},\end{aligned} \tag{152}$$

the properties of the spin $Q$-plet (the values of $E_{(0,I)}^{[v]}$, $\mathcal{N}_{(I)}$ and $S_{(I)}^{[v]}$ are the same for all spin $Q$-plet macrostates) and $a_{\mathrm{S}} = 0$ were used. Based on these conditional probabilities, the average spin number for the spin $Q$-plet can be expressed as

$$\mathcal{S}[Q,b_{\mathrm{s}}] = \sum_{q=-(Q-1)}^{(Q-1),2} q\exp(qb_{\mathrm{s}}) \Big/ \sum_{q=-(Q-1)}^{(Q-1),2}\exp(qb_{\mathrm{s}}), \tag{153}$$

so $\mathcal{S}^0$ and $\mathcal{S}^\pm$ are

$$\begin{aligned}\mathcal{S}^0 &= \mathcal{S}[2,b_{\mathrm{S}}] = \frac{\exp(b_{\mathrm{S}}) - \exp(-b_{\mathrm{S}})}{\exp(b_{\mathrm{S}}) + \exp(-b_{\mathrm{S}})} = \tanh(b_{\mathrm{S}}), \\ S^\pm &= \mathcal{S}[3,b_{\mathrm{S}}] = \frac{2\left(\exp(2b_{\mathrm{S}}) - \exp(-2b_{\mathrm{S}})\right)}{1+\exp(2b_{\mathrm{S}}) + \exp(-2b_{\mathrm{S}})} = \frac{4\sinh(2b_{\mathrm{S}})}{1+2\cosh(2b_{\mathrm{S}})}\end{aligned}, \tag{154}$$

Finally, after inserting $\mathcal{S}^0 = \mathcal{S}\left[2, b_\mathrm{S}\right]$ and $S^\pm = \mathcal{S}\left[3, b_\mathrm{S}\right]$ into Eq.(148), the value $b_\mathrm{S}\left[\boldsymbol{\mathcal{N}}\right]$ can be found numerically as the solution of the Eq.(148) for fixed $\left(\mathcal{N}, \mathcal{S}\right)$

Using Eq.(119), the relation between $\omega^\pm$ and $1-\omega^\pm$ yields

$$\frac{\omega^\pm}{\left(1-\omega^\pm\right)} = \frac{\sum_{k=3}^{5} \omega_{\{k\}}^{[\mathcal{N};v]}}{\sum_{l=1}^{2} \omega_{\{l\}}^{[\mathcal{N};v]}} = \exp\left(\pm b_\mathrm{N}^{[\mathcal{N};v]}\right) \frac{\sum_{k=3}^{5} \exp\left(\mathcal{S}_{\{k\}} b_\mathrm{S}^{[\mathcal{N};v]} + S_{\{k\}}^{[v]}\right)}{\sum_{l=1}^{2} \exp\left(\mathcal{S}_{\{l\}} b_\mathrm{S}^{[\mathcal{N};v]} + S_{\{l\}}^{[v]}\right)}, \tag{155}$$

allowing finally to calculate $b_\mathrm{N}$ as

$$b_\mathrm{N}^{[\mathcal{N};v]} = \pm\left( \ln \frac{\omega^\pm}{\left(1-\omega^\pm\right)} \frac{d_0^{[v]}}{d_\pm^{[v]}} - \ln\left( \frac{\sum_{k=3}^{5} \exp\left(\mathcal{S}_{\{k\}} b_\mathrm{S}^{[\mathcal{N};v]}\right)}{\sum_{l=1}^{2} \exp\left(\mathcal{S}_{\{l\}} b_\mathrm{S}^{[\mathcal{N};v]}\right)} \right)\right), \tag{156}$$

(here $d_\pm^{[v]}$ and $d_0^{[v]}$ denote spatial degeneracies of the triplet and the doublet, respectively, see Eq.(55)) i.e., as a function $\boldsymbol{\mathcal{N}}$ (via $\omega^\pm / \left(1-\omega^\pm\right) = \left(\mathcal{N} - \mathcal{Z}\right) / \left(\left(\mathcal{Z} - \mathcal{N}\right) \mp 1\right)$ and $b_\mathrm{S}$). Using $b_\mathrm{N}$ and $b_\mathrm{S}$, the zero-temperature weights for the point $\left(\mathcal{N}, \mathcal{S}\right)$ inside the doublet-triplet trapezoid can be evaluated

$$\omega_{\{k\}}^{[\mathcal{N};v]} = \frac{\exp\left(\boldsymbol{b}^{[\mathcal{N},v]} \boldsymbol{\mathcal{N}}_{\{k\}} + S_{\{k\}}^{[v]}\right)}{\sum_{l=1}^{5} \exp\left(\boldsymbol{b}^{[\mathcal{N},v]} \boldsymbol{\mathcal{N}}_{\{l\}} + S_{\{l\}}^{[v]}\right)}, \quad k \in \{1,2,3,4,5\}. \tag{157}$$

Calculating weights at macrostate points belongings to trapezoid as limit w.r.t $\boldsymbol{\mathcal{N}}$ lying inside the trapezoid, and denoting $\omega_{\{l\}}\left[\boldsymbol{\mathcal{N}}_{\{k\}}; \boldsymbol{v}\right] = \lim_{\boldsymbol{\mathcal{N}} \to \boldsymbol{\mathcal{N}}_{\{k\}}} \omega_{\{l\}}\left[\boldsymbol{\mathcal{N}}; \boldsymbol{v}\right]$, we find from equation above for $k \neq 4$, $\omega_{\{l\}}\left[\boldsymbol{\mathcal{N}}_{\{k\}}; \boldsymbol{v}\right] = \delta_{kj}$; for $k = 4$, $\omega_{\{l\}}\left[\boldsymbol{\mathcal{N}}_{\{4\}}; \boldsymbol{v}\right] = 1/3$ when $l \in \{3,4,5\}$ and $\omega_{\{l\}}\left[\boldsymbol{\mathcal{N}}_{\{4\}}; \boldsymbol{v}\right] = 0$ when $l \in \{1,2\}$, i.e., the weight for the trapezoid vertex at this vertex is 1, but at $\boldsymbol{\mathcal{N}}_{\{k\}}$ – the internal macrostate point of the triplet, the weights of all triplet points are equal $1/3$, while the weights of the doublet points are zero.

The case of trapezoids with higher multiplets can be investigeted similarly and the flat-plane energy surface for $(\mathcal{N},\mathcal{S})$ points inside any trapezoid is similar to Eq.(149).

Other cases are not considered in this paper. In general, to obtain $\boldsymbol{a}[\boldsymbol{\mathcal{N}};\boldsymbol{v}]$ and $\boldsymbol{b}[\boldsymbol{\mathcal{N}};\boldsymbol{v}]$ contributions of additional macrostates at low temperature are to be taken into account and determination of $\boldsymbol{a}[\boldsymbol{\mathcal{N}};\boldsymbol{v}]$ and $\boldsymbol{b}[\boldsymbol{\mathcal{N}};\boldsymbol{v}]$ needs special care. All details and results for $\boldsymbol{\mathcal{N}}$ lying on an open trapezoid base or on the common side of a triangle and a trapezoid (or another triangle), or for $\boldsymbol{\mathcal{N}}$ at a vertex will be published separately.[34] Here, only the case of $\boldsymbol{\mathcal{N}}$ equal a singlet macrostate will be discussed. In general, for an arbitrary vertex, i.e., for $(\mathcal{N},\mathcal{S})=(\mathcal{N}_{\{1\}},\mathcal{S}_{\{1\}})=(\mathcal{Z},\Sigma)$, as the first step of analysis the "neighboring" macrostates are to be recognized for which weights decrease with temperature more slowly than for other macrostates. The $\omega_{\{1\}}$ and these neighboring weights must satisfy the normalization and spin-canonical conditions at this $\boldsymbol{\mathcal{N}}_{\{1\}}$; from this requirement $\left(a_{\mathrm{N}}^{[\mathcal{N};v]},a_{\mathrm{S}}^{[\mathcal{N};v]}\right)$ can be found. This procedure is quite simple for the singlet state $(\mathcal{N},\mathcal{S})=(\mathcal{Z},0)$. In this case, the maximum number of the discontinuity lines departing from this point is ten[34] (e.g. see point $(4,0)$ at Fig. 2: there are four lines for the electron addition, two for the singlet-triplet excitation, and four lines for the electron removing, but not shown). Due to a convexity of the ground-state energy as the spin-number function, excited ionic states can be excluded. So, the six-"neighboring" states consist of two doublets $[\mathcal{Z}-1,1]$, $[\mathcal{Z}-1,-1]$ and $[\mathcal{Z}+1,1]$, $[\mathcal{Z}+1,-1]$, and two members of the excited triplet $[\mathcal{Z},2]$, $[\mathcal{Z},-2]$. The contribution from doublets to the electron number $\mathcal{N}=\mathcal{Z}$ has to equalize (at low temperature) and vanish at 0K (the triplet states are not involved here).

$$\frac{\omega_{[\mathcal{Z}+1,-1]}^{[\beta,\mathcal{N};v]} + \omega_{[\mathcal{Z}+1,1]}^{[\beta,\mathcal{N};v]}}{\omega_{[\mathcal{Z}-1,-1]}^{[\beta,\mathcal{N};v]} + \omega_{[\mathcal{Z}-1,1]}^{[\beta,\mathcal{N};v]}} = 1. \tag{158}$$

Using Eq.(137), we find at 0K limit

$$a_{\mathrm{N}}^{[\mathcal{N};v]} = \frac{1}{2}\left(E_{(0,[\mathcal{Z}+1,1])}^{[v]} - E_{(0,[\mathcal{Z}-1,1])}^{[v]}\right) = -\frac{1}{2}(I + A), \tag{159}$$

$$b_{\mathrm{N}}^{[\mathcal{N};v]} = -\frac{1}{2}\frac{d_{(0,[\mathcal{Z}+1,1])}^{[v]}}{d_{(0,[\mathcal{Z}-1,1])}^{[v]}}, \tag{160}$$

where $I$ and $A$ are the ionization potential and the electron affinity, respectively. These results are the same as obtained in Ref.[35] (see Eq.(4.3.8) therein). The spin canonical condition for the spin number $\mathcal{S} = 0$ yields

$$\omega_{[\mathcal{Z}+1,1]}^{[\beta,\mathcal{N};v]} + \omega_{[\mathcal{Z}-1,1]}^{[\beta,\mathcal{N};v]} + 2\omega_{[\mathcal{Z},2]}^{[\beta,\mathcal{N};v]} = \omega_{[\mathcal{Z}+1,-1]}^{[\beta,\mathcal{N};v]} + \omega_{[\mathcal{Z}-1,-1]}^{[\beta,\mathcal{N};v]} + 2\omega_{[\mathcal{Z},-2]}^{[\beta,\mathcal{N};v]}, \tag{161}$$

and using Eq.(137), we find at 0K limit

$$1 = \frac{\omega_{[\mathcal{Z}+1,1]}^{[\mathcal{N};v]}}{\omega_{[\mathcal{Z}+1,-1]}^{[\mathcal{N};v]}} = \frac{\omega_{[\mathcal{Z}-1,1]}^{[\mathcal{N};v]}}{\omega_{[\mathcal{Z}-1,-1]}^{[\mathcal{N};v]}} = \left(\frac{\omega_{[\mathcal{Z},2]}^{[\mathcal{N};v]}}{\omega_{[\mathcal{Z},-2]}^{[\mathcal{N};v]}}\right)^{1/2} \doteq \exp\left(2\beta a_{\mathrm{S}}^{[\mathcal{N};v]}\right)\exp\left(2b_{\mathrm{S}}^{[\mathcal{N};v]}\right), \tag{162}$$

so these equations give

$$a_{\mathrm{S}}^{[\mathcal{N};v]} = 0, \qquad b_{\mathrm{S}}^{[\mathcal{N};v]} = 0. \tag{163}$$

But note that the values of $\boldsymbol{a}$ and $\boldsymbol{b}$ for the $\boldsymbol{\mathcal{N}} = (\mathcal{Z} + \delta_{\mathcal{N}}, \Sigma + \delta_{\mathcal{S}})$ at 0K limit depend on $\boldsymbol{\mathcal{N}}$ belonging to a particular open polygon or segment. This means that

$$\lim_{\substack{\delta_{\mathcal{N}}\to 0\\ \delta_{\mathcal{S}}\to 0}}\left(\lim_{\beta\to\infty}\breve{\boldsymbol{\mu}}^{[\beta,(\mathcal{Z}+\delta_{\mathcal{N}},\Sigma+\delta_{\mathcal{S}});v]}\right) \neq \lim_{\beta\to\infty}\breve{\boldsymbol{\mu}}^{[\beta,(\mathcal{Z},\Sigma);v]}. \tag{164}$$

To conclude this subsection, the analysis presented here for $\boldsymbol{\mathcal{N}}$ restricted to integral electron and spin numbers removes the fuzziness in $\breve{\boldsymbol{\mu}}[\boldsymbol{\mathcal{N}};\boldsymbol{v}]$, Eq.(95), (see Ref.[36] for discussion about $\mu_{\mathrm{N}}[\mathcal{N},0;\boldsymbol{v}]$ at 0K limit). In the spin-grand-canonical ensemble, there is a set of $\boldsymbol{\mu} = (\mu_{\mathrm{N}}, \mu_{\mathrm{S}})$ points which satisfy $\mathcal{N}\left[\beta,\boldsymbol{\mu};\mathbf{v}\right] \doteq \left(\mathcal{Z},\Sigma\right)$, but only one point satisfies $\boldsymbol{\mu}\left[\beta,\mathcal{N}=\left(\mathcal{Z},\Sigma\right);\mathbf{v}\right] \doteq \boldsymbol{\mu}\left[\left(\mathcal{Z},\Sigma\right);\mathbf{v}\right]$.

# VI. THE HOHENBERG-KOHN THEOREM AT 0K LIMIT

In this section, we are going to formulate the 0K limit of the thermodynamic extension of the Hohnberg-Konh (HK) theorem of Paper II. The HK theorem at finite temperature is formulated there in two equivalent forms, depending on the type of the applied representation. In the entropy representation the HK theorem has a form of the maximum entropy principle, while in the energy representation – the minimum energy principle. Both forms are equivalent at finite temperature.

All state functions introduced in Paper II are suitable for describing the thermodynamic systems at finite temperature. Due to the variables transformations, any state function may be transformed into another state function (see Fig. 1). For the $\left\{\hat{H},\hat{\mathcal{N}}\right\}$ system we have the following variables transformations and mappings

$$\left\{\mathcal{N}\right\} \underset{\mathcal{L}}{\overset{\mathcal{L}^{-1}}{\rightleftarrows}} \left\{\boldsymbol{\alpha}\right\} \underset{\mathcal{M}^{-1}}{\overset{\mathcal{M}}{\rightleftarrows}} \left\{\boldsymbol{\mu}\right\} \tag{165}$$

where $\mathcal{L}$ and $\mathcal{M}$ denote the Legendre transformation and the Massieu-Planck transformation, respectively.[1,2] The superscript "-1" at transformation denotes the reverse one. At low temperature, these maps in the spin-canonical ensemble can be rewritten as

$$\left\{\mathcal{N}\right\} \xrightarrow{\mathcal{L}^{-1}} \left\{\left(\mathbf{a}^{\left[\mathcal{N};\mathbf{v}\right]},\mathbf{b}^{\left[\mathcal{N};\mathbf{v}\right]}\right)\right\} \tag{166}$$

and in the spin-grand-canonical ensemble as

$$\left\{\mathcal{N}^{\left[\beta,\boldsymbol{\alpha};\mathbf{v}\right]}\right\} \xleftarrow[\mathcal{L}]{} \left\{\boldsymbol{\alpha}\right\} \underset{\mathcal{M}^{-1}}{\overset{\mathcal{M}}{\rightleftarrows}} \left\{\boldsymbol{\mu}\right\} \xrightarrow[\mathcal{L}]{} \left\{\mathcal{N}^{\left[\beta,\boldsymbol{\mu};\mathbf{v}\right]}\right\}, \tag{167}$$

with identities

$$\mathcal{N} = \breve{\mathcal{N}}[\beta,\boldsymbol{\alpha};\boldsymbol{v}] = \breve{\mathcal{N}}[\beta,\boldsymbol{\alpha}=-\beta\boldsymbol{\mu};\boldsymbol{v}] = \breve{\mathcal{N}}[\beta,\boldsymbol{\mu};\boldsymbol{v}] = \\ \simeq \mathcal{N}\left[\beta,\boldsymbol{\alpha}=-\beta\boldsymbol{a}^{[\mathcal{N};v]}-\boldsymbol{b}^{[\mathcal{N};v]};\boldsymbol{v}\right] = \mathcal{N}\left[\beta,\boldsymbol{\mu}=\boldsymbol{a}^{[\mathcal{N};v]}+\beta^{-1}\boldsymbol{b}^{[\mathcal{N};v]};\boldsymbol{v}\right], \tag{168}$$

where the low-temperature asymptotic forms of $\boldsymbol{\mu}$ defined in Eq.(134), were used (see Eqs.(135) and (136)).

The bijective (one-to-one) maps at finite temperature, Eqs.(165)–(167), lose this property at 0K limit. The map $\boldsymbol{\alpha}\to\boldsymbol{\mu}=-\beta^{-1}\boldsymbol{\alpha}$ by the definition of the Massieu-Planck transformation[1,37] has one element codomain, $(\mu_{\mathrm{N}},\mu_{\mathrm{S}})=(0,0)$. The reverse transformation is undefined at 0K limit.

As was discussed in Sec. IV.A (degenerate case), $\breve{\mathcal{N}}[\beta,\boldsymbol{\alpha};\boldsymbol{v}]\doteq\left(\mathcal{N}_{(0)},\bar{\mathcal{S}}[\alpha_{\mathrm{S}}]\right)$, $\bar{\mathcal{S}}\in\left[\mathcal{S}_{(0)},-\mathcal{S}_{(0)}\right]$ for entropy representation. In energy representation, $\breve{\mathcal{N}}[\beta,\boldsymbol{\mu};\boldsymbol{v}]\doteq\breve{\mathcal{N}}[\boldsymbol{\mu};\boldsymbol{v}]$ and, for sets, $\left\{\breve{\mathcal{N}}[\boldsymbol{\mu};\boldsymbol{v}]\right\}$ is equal to $\left\{\breve{\mathcal{N}}\left[\boldsymbol{a}^{[\mathcal{N};v]};\boldsymbol{v}\right]\right\}$. The maps $\{\boldsymbol{\alpha}\}\to\left\{\breve{\mathcal{N}}^{[\alpha;v]}\right\}$ and, $\{\boldsymbol{\mu}\}\to\left\{\breve{\mathcal{N}}^{[\mu;v]}\right\}$ have different codomains, $\left\{\breve{\mathcal{N}}^{[\beta,\alpha;v]}\right\}\subset\left\{\breve{\mathcal{N}}^{[\beta,\mu;v]}\right\}\subset\{\mathcal{N}\}$, so the map between $\{\boldsymbol{\mu}\}$ and $\{\boldsymbol{\alpha}\}$ via $\{\mathcal{N}\}$ does not exist at 0K limit. The map $\left\{\left(\boldsymbol{a}^{[\mathcal{N};v]},\boldsymbol{b}^{[\mathcal{N};v]}\right)\right\}\to\{\boldsymbol{\mu}\}$,(using Eq.(134)) is non-surjective at 0K limit, so is not reversible, and $\left\{\left(\boldsymbol{a}^{[\mathcal{N};v]},\boldsymbol{b}^{[\mathcal{N};v]}\right)\right\}\to\{\boldsymbol{\alpha}\}$ is undefined at 0K limit.

The set $\left\{\left(\boldsymbol{a}^{[\mathcal{N};v]},\boldsymbol{b}^{[\mathcal{N};v]}\right)\right\}$ can be extended to a set of admissible pair of real-number 2vecs, $(\boldsymbol{a},\boldsymbol{b})=\left((a_{\mathrm{N}},a_{\mathrm{S}}),(b_{\mathrm{N}},b_{\mathrm{S}})\right)$. Replacing $\left(\boldsymbol{a}^{[\mathcal{N};v]},\boldsymbol{b}^{[\mathcal{N};v]}\right)$ in Eq.(A10) by an element $(\boldsymbol{a},\boldsymbol{b})$ of this set, yields the macrostate weights, which can be used to calculate $\mathcal{N}\left[(\boldsymbol{a},\boldsymbol{b})\right]$ from Eq.(109). The $\{(\boldsymbol{a},\boldsymbol{b})\}\to\{\mathcal{N}\}$ map is non-injective and surjective at 0K limit. Only $\{\mathcal{N}\}\leftrightarrow\left\{\left(\boldsymbol{a}^{[\mathcal{N};v]},\boldsymbol{b}^{[\mathcal{N};v]}\right)\right\}$ is bijective map, and only elements of this map codomain are equilibrium characteristics of the system. The 2vecs $\boldsymbol{a}^{[\mathcal{N};\boldsymbol{v}]}$ and $\boldsymbol{b}^{[\mathcal{N};\boldsymbol{v}]}$ are the 0K limits of the derivative w.r.t. $\mathcal{N}$ of the average energy and average entropy at equilibrium, respectively. In the spinless DFT, the value of the chemical potential at an integer electron number is indeed a thorny subject (see Ref.[38] for discussion). Based on Sec. IV. analysis for the spin-grand-canonical ensemble, the average electron number for $\mu_{\mathrm{N}} \in (-I,-A)$ ($\mu_{\mathrm{S}} = 0$, since the system is assumed to be optimized with respect to spin number to describe the spinless theory) is exactly equal to chosen integer, $\breve{\mathcal{N}}\left[(\mu_{\mathrm{N}},0);\boldsymbol{v}\right] = (\mathcal{Z},0)$. But from Eq.(159), $a_{\mathrm{N}}\left[\breve{\mathcal{N}}\left[(\mu_{\mathrm{N}},0);\boldsymbol{v}\right];\boldsymbol{v}\right] = a_{\mathrm{N}}\left[(\mathcal{Z},0);\boldsymbol{v}\right] = -(I+A)/2$ — this and only this value characterizes the equilibrium state at 0K limit.

For the $\left\{\hat{F}_{\mathrm{int}},\hat{\boldsymbol{\rho}}\right\}$ system we have the following maps between the set of the equilibrium densities and the set of the conjugate sources (the Lagrange multipliers)

$$\{\boldsymbol{\rho}_{\mathrm{eq}}\} \underset{\mathcal{L}}{\overset{\mathcal{L}^{-1}}{\rightleftharpoons}} \{\boldsymbol{w}\} \underset{\mathcal{M}^{-1}}{\overset{\mathcal{M}}{\rightleftharpoons}} \{\boldsymbol{u}\}. \tag{169}$$

Using the equivalence conditions between the $\left\{\hat{H},\hat{\mathcal{N}}\right\}$ system and the $\left\{\hat{F}_{\mathrm{int}},\hat{\boldsymbol{\rho}}\right\}$ system found in Paper II, $\int[\boldsymbol{\rho}] = \mathcal{N}$ and $\boldsymbol{w}(\mathbf{r}) = \beta\boldsymbol{v}(\mathbf{r}) + \boldsymbol{\alpha}$, these maps can be rewritten as

$$\begin{array}{ccc} \{\boldsymbol{\rho}_{\mathrm{eq}}\} \leftrightarrow \{\{\boldsymbol{v}\},\{\boldsymbol{\alpha}\}\} \leftrightarrow \{\{\boldsymbol{v}\},\{\boldsymbol{\mu}\}\} \\ \updownarrow \qquad\qquad \searrow \qquad \swarrow \\ \left\{\{\boldsymbol{\rho}_{\mathrm{eq},(I)}\},\{\mathcal{N}\}\right\} \leftrightarrow \{\{\boldsymbol{v}\},\{\mathcal{N}\}\} \end{array} \tag{170}$$

where the $\{\boldsymbol{\rho}_{\mathrm{eq},(I)}\} \leftrightarrow \{\boldsymbol{v}\}$ map was used. The $\{\boldsymbol{\rho}_{\mathrm{eq},(I)}\} \leftrightarrow \{\boldsymbol{v}\}$ map is a macrostate extension of the well-know identity between the density-matrix constrained search functional introduced by Valone[39] and the Lieb's Legendre transform functional[40,41] in the Hilbert space. As was discussed in Sec.III.C, the $\{\boldsymbol{\rho}_{\mathrm{eq},(I)}\} \leftrightarrow \{\boldsymbol{v}\}$ map exists at 0K limit. The maps $\{\{\boldsymbol{v}\},\{\boldsymbol{\mathcal{N}}\}\} \leftrightarrow \{\{\boldsymbol{\rho}_{\mathrm{eq},(I)}\},\{\boldsymbol{\mathcal{N}}\}\} \leftrightarrow \{\boldsymbol{\rho}_{\mathrm{eq}}\}$ can be established through the spin Massieu function maximization or the spin Helmholtz function minimization. The constrained-search formalism can be written in the energy representation as

$$\begin{aligned} A[\beta,\boldsymbol{\mathcal{N}};\boldsymbol{v}] &\equiv \underset{\hat{\Gamma}}{\mathrm{Min}}\left\{\mathrm{Tr}\,\hat{\Gamma}\left(\hat{H}[\boldsymbol{v}]-\beta^{-1}\ln\hat{\Gamma}+\right)\middle|\mathrm{Tr}\,\hat{\Gamma}\hat{\boldsymbol{\mathcal{N}}}=\boldsymbol{\mathcal{N}}\right\} \\ &= \underset{\boldsymbol{\rho}}{\mathrm{Min}}\left\{\underset{\hat{\Gamma}}{\mathrm{Min}}\left\{\mathrm{Tr}\,\hat{\Gamma}\left(\hat{F}_{\mathrm{int}}-\beta^{-1}\ln\hat{\Gamma}\right)\middle|\mathrm{Tr}\,\hat{\Gamma}\hat{\boldsymbol{\rho}}=\boldsymbol{\rho}\right\}+\int[\boldsymbol{v}\boldsymbol{\rho}]\middle|\int[\boldsymbol{\rho}]=\boldsymbol{\mathcal{N}}\right\} \\ &= \underset{\boldsymbol{\rho}}{\mathrm{Min}}\left\{F[\beta,\boldsymbol{\rho}]+\int[\boldsymbol{v}\boldsymbol{\rho}]\middle|\int[\boldsymbol{\rho}]=\boldsymbol{\mathcal{N}}\right\} \\ &= E^{[\beta,\boldsymbol{\mathcal{N}};\boldsymbol{v}]}-\beta^{-1}S^{[\beta,\boldsymbol{\mathcal{N}};\boldsymbol{v}]}, \end{aligned} \tag{171}$$

and can be rewritten in terms of the macrostates as

$$\begin{aligned} A[\beta,\boldsymbol{\mathcal{N}};\boldsymbol{v}] &\equiv \underset{\hat{\Gamma}}{\mathrm{Min}}\left\{\mathrm{Tr}\,\hat{\Gamma}\left(\hat{H}[\boldsymbol{v}]+\beta^{-1}\ln\hat{\Gamma}\right)\middle|\mathrm{Tr}\,\hat{\Gamma}\hat{\boldsymbol{\mathcal{N}}}=\boldsymbol{\mathcal{N}}\right\} \\ &= \underset{\{p_{(J)}\}}{\mathrm{Min}}\left\{\sum_I p_{(I)}\left(\underset{\boldsymbol{\rho}\in\{\boldsymbol{\rho}_{(I)}\}}{\mathrm{Min}}\left\{F_{(I)}[\beta,\boldsymbol{\rho}]+\int[\boldsymbol{\rho}\boldsymbol{v}]\right\}+\beta^{-1}\ln p_{(I)}\right)\middle|\sum_I p_{(I)}\boldsymbol{\mathcal{N}}_{(I)}=\boldsymbol{\mathcal{N}}\right\} \\ &= \underset{\{p_{(J)}\}}{\mathrm{Min}}\left\{\sum_I p_{(I)}\left(A_{(I)}^{[\beta;\boldsymbol{v}]}+\beta^{-1}\ln p_{(I)}\right)\middle|\sum_I p_{(I)}\boldsymbol{\mathcal{N}}_{(I)}=\boldsymbol{\mathcal{N}}\right\} \\ &= \sum_I \omega_{(I)}^{[\beta,\boldsymbol{\mathcal{N}};\boldsymbol{v}]}\left(E_{(I)}^{[\beta;\boldsymbol{v}]}+\beta^{-1}\left(\ln\omega_{(I)}^{[\beta,\boldsymbol{\mathcal{N}};\boldsymbol{v}]}-S_{(I)}^{[\beta;\boldsymbol{v}]}\right)\right)=E^{[\beta,\boldsymbol{\mathcal{N}};\boldsymbol{v}]}-\beta^{-1}S^{[\beta,\boldsymbol{\mathcal{N}};\boldsymbol{v}]} \end{aligned} \tag{172}$$

where, for the inner minimization, the definition of the macrostate spin Massieu function was used, Eq.(47). At low temperature, the spin Helmholtz function can be approximated by its slant asymptote (see Eq.(56) for its macrostate counterpart) as

$$A[\beta,\boldsymbol{\mathcal{N}};\boldsymbol{v}] \simeq E^{[\boldsymbol{\mathcal{N}};\boldsymbol{v}]}-\beta^{-1}S^{[\boldsymbol{\mathcal{N}};\boldsymbol{v}]} \tag{173}$$

and its 0K limit is obviously the ground-state energy. Recalling the definition of the ground-state energy at 0K, we know that there exist a set of the ground-state densities, $\{\boldsymbol{\rho}_{\mathrm{gs}}\}$, defined as a set of minimizers of the ground-state energy

$$\{\boldsymbol{\rho}_{\mathrm{gs}}[\boldsymbol{\mathcal{N}},\boldsymbol{v}]\} \equiv \arg\underset{\boldsymbol{\rho}}{\mathrm{Min}}\left\{F[\boldsymbol{\rho}]+\int[\boldsymbol{\rho}\boldsymbol{v}]\middle|\boldsymbol{\rho}\in\{\boldsymbol{\rho}_{\boldsymbol{\mathcal{N}}}\}\right\}, \tag{174}$$

where $\{\boldsymbol{\rho}_{\boldsymbol{\mathcal{N}}}\}$ is a set of the densities which integrate to the given $\boldsymbol{\mathcal{N}}$, $\int[\boldsymbol{\rho}_{\boldsymbol{\mathcal{N}}}]=\boldsymbol{\mathcal{N}}$.

Each element of this set is associated with a set of the DM operators for which $F_L[\rho]$ is minimized. In general, for a given density, the set of the Lieb-Levy functional minimizers is defined as

$$\left\{\hat{\Gamma}_{\mathrm{L}}[\rho]\right\} \equiv \arg \underset{\hat{\Gamma}}{\operatorname{Min}}\left\{\operatorname{Tr} \hat{\Gamma} \hat{F}_{\mathrm{int}} \middle| \operatorname{Tr} \hat{\Gamma} \hat{\rho}=\rho\right\}, \qquad (175)$$

and the entropy functional associated with it is defined as

$$S[\rho] \equiv \underset{\hat{\Gamma}}{\operatorname{Max}}\left\{-\operatorname{Tr} \hat{\Gamma} \ln \hat{\Gamma} \middle| \hat{\Gamma} \in\left\{\hat{\Gamma}_{\mathrm{L}}[\rho]\right\}\right\}. \qquad (176)$$

On the other hand, the entropy of the system with the given $\mathcal{N}$ and $\boldsymbol{v}$ at 0K is defined as

$$S[\mathcal{N};\boldsymbol{v}]=\underset{\rho}{\operatorname{Max}}\left\{S[\rho] \middle| \rho \in\left\{\rho_{\mathrm{gs}}[\mathcal{N},\boldsymbol{v}]\right\}\right\}=S\left[\rho_{\mathrm{eq}}[\mathcal{N};\boldsymbol{v}]\right], \qquad (177)$$

$\rho_{\mathrm{eq}}(\mathbf{r})[\mathcal{N};\boldsymbol{v}]$ – the maximizer, is the equilibrium density associated with the given $\mathcal{N}$ and $\boldsymbol{v}$ at 0K limit. From the Ritz variational principle, at the 0K temperature an inequality for the ground state energy can be rewritten as

$$F\left[\rho_{\mathcal{N}}\right]+\int\left[\rho_{\mathcal{N}} \boldsymbol{v}\right] \geq F\left[\rho_{\mathrm{gs}}^{[\mathcal{N},v]}\right]+\int\left[\rho_{\mathrm{gs}}^{[\mathcal{N},v]} \boldsymbol{v}\right]=F\left[\rho_{\mathrm{eq}}^{[\mathcal{N},v]}\right]+\int\left[\rho_{\mathrm{eq}}^{[\mathcal{N},v]} \boldsymbol{v}\right], \qquad (178)$$

and accompanied by an inequality for the entropy

$$S\left[\rho_{\mathrm{gs}}^{[\mathcal{N},v]}\right] \leq S\left[\rho_{\mathrm{eq}}^{[\mathcal{N},v]}\right]. \qquad (179)$$

Note that, in general, $S\left[\rho_{\mathcal{N}}\right] \neq S\left[\rho_{\mathrm{eq}}^{[\mathcal{N},v]}\right]$, and it may even happen that $S\left[\rho_{\mathcal{N}}\right]>S\left[\rho_{\mathrm{eq}}^{[\mathcal{N},v]}\right]$ for some $\rho_{\mathcal{N}}$.

It is clear that only $\rho_{\mathrm{eq}}[\mathcal{N},\boldsymbol{v}]$ is simultaneously a minimizer of the density functional of the ground-state energy and a maximizer of the entropy at the 0K temperature, and also that it is the 0K limit of the equilibrium density associated with the given $\mathcal{N}$ and $\boldsymbol{v}$ in the spin-canonical ensemble $\rho_{\mathrm{eq}}^{[\mathcal{N};v]}(\mathbf{r}) \doteq \rho_{\mathrm{eq}}^{[\beta,\mathcal{N};v]}(\mathbf{r})$

$$\begin{aligned}\boldsymbol{\rho}_{\text{eq}}^{[\beta,\mathcal{N};\boldsymbol{v}]}(\mathbf{r}) &= \operatorname{Tr}\hat{\Gamma}_{\text{eq}}^{[\beta,\mathcal{N};\boldsymbol{v}]}\hat{\boldsymbol{\rho}}(\mathbf{r}) = \sum_I \omega_{(I)}^{[\beta,\mathcal{N};\boldsymbol{v}]}\operatorname{Tr}\hat{\Gamma}_{(I)}^{[\beta;\boldsymbol{v}]}\hat{\boldsymbol{\rho}}(\mathbf{r}) \\ &= \sum_I \omega_{(I)}^{[\beta,\mathcal{N};\boldsymbol{v}]}\boldsymbol{\rho}_{\text{eq},(I)}^{[\beta;\boldsymbol{v}]}(\mathbf{r}).\end{aligned} \tag{180}$$

We can define the following map $\{\{\boldsymbol{\mathcal{N}}\},\{\boldsymbol{v}\}\} \leftrightarrow \{\boldsymbol{\rho}_{\text{eq}}^{[\boldsymbol{\mathcal{N}},\boldsymbol{v}]}\}$ which is a bijective map, in contradistinction to the $\{\boldsymbol{\rho}_{\text{gs}}^{[\boldsymbol{\mathcal{N}},\boldsymbol{v}]}\} \rightarrow \{\{\boldsymbol{\mathcal{N}}\},\{\boldsymbol{v}\}\}$ map which is a non-injective and surjective map. Combining the $\{\{\boldsymbol{\mathcal{N}}\},\{\boldsymbol{v}\}\} \leftrightarrow \{\boldsymbol{\rho}_{\text{eq}}^{[\boldsymbol{\mathcal{N}},\boldsymbol{v}]}\}$ map with the $\{\boldsymbol{\mathcal{N}}\} \leftrightarrow \{(\boldsymbol{a}^{[\boldsymbol{\mathcal{N}};\boldsymbol{v}]},\boldsymbol{b}^{[\boldsymbol{\mathcal{N}};\boldsymbol{v}]})\}$ map yields at 0K limit

$$\{\{\boldsymbol{\mathcal{N}}\},\{\boldsymbol{v}\}\} \leftrightarrow \left\{\left\{\left(\boldsymbol{a}^{[\boldsymbol{\mathcal{N}};\boldsymbol{v}]},\boldsymbol{b}^{[\boldsymbol{\mathcal{N}};\boldsymbol{v}]}\right)\right\},\left\{\boldsymbol{\rho}_{\text{eq},(I)}^{[\boldsymbol{v}]}\right\}\right\} \leftrightarrow \left\{\boldsymbol{\rho}_{\text{eq}}\left[\boldsymbol{\mathcal{N}};\boldsymbol{v}\right]\right\}. \tag{181}$$

Based on these maps, the *Hohenberg-Kohn theorem* for a system with the given $\boldsymbol{\mathcal{N}}$ and $\boldsymbol{v}$ at 0K limit can be formulated as follows:

*In the spin-canonical ensemble characterized by $\boldsymbol{\mathcal{N}} = (\mathcal{N},\mathcal{S})$ at 0K temperature, the 2vec equilibrium electron density $\boldsymbol{\rho}_{\text{eq}}(\mathbf{r})$ is determined uniquely by the 2vec potential $\boldsymbol{v}(\mathbf{r}) = (v_{\text{ext}}(\mathbf{r}), B_{\text{z}}(\mathbf{r}))$ and pair of real-number 2vecs $(\boldsymbol{a},\boldsymbol{b}) = ((a_{\text{N}},a_{\text{S}}),(b_{\text{N}},b_{\text{S}}))$ which are the 0K limits of the derivative w.r.t. $\boldsymbol{\mathcal{N}}$ of the average energy and the minus average entropy at equilibrium, respectively. The correct equilibrium density, for the given $\boldsymbol{v}(\mathbf{r})$ and $\boldsymbol{\mathcal{N}}$, minimizes the average electron energy and maximizes the entropy, which are the asymptotic characteristics of the spin Massieu function and the spin Helmholtz function in the energy and the entropy representation, respectively.*

The additional inequality for the entropy, Eq.(179) is consistent with the theory symmetrized with respect to the symmetry group of the Hamiltonian operator.[20] Thus necessity of using symmetry-dependent functionals and potentials arises not only due the formalism proposed there. The common application of the Kohn-Sham treatment of the lowest-energy state of each symmetry also requires symmetry-dependent functionals and potentials.[42-48]

## VII. CONCLUSIONS

In this work, the zero-temperature limit of the thermodynamic spin-density functional theory was investigated. After deriving properties of the entropy in the coarse-grained approach and applying this to the equilibrium density operator, it was shown that the "grouping" properties play an important role in the relations between different descriptions of the equilibrium state. In order to derive the 0K limit of the TSDFT, the characteristic functions of a macrostate were introduced and their zero-temperature limits were investigated. A detailed discussion of the spin-grand-canonical ensemble in the entropy and the energy representations was performed. It was shown for both representations that the map between the external sources (conjugate to variables $(\mathcal{N},\mathcal{S})$), $\boldsymbol{\alpha}=(\alpha_{\mathrm{N}},\alpha_{\mathrm{S}})$ or $\boldsymbol{\mu}=(\mu_{\mathrm{N}},\mu_{\mathrm{S}})$ and the variables $(\mathcal{N},\mathcal{S})$ is non-injective (many-to-one) and non-surjective (not all values of $(\mathcal{N},\mathcal{S})$ are accessible at 0K limit). To illustrate this behavior, the values of the average electron number and the average spin number as functions of $\boldsymbol{\mu}=-\beta^{-1}\boldsymbol{\alpha}$ for the carbon atom in the spin-grand-canonical ensemble at 0K limit were presented in Fig. 4 and 5. In the spin-canonical ensemble, where, in addition to the inverse temperature, the independent variables are the electron and spin numbers, the energy surface and the discontinuity pattern were investigated. The energy surface at 0K limit is a convex function of $\boldsymbol{\mathcal{N}}$ — a collection of the triangles and trapezoids joined at their edges (see Fig. 2 for illustrations). The polygon sides of the projection of this surface on $(\mathcal{N},\mathcal{S})$ were recognized as the discontinuity lines of the energy derivatives. Based on the macrostates contributions to the ensemble with a given $(\mathcal{N},\mathcal{S})$ at 0K limit, the discontinuity pattern in the vicinity of the $(\mathcal{N},\mathcal{S})$ point was identified. Knowing it (for the given external potential $\boldsymbol{v}$), the macrostate weights and the external sources conjugate to $(\mathcal{N},\mathcal{S})$ variables at 0K limit are accessible.

Finally, the maps between the state function variables at 0K limit were rigorously studied. The most fundamental conclusion which follows from this study is that at the 0K temperature only $\rho_{\mathrm{eq}}\left[\mathcal{N},\boldsymbol{v}\right]$ is simultaneously a minimizer of the density functional of the ground-state energy and a maximizer of the entropy. When the formalism for zero temperature is used, the electron-number density and the spin-number density, on which this functional depends, have to be the 0K limit of the 2vec equilibrium density associated with the given $\mathcal{N}$ and $\boldsymbol{v}$ in the spin-canonical ensemble. This is often overlooked in the zero-temperature limit of the spin-density functional theory.

## ACKNOWLEDGMENT

This work was partially supported by the Ministry of Science and Higher Education of Poland through Grant No. N204275939 and the International PhD Projects Programme of the Foundation for Polish Science, co-financed from European Regional Development Fund within Innovative Economy Operational Programme "Grants for innovation". I am grateful to A. Holas for helpful discussions. Special thanks are due to Professor Paul W. Ayers for his valuable comments and beneficial remarks on the manuscript.

## APPENDIX

The relation between the state function and the partition function is used to prove the identities concerning the weights of the macrostates. In the entropy representation, the state function is equal to the logarithm of the partition function, Eq.(I.11), e.g. in the spin grand canonical ensemble: $K^{[\beta,\alpha;v]} = \ln \Xi^{[\beta,\alpha;v]}$, Eq.(II.8). This relation is also true for the macrostate functions, $Y_{(I)}^{[\beta;v]} = \ln \Xi_{(I)}^{[\beta;v]}$. The macrostate weight factor defined in Eq.(22) can be rearranged as

$$\begin{aligned}\omega_{(I)}^{[\beta,\alpha;v]} &= \frac{\operatorname{Tr} \hat{g}_{(I)} \exp\left(-\beta \hat{H}\left[\boldsymbol{v}\right] - \boldsymbol{\alpha}\hat{\boldsymbol{\mathcal{N}}}\right)}{\operatorname{Tr} \exp\left(-\beta \hat{H}\left[\boldsymbol{v}\right] - \boldsymbol{\alpha}\hat{\boldsymbol{\mathcal{N}}}\right)} \\ &= \frac{\exp\left(-\boldsymbol{\alpha}\boldsymbol{\mathcal{N}}_{(I)}\right) \operatorname{Tr} \hat{g}_{(I)} \exp\left(-\beta \hat{H}\left[\boldsymbol{v}\right]\right)}{\operatorname{Tr} \exp\left(-\beta \hat{H}\left[\boldsymbol{v}\right] - \boldsymbol{\alpha}\hat{\boldsymbol{\mathcal{N}}}\right)} = \frac{\exp\left(-\boldsymbol{\alpha}\boldsymbol{\mathcal{N}}_{(I)}\right) \Xi_{(I)}^{[\beta;v]}}{\Xi^{[\beta,\alpha;v]}}.\end{aligned} \tag{A1}$$

The global partition function can be rewritten as

$$\begin{aligned}\Xi^{[\beta,\alpha;v]} &= \operatorname{Tr} \exp\left(-\beta \hat{H}\left[\boldsymbol{v}\right] - \boldsymbol{\alpha}\hat{\boldsymbol{\mathcal{N}}}\right) = \sum_K \exp\left(-\beta E_K^{[v]} - \boldsymbol{\alpha}\hat{\boldsymbol{\mathcal{N}}}_K\right) \\ &= \sum_I \exp\left(-\beta E_{(I)}^{[\beta;v]} + S_{(I)}^{[\beta;v]} - \boldsymbol{\alpha}\boldsymbol{\mathcal{N}}_{(I)}\right) = \sum_I \exp\left(Y_{(I)}^{[\beta;v]} - \boldsymbol{\alpha}\boldsymbol{\mathcal{N}}_{(I)}\right) \\ &= \exp\left(-\beta E^{[\beta,\alpha;v]} + S^{[\beta,\alpha;v]} - \boldsymbol{\alpha}\boldsymbol{\mathcal{N}}^{[\beta,\alpha;v]}\right) = \exp\left(K^{[\beta,\alpha;v]}\right).\end{aligned} \tag{A2}$$

After inserting this $\Xi^{[\beta,\alpha;v]} = \sum_I \exp\left(Y_{(I)}^{[\beta;v]} - \boldsymbol{\alpha}^{\mathrm{T}}\hat{\boldsymbol{\mathcal{N}}}_{(I)}\right)$ and $\Xi_{(I)}^{[\beta;v]} = \exp\left(Y_{(I)}^{[\beta;v]}\right)$ into (A1), we find Eq.(81) proven. Other identities, similar to Eq.(81), can be obtained using the previous steps, but in the energy representation the state function have to be multiplied by $-\beta$, e.g. $\ln \Xi^{[\beta,\mu;v]} = -\beta \Omega^{[\beta,\alpha;v]}$ and $\boldsymbol{\alpha} = -\beta\boldsymbol{\mu}$ inserted. Using these relations, Eq.(89) and Eq.(107) can be proven.

The weights in the spin canonical ensemble are obtained by replacing $\boldsymbol{\alpha}\hat{\boldsymbol{\mathcal{N}}}$ with $\breve{\boldsymbol{\alpha}}^{[\beta,\boldsymbol{\mathcal{N}};\boldsymbol{v}]}(\hat{\boldsymbol{\mathcal{N}}}-\boldsymbol{\mathcal{N}})$ in Eq.(A1). Here the identity $\breve{\boldsymbol{\mathcal{N}}}\left[\beta,\boldsymbol{\alpha}=\breve{\boldsymbol{\alpha}}\left[\beta,\boldsymbol{\mathcal{N}};\boldsymbol{v}\right];\boldsymbol{v}\right]=\boldsymbol{\mathcal{N}}$ holds (see discussion after Eq.(II.15)). The weights can be written as

$$
\begin{aligned}
\omega_{(I)}^{[\beta,\boldsymbol{\mathcal{N}};\boldsymbol{v}]} &= \frac{\exp\left(-\boldsymbol{\alpha}\boldsymbol{\mathcal{N}}_{(I)}\right)\Xi_{(I)}^{[\beta;\boldsymbol{v}]}}{\Xi^{[\beta,\boldsymbol{\mathcal{N}};\boldsymbol{v}]}} = \frac{\exp\left(Y_{(I)}^{[\beta;\boldsymbol{v}]}-\breve{\boldsymbol{\alpha}}^{[\beta,\boldsymbol{\mathcal{N}};\boldsymbol{v}]}\left(\boldsymbol{\mathcal{N}}_{(I)}-\boldsymbol{\mathcal{N}}\right)\right)}{\exp Y^{[\beta,\boldsymbol{\mathcal{N}};\boldsymbol{v}]}} \\
&= \frac{\exp\left(-\beta\left(A_{(I)}^{[\beta;\boldsymbol{v}]}-\boldsymbol{\mu}^{[\beta,\boldsymbol{\mathcal{N}};\boldsymbol{v}]}\left(\boldsymbol{\mathcal{N}}_{(I)}-\boldsymbol{\mathcal{N}}\right)\right)\right)}{\exp\left(-\beta A^{[\beta,\boldsymbol{\mathcal{N}};\boldsymbol{v}]}\right)} \\
&= \frac{\exp\left(-\beta\left(A_{(I)}^{[\beta;\boldsymbol{v}]}-\boldsymbol{\mu}^{[\beta,\boldsymbol{\mathcal{N}};\boldsymbol{v}]}\left(\boldsymbol{\mathcal{N}}_{(I)}-\boldsymbol{\mathcal{N}}\right)\right)\right)}{\sum_J \exp\left(-\beta\left(A_{(J)}^{[\beta;\boldsymbol{v}]}-\boldsymbol{\mu}^{[\beta,\boldsymbol{\mathcal{N}};\boldsymbol{v}]}\left(\boldsymbol{\mathcal{N}}_{(J)}-\boldsymbol{\mathcal{N}}\right)\right)\right)} \quad .
\end{aligned} \tag{A3}
$$

The logarithmic form of this equation,

$$
\ln\omega_{(I)}^{[\beta,\boldsymbol{\mathcal{N}};\boldsymbol{v}]} = -\beta\left(\left(A_{(I)}^{[\beta;\boldsymbol{v}]}-A^{[\beta,\boldsymbol{\mathcal{N}};\boldsymbol{v}]}\right)-\breve{\boldsymbol{\mu}}^{[\beta,\boldsymbol{\mathcal{N}};\boldsymbol{v}]}\left(\boldsymbol{\mathcal{N}}_{(I)}-\boldsymbol{\mathcal{N}}\right)\right), \tag{A4}
$$

is used in Eq.(119).

From Eq.(134) follows immediately

$$
\breve{\boldsymbol{\alpha}}^{[\beta,\boldsymbol{\mathcal{N}};\boldsymbol{v}]} = -\beta\breve{\boldsymbol{\mu}}^{[\beta,\boldsymbol{\mathcal{N}};\boldsymbol{v}]} = -\beta\boldsymbol{a}^{[\beta,\boldsymbol{\mathcal{N}};\boldsymbol{v}]} - \boldsymbol{b}^{[\beta,\boldsymbol{\mathcal{N}};\boldsymbol{v}]}. \tag{A5}
$$

The large-$\beta$ (low-temperature) behavior of $\breve{\boldsymbol{\alpha}}^{[\beta,\boldsymbol{\mathcal{N}};\boldsymbol{v}]}$ can be conveniently described in terms of a slant asymptote. In general, a function $f[\beta]$ is represented by a slant asymptote in the large $\beta$ region ( the "asymptotic equal" notation $\simeq$ is used)

$$
f[\beta] \simeq f_{\mathrm{a}}\beta + f_{\mathrm{b}}, \tag{A6}
$$

when the following two finite limits exist

$$
\lim_{\beta\to\infty}\frac{f[\beta]}{\beta} = f_{\mathrm{a}} \qquad \text{and} \qquad \lim_{\beta\to\infty}\left(f[\beta]-f_{\mathrm{a}}\beta\right) = f_{\mathrm{b}}. \tag{A7}
$$

This means that the difference between $f[\beta]$ and its slant asymptote tends to zero when $\beta \to \infty$. The constants $f_{\mathrm{a}}$ and $f_{\mathrm{b}}$ (i.e. independent of $\beta$ objects) are named the tangent and the shift of the slant asymptote.

The following slant asymptote of $\breve{\boldsymbol{\alpha}}^{[\beta,\mathcal{N};\boldsymbol{v}]}$ follows from (A5)

$$\breve{\boldsymbol{\alpha}}^{[\beta,\mathcal{N};\boldsymbol{v}]} \simeq -\beta \boldsymbol{a}^{[\mathcal{N};\boldsymbol{v}]} - \boldsymbol{b}^{[\mathcal{N};\boldsymbol{v}]}, \tag{A8}$$

provided the finite limits exist: $\lim_{\beta\to\infty} \boldsymbol{a}^{[\beta,\mathcal{N};\boldsymbol{v}]} = \boldsymbol{a}^{[\mathcal{N};\boldsymbol{v}]}$ and $\lim_{\beta\to\infty} \mathbf{b}^{[\beta,\mathcal{N};\boldsymbol{v}]} = \boldsymbol{b}^{[\mathcal{N};\boldsymbol{v}]}$. However $\lim_{\beta\to\infty} \breve{\boldsymbol{\alpha}}^{[\beta,\mathcal{N};\boldsymbol{v}]} = \pm\boldsymbol{\infty}$. In the case of $\breve{\boldsymbol{\mu}}^{[\infty,\mathcal{N};\boldsymbol{v}]}$, the following limit exist

$$\breve{\boldsymbol{\mu}}^{[\beta,\mathcal{N};\boldsymbol{v}]} = \boldsymbol{a}^{[\beta,\mathcal{N};\boldsymbol{v}]} + \beta^{-1}\boldsymbol{b}^{[\beta,\mathcal{N};\boldsymbol{v}]} \simeq \boldsymbol{a}^{[\mathcal{N};\boldsymbol{v}]} + \beta^{-1}\boldsymbol{b}^{[\mathcal{N};\boldsymbol{v}]} \doteq \boldsymbol{a}^{[\mathcal{N};\boldsymbol{v}]}. \tag{A9}$$

As noted in the Introduction, the symbol $\doteq$ is used as the short notation for the 0K limit, $f[\beta] \doteq g$ means $\lim_{\beta\to\infty} f[\beta] = g$.

At low-temperatures, Eq.(A3) can be rewritten as

$$\omega_{(I)}^{[\beta,\mathcal{N};\boldsymbol{v}]} = \frac{\exp\left(-\beta\left(E_{(I)}^{[\boldsymbol{v}]} - \boldsymbol{a}^{[\mathcal{N};\boldsymbol{v}]}\boldsymbol{\mathcal{N}}_{(I)}\right) + \boldsymbol{b}^{[\mathcal{N};\boldsymbol{v}]}\boldsymbol{\mathcal{N}}_{(I)} + S_{(I)}^{[\boldsymbol{v}]}\right)}{\sum_J \exp\left(-\beta\left(E_{(J)}^{[\boldsymbol{v}]} - \boldsymbol{a}^{[\mathcal{N};\boldsymbol{v}]}\boldsymbol{\mathcal{N}}_{(J)}\right) + \boldsymbol{b}^{[\mathcal{N};\boldsymbol{v}]}\boldsymbol{\mathcal{N}}_{(J)} + S_{(J)}^{[\boldsymbol{v}]}\right)}. \tag{A10}$$

## FIGURE LEGENDS

**FIG. 1.** Mnemonic cube – the state functions with their arguments are at the corners, transformations of variables are along edges: L – Legendre, MP – Masieu-Planck, EC – equivalence condition. The state functions in the entropy representations – the front surface, in the energy representations – the back surface, in the spin-grand-canonical ensemble – the left surface, in the spin-canonical ensemble – the right surface.

**FIG. 2.** An illustrative example – the energy surface at 0K limit (left panel) and its projection on the $(\mathcal{N},\mathcal{S})$ plane (right panel) for the carbon-like atom and its ions, in absence of magnetic field. All data from aug-cc-pVTZ/PBE calculations, in atomic units.

**FIG. 3.** The mapping of $(\mu_{\mathrm{N}},\mu_{\mathrm{S}})$ into $(\mathcal{N},\mathcal{S})$ at $\beta = 1/300\mathrm{K}$. The functions $\mathcal{N} = \mathcal{N}[\beta,\mu_{\mathrm{N}},\mu_{\mathrm{S}}]$ and $\mathcal{S} = \mathcal{S}[\beta,\mu_{\mathrm{N}},\mu_{\mathrm{S}}]$ are represented as axis values and a coloring scale, respectively, on the left panel, and inversely on the right panel. An illustrative example, see Fig. 2 for its details.

**FIG. 4.** The zero-temperature limit of the average electron and spin numbers. The functions $\mathcal{N} = \mathcal{N}[\mu_{\mathrm{N}},\mu_{\mathrm{S}}]$ and $\mathcal{S} = \mathcal{S}[\mu_{\mathrm{N}},\mu_{\mathrm{S}}]$ are represented as axis values and a coloring scale, respectively, on the left panel. The projection of $\mathcal{S} = \mathcal{S}[\mu_{\mathrm{N}},\mu_{\mathrm{S}}]$ on the $(\mu_{\mathrm{N}},\mu_{\mathrm{S}})$ plane on the right plane. An illustrative example, see Fig. 2 for its details.

**FIG. 5.** The zero-temperature limit of the average electron and spin numbers. The functions $\mathcal{S} = \mathcal{S}[\mu_{\mathrm{N}},\mu_{\mathrm{S}}]$ and $\mathcal{N} = \mathcal{N}[\mu_{\mathrm{N}},\mu_{\mathrm{S}}]$ are represented as axis values and a coloring scale, respectively, on the left panel. The projection of $\mathcal{N} = \mathcal{N}[\mu_{\mathrm{N}},\mu_{\mathrm{S}}]$ on the $(\mu_{\mathrm{N}},\mu_{\mathrm{S}})$ plane on the right plane. An illustrative example, see Fig. 2 for its details.

**FIG. 6.** The zero-temperature limit of the external sources conjugate to $(\mathcal{N},\mathcal{S})$ in the energy representation. The function $\mu_{\mathrm{N}} = \mu_{\mathrm{N}}[\mathcal{N},\mathcal{S}]$ on the left panel and the function $\mu_{\mathrm{S}} = \mu_{\mathrm{S}}[\mathcal{N},\mathcal{S}]$ on the right panel, represented as axis values and a coloring scale. An illustrative example, see Fig. 2 for its details.

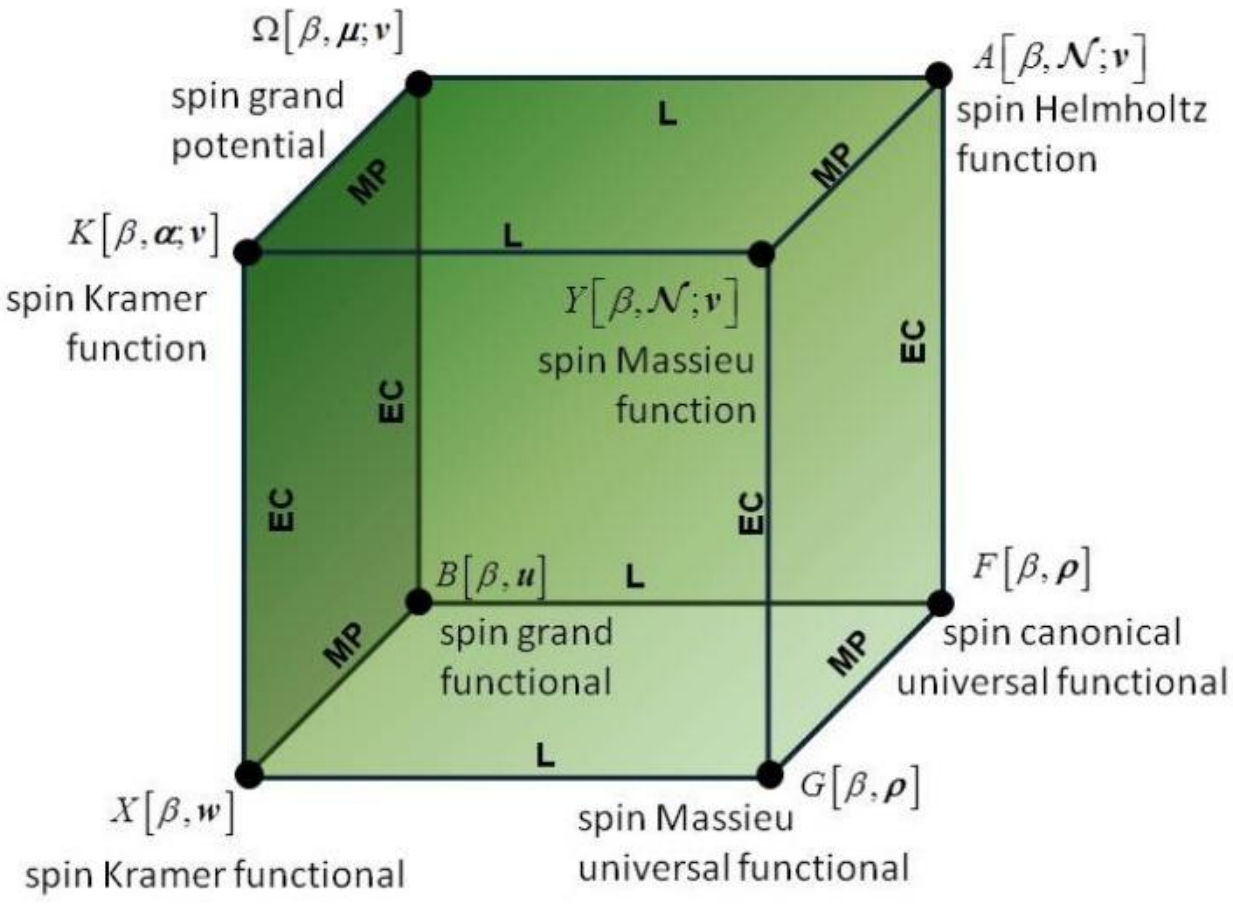

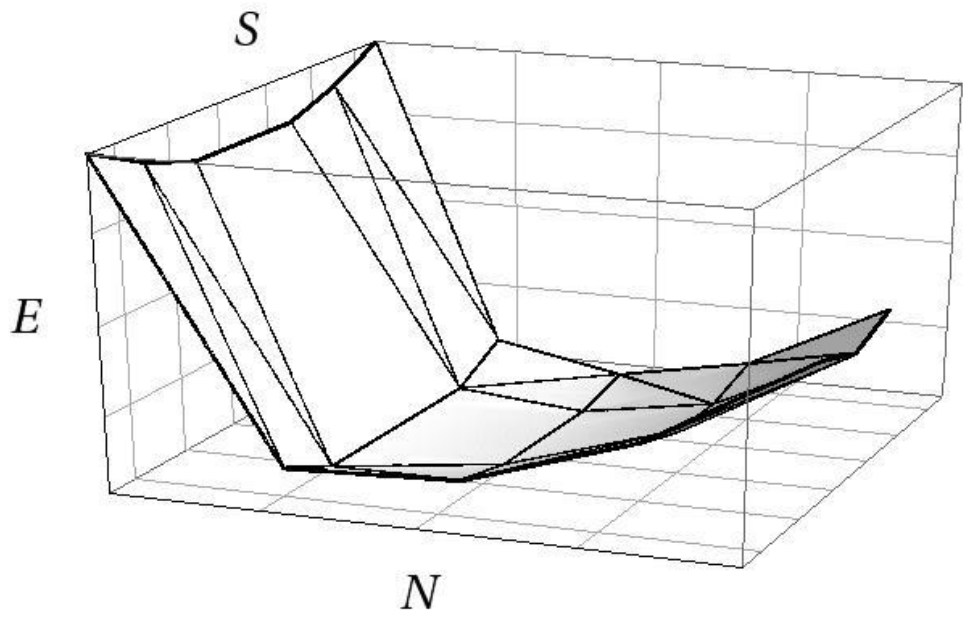
S
E
N

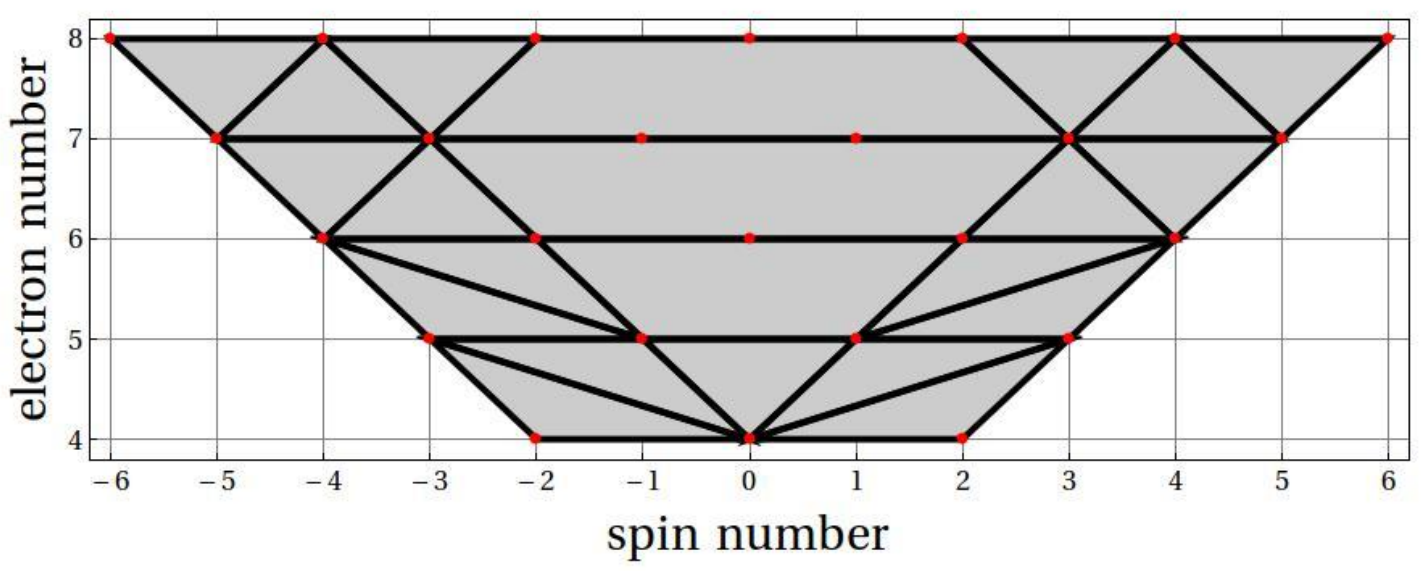
electron number
spin number
8
7
6
5
4
−6 −5 −4 −3 −2 −1 0 1 2 3 4 5 6

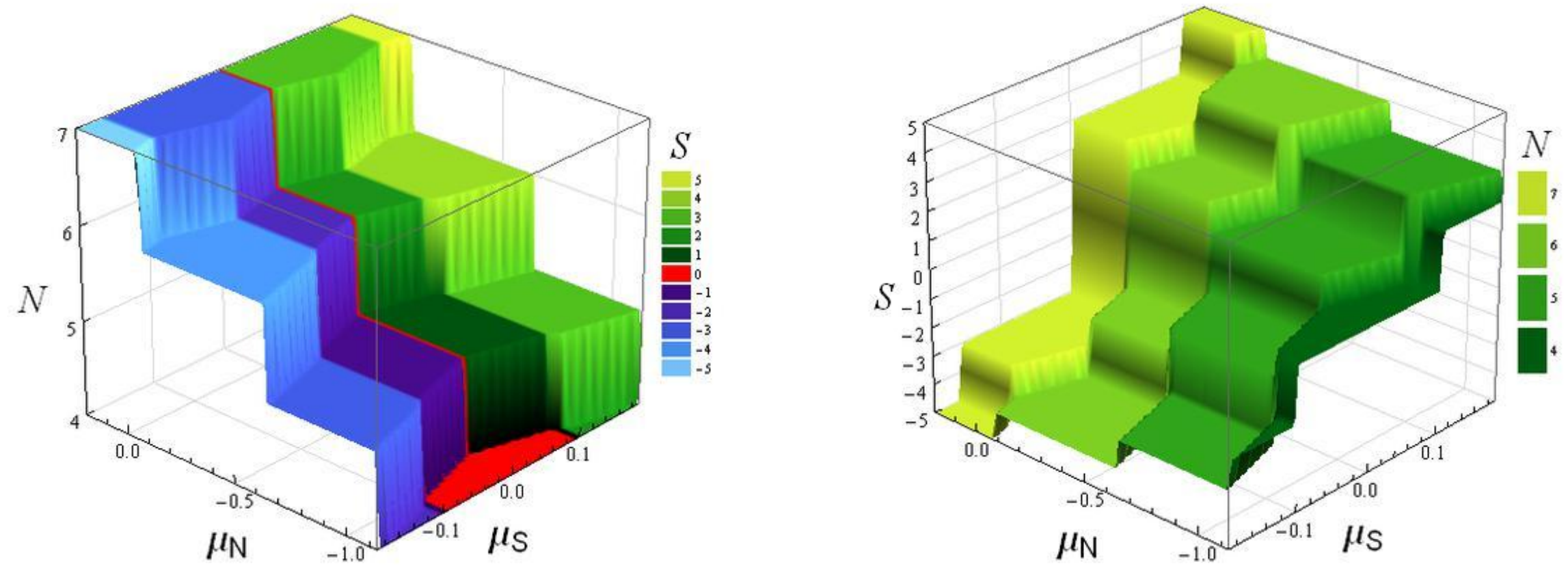

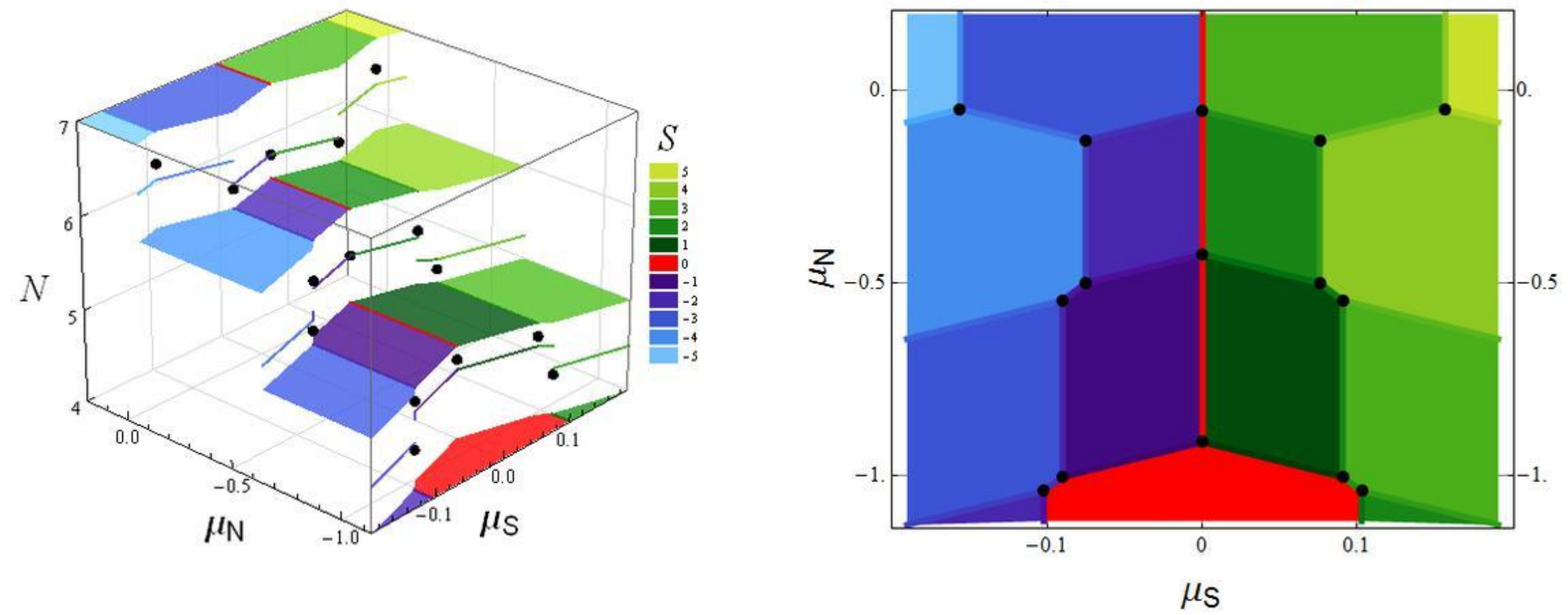

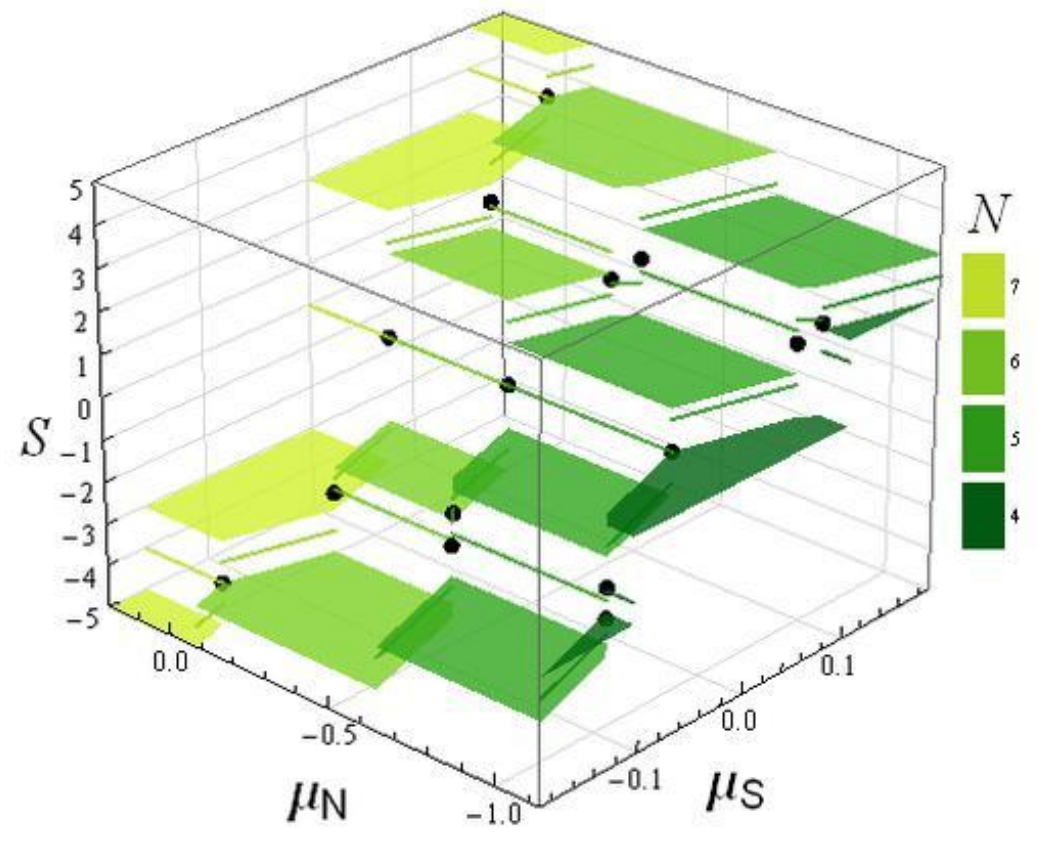

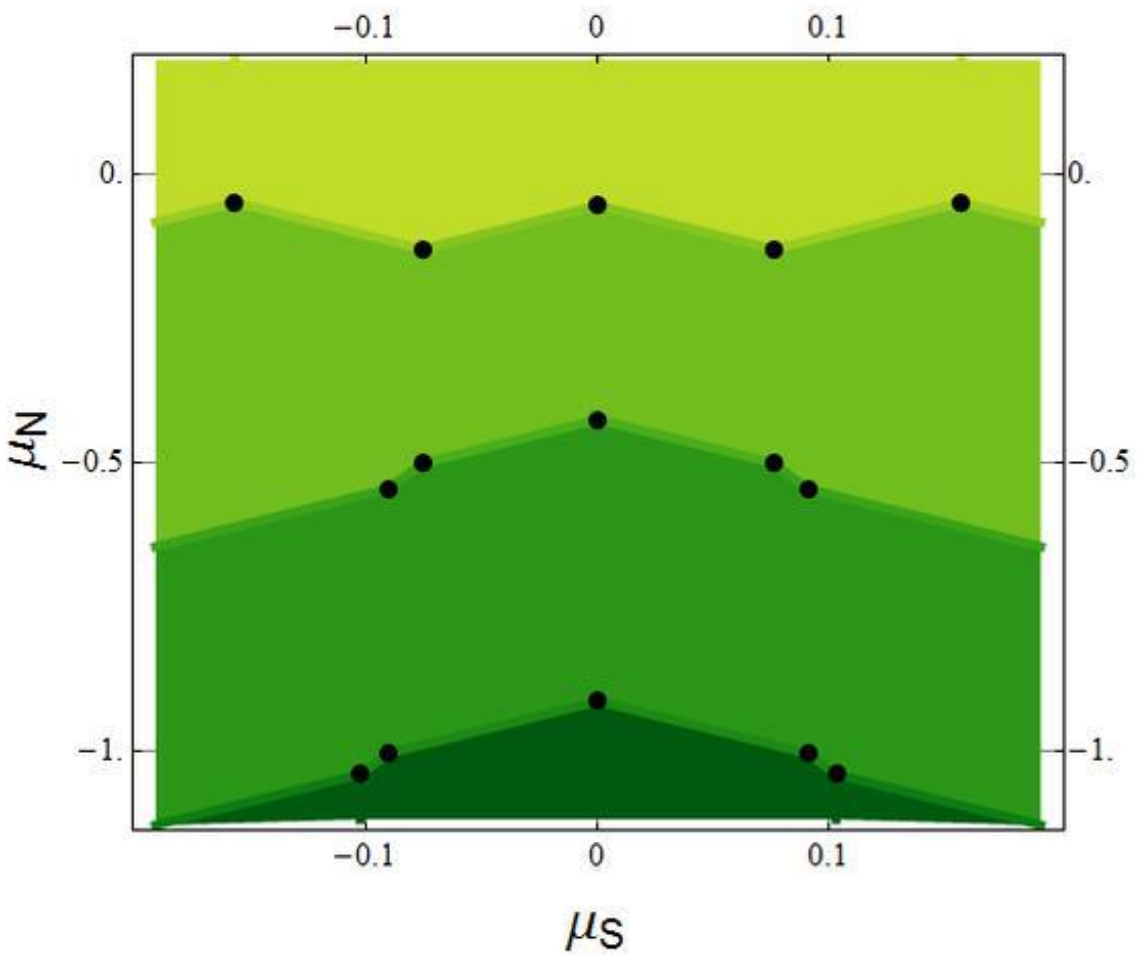
μN
μS

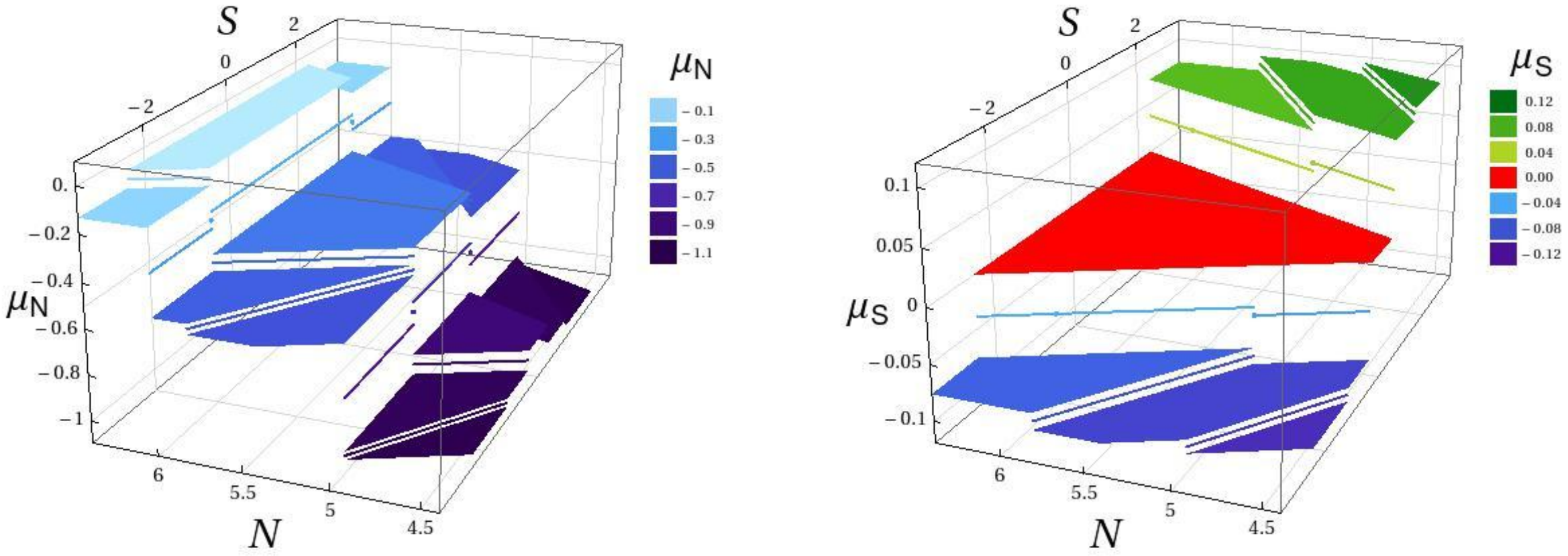